\documentclass [PhD] {uclathes}
\newcommand{ \be }{\begin{equation}}
\newcommand{ \ee }{\end{equation}}
\newcommand{ \bea }{\begin{eqnarray}}
\newcommand{ \eea }{\end{eqnarray}}

\usepackage{revsymb}
\usepackage{rotate}
\usepackage{epsfig}
\usepackage{eepic}
\usepackage{epic}
\usepackage{graphics}
\usepackage{doublespace}
\usepackage{multicols}
\usepackage{multirow}
\usepackage{amsmath}

\title{
Kaon and Lambda Production at Intermediate $p_T$: Insights into
the Hadronization of the Bulk Partonic Matter Created in Au+Au
Collisions at RHIC
}

\author         {Paul Richard Sorensen}
\department     {Physics}
\degreeyear     {2003}

\member  {Charles A.\ Whitten\ Jr.} \chair  {Huan Z.\ Huang}
\member {Charles\ Buchanan} \member {An\ Yin} \member {Nu\ Xu}

\dedication{ \textsl{To my family, whose unceasing hard work,
effort and diligence have given me advantages that made it
possible for me to pursue this goal and to my friends, whose
support throughout my life gave me the strength, confidence and
character to complete it. } }

\acknowledgments{ I feel especially compelled to thank all the
members --- past and present --- of the UCLA Relativistic
Heavy-Ion Group who, together, have created an environment
conducive to learning, excellence and success.  The work presented
in this thesis builds on the pioneering work of its members. A
graduate student cannot hope for better advisors and colleagues.
It is also my great fortune to have worked closely with the
Relativistic Nuclear Collisions group at Lawrence Berkeley
National Laboratory. Their assistance and vision was of great
value to me. The completion of this thesis was made possible by
the many members of the STAR collaboration and the RHIC operations
group who work tirelessly for the advancement of the field of
relativistic heavy-ion collisions. I am grateful that I've had the
opportunity to be a part of this historic scientific endeavor.

I would like to thank my parents Marilyn and Herb Sorensen, and my
brother Jon Sorensen for the support they offered during my
graduate career. I would also like to acknowledge the support and
sacrifices made by B\'eatrice Sorensen and all who were close to
me while I dedicated myself to the completion of my graduate
degree.

}

\vitaitem{8 Nov. 1972}{Born, Portland, Oregon, USA.}

\vitaitem{1996}{B.S. Physics \vspace{-0.5em}\\University of
Nebraska - Lincoln \vspace{-0.5em}\\Lincoln, NE}

\vitaitem {1997 -- 2000}{Teaching
Assistant\vspace{-0.5em}\\Department of
Physics\vspace{-0.5em}\\University of California - Los Angeles}

\vitaitem{1998}{M.S. Physics\vspace{-0.5em}\\University of
California - Los Angeles\vspace{-0.5em}\\Los Angeles, CA}

\vitaitem{1998 -- 2000}{Teaching Assistant
Coordinator\vspace{-0.5em}\\Department of
Physics\vspace{-0.5em}\\University of California - Los Angeles}

\vitaitem{2000 -- 2003}{Graduate Research
Assistant\vspace{-0.5em}\\Intermediate Energy and Relativistic
Heavy-Ion Group\vspace{-0.5em}\\University of California - Los
Angeles}

\publication{
\begin{singlespace}
\noindent
P.~Sorensen\\
Azimuthal Anisotropy of $K_S^0$ and $\Lambda + \overline{\Lambda}$ Production at Mid-rapidity from Au + Au Collisions at $\sqrt{s_{_{NN}}} = 130$~GeV.\\
\textit{J.\ Phys.\ G: Nucl.\ Part.\ Phys.} \textbf{28}:2089, 2002.
\end{singlespace}
}

\publication{
\begin{singlespace}
\noindent
P.~Sorensen \textit{et al.}\\
Azimuthal Anisotropy of $K_S^0$ and $\Lambda$ Production at Mid-rapidity from Au + Au Collisions at $\sqrt{s_{_{NN}}} = 130$~GeV.\\
\textit{AIP Conf. Proc.} \textbf{631}:366, 2002.
\end{singlespace}
}

\publication{
\begin{singlespace}
\noindent
K.~\v{S}afa\v{r}\'{i}k, I.~Kraus, J.~Newby and P.~Sorensen\\
Particle Tracking.\\
\textit{AIP Conf. Proc.} \textbf{631}:377, 2002.
\end{singlespace}
}

\publication{
\begin{singlespace}
\noindent
C.~Adler \textit{et al.}\\
Mid-rapidity $\Lambda$ and $\overline{\Lambda}$ Production in Au + Au Collisions at $\sqrt{s_{_{NN}}} = 130$~GeV.\\
\textit{Phys.\ Rev.\ Lett.} \textbf{89}:092301, 2002.
\end{singlespace}
}

\publication{
\begin{singlespace}
\noindent
C.~Adler \textit{et al.}\\
Azimuthal Anisotropy of $K_S^0$ and $\Lambda + \overline{\Lambda}$ Production at Mid-rapidity from Au + Au Collisions at $\sqrt{s_{_{NN}}} = 130$~GeV.\\
\textit{Phys.\ Rev.\ Lett.} \textbf{89}:132301, 2002.
\end{singlespace}
}

\publication{
\begin{singlespace}
\noindent
C.~Adler \textit{et al.}\\
$K^*(892)^0$ Production in Relativistic Heavy-Ion Collisions at $\sqrt{s_{_{NN}}} = 130$~GeV.\\
 \textit{Phys.\ Rev.\ C} \textbf{66}:061901, 2002.
\end{singlespace}
}

\publication{
\begin{singlespace}
\noindent
C.~Adler \textit{et al.}\\
Elliptic Flow from Two- and Four-particle Correlations in Au + Au Collisions at $\sqrt{s_{_{NN}}} = 130$~GeV.\\
\textit{Phys.\ Rev.\ C} \textbf{66}:034904, 2002.
\end{singlespace}
}

\publication{
\begin{singlespace}
\noindent
C.~Adler \textit{et al.}\\
Coherent $\rho^0$ Production in Ultra-peripheral Heavy-Ion Collisions.\\
\textit{Phys.\ Rev.\ Lett.} \textbf{89}:270302, 2002.
\end{singlespace}
}

\publication{
\begin{singlespace}\noindent
C.~Adler \textit{et al.}\\
Azimuthal Anisotropy and Correlations in the Hard Scattering Regime at RHIC.\\
\textit{Phys.\ Rev.\ Lett.} \textbf{90}:032301, 2003.
\end{singlespace}
}

\publication{
\begin{singlespace}\noindent
C.~Adler \textit{et al.}\\
Kaon Production and Kaon to Pion Ratio in Au + Au Collisions at $\sqrt{s_{_{NN}}} = 130$~GeV.\\
e-print:nucl-ex/0206008.
\end{singlespace}
}

\publication{
\begin{singlespace}\noindent
C.~Adler \textit{et al.}\\
Disappearance of Back-to-back High $p_T$ Hadron Correlations in Central Au + Au Collisions at $\sqrt{s_{_{NN}}} = 200$~GeV.\\
\textit{Phys.\ Rev.\ Lett.} \textbf{90}:082302, 2003.
\end{singlespace}
}

\publication{
\begin{singlespace}\noindent
J.~Adams \textit{et al.}\\
Narrowing of the Balance Function with Centrality in Au + Au Collisions at $\sqrt{s_{_{NN}}} = 130$~GeV.\\
\textit{Phys.\ Rev.\ Lett.} \textbf{90}:172301, 2003.
\end{singlespace}
}

\publication{
\begin{singlespace}\noindent
J.~Adams \textit{et al.}\\
Strange Anti-particle to Particle Ratios at Mid-rapidity in $\sqrt{s_{_{NN}}} = 130$~GeV Au + Au Collisions.\\
\textit{Phys.\ Lett.} \textbf{B567}:167-174, 2003.
\end{singlespace}
}

\publication{
\begin{singlespace}\noindent
P.~Sorensen \textit{et al.}\\
Particle Dependence of Elliptic Flow in Au + Au Collisions at $\sqrt{s_{_{NN}}} = 200$~GeV.\\
\textit{J.\ Phys.\ G} \textbf{29}:1-6, 2003.
\end{singlespace}
}

\publication{
\begin{singlespace}\noindent
J.~Adams \textit{et al.}\\
Transverse Momentum and Collision Energy Dependence of High $p_T$ Hadron Suppression in Au + Au Collisions at Ultra-relativistic Energies.\\
\textit{Phys.\ Rev.\ Lett.} \textbf{90}:082302, 2003.
\end{singlespace}
}

\publication{
\begin{singlespace}\noindent
J.~Adams \textit{et al.}\\
Particle Dependence of Azimuthal Anisotropy and Nuclear Modification of Particle Production at Moderate $p_T$ in Au + Au Collisions at $\sqrt{s_{_{NN}}} = 200$~GeV.\\
submitted to \textit{Phys.\ Rev.\ Lett.}, e-print:nucl-ex/0306007.
\end{singlespace}
}

\publication{
\begin{singlespace}\noindent
J.~Adams \textit{et al.}\\
Evidence from d + Au Measurements for Final-state Suppression of High $p_T$ Hadrons in Au + Au Collisions at RHIC.\\
\textit{Phys.\ Rev.\ Lett.} \textbf{91}:072304, 2003.
\end{singlespace}
}

\publication{
\begin{singlespace}\noindent
J.~Adams \textit{et al.}\\
Three-pion HBT Correlations in Relativistic Heavy-Ion Collisions from the STAR Experiment.\\
submitted to \textit{Phys.\ Rev.\ Lett.}, e-print:nucl-ex/0306028.
\end{singlespace}
}

\publication{
\begin{singlespace}\noindent
J.~Adams \textit{et al.}\\
Rapidity and Centrality Dependence of Proton and Anti-proton Production from Au + Au Collisions at $\sqrt{s_{_{NN}}} = 130$~GeV.\\
submitted to \textit{Phys.\ Rev.\ Lett.}, e-print:nucl-ex/0306029.
\end{singlespace}
}

\publication{
\begin{singlespace}\noindent
J.~Adams \textit{et al.}\\
$\rho^0$ Production and Possible Modification in Au + Au and p + p Collisions at $\sqrt{s_{_{NN}}} = 200$~GeV.\\
submitted to \textit{Phys.\ Rev.\ Lett.}, e-print:nucl-ex/0307023.
\end{singlespace}
}

\publication{
\begin{singlespace}\noindent
J.~Adams \textit{et al.}\\
Multi-strange Baryon Production in Au + Au Collisions at $\sqrt{s_{_{NN}}} = 130$~GeV.\\
submitted to \textit{Phys.\ Rev.\ Lett.}, e-print:nucl-ex/0307024.
\end{singlespace}
}

\presentation{
\begin{singlespace}\noindent
P. Sorensen\\
Measurement of Elliptic Flow of $K_S^0$ and $\Lambda$ at $\sqrt{s_{_{NN}}} = 130$~GeV.\\
\textit{Strange Quarks in Matter--Frankfurt, Germany}, 2001.
\end{singlespace}
}

\presentation{
\begin{singlespace}\noindent
P. Sorensen\\
Measurement of Elliptic Flow of $K_S^0$ and $\Lambda$ at $\sqrt{s_{_{NN}}} = 130$~GeV.\\
\textit{Pan American Advanced Studies Institute on New States of
Matter in Hadronic Interactions--Campos do Jordao, Brazil}, 2002.
\end{singlespace}
}

\presentation{
\begin{singlespace}\noindent
P. Sorensen\\
Measurement of Elliptic Flow of $K_S^0$ and $\Lambda$ at $\sqrt{s_{_{NN}}} = 130$~GeV. \\
\textit{American Physical Society--Davis, California} 2002.
\end{singlespace}
}

\presentation{
\begin{singlespace}\noindent
P. Sorensen\\
Recent Results on $K_S^0$ and $\Lambda$ Production at Large Transverse Momentum. \\
\textit{INT Winter Workshop--Seattle, Washington} 2002.
\end{singlespace}
}

\presentation{
\begin{singlespace}\noindent
P. Sorensen\\
Azimuthal Anisotropy of $K_S^0$ and $\Lambda$ Production at Mid-rapidity from Au+Au Collisions at $\sqrt{s_{_{NN}}} = 200$~GeV.\\
\textit{Strange Quarks in Matter--Atlantic Beach, North Carolina}
2003.
\end{singlespace}
}

\presentation{
\begin{singlespace}\noindent
P. Sorensen\\
Azimuthal Anisotropy and Suppression of $K_S^0$ and $\Lambda + \overline{\Lambda}$ Production at High Transverse Momentum in Au + Au Collisions at $\sqrt{s_{_{NN}}} = 200$~GeV.\\
\textit{American Physical Society--Philadelphia, Pennsylvania}
2003.
\end{singlespace}
}

\presentation{
\begin{singlespace}\noindent
P. Sorensen\\
Particle Production at Low, Intermediate and High $p_T$: What We've Learned About Heavy-Ion Collisions, and Hadronization of Bulk Partonic Matter from Measurements of $K_S^0$ and $\Lambda$ Production.\\
\textit{Oak Ridge National Laboratory Seminar--Oak Ridge,
Tennessee} 2003.
\end{singlespace}
}

\abstract { Measurements of identified particles over a broad
transverse momentum $p_T$ range may provide particularly strong
evidence for the existence of a thermalized partonic state in
heavy-ion collisions (\textit{i.e.} a quark-gluon plasma). Of
particular interest are the centrality dependence and the
azimuthal anisotropy in the yield of baryons and mesons at
intermediate $p_T$. The first measurements of $v_2$ --- an
event-by-event azimuthal anisotropy parameter --- and the nuclear
modification factor $R_{CP}$ for mid-rapidity $K_S^0$ and
$\Lambda+\overline{\Lambda}$ production in Au+Au collisions at
ultra-relativistic energy are presented. The $K_S^0$, $\Lambda$,
and $\overline{\Lambda}$ candidates are selected based on
characteristics of their decays in the STAR Time Projection
Chamber (TPC). A statistical treatment is used to extract
$v_2(p_T)$ and $R_{CP}(p_T)$ from their invariant mass
distributions. These measurements establish the particle type
dependence of $v_2$ and $R_{CP}$ in the kinematic region $0.4 <
p_T < 6.0$ and $|y| < 1.0$.

In the low $p_T$ region ($p_T < 1.0$~GeV/c) the $v_2$ values for
different particles are increasing with $p_T$ and follow a mass
dependence similar to that expected from hydrodynamical models of
Au+Au collisions --- where, at a given $p_T$, the particle with
the larger mass will have a smaller $v_2$.  At higher $p_T$
however, $v_2$ of the heavier $\Lambda$ hyperon continues to
increase while $v_2$ of the lighter $K_S^0$ meson saturates at
$v_2 \sim 0.13$ for $2.0 < p_T < 5.0$~GeV/c. At intermediate $p_T$
the $v_2$ of $K_S^0$ and $\Lambda+\overline{\Lambda}$ are shown to
follow a number-of-constituent-quark scaling with
$\frac{v_2^{kaon}(p_T/2)}{2} \approx
\frac{v_2^{lambda}(p_T/3)}{3}$.

The binary collision scaled centrality ratio $R_{CP}$ shows that
$\Lambda+\overline{\Lambda}$ production at intermediate $p_T$
increases more rapidly with system size than kaon production: This
is consistent with a scenario where multi-parton dynamics play an
important role in particle production. At $p_T \approx 5.5$~GeV/c
$\Lambda+\overline{\Lambda}$, $K_S^0$, and charged hadron
production are all suppressed by a similar amount: a factor of
three below expectations from binary nucleon-nucleon collision
scaling (\textit{i.e.} $R_{CP} \approx 0.33$). This $p_T$ value
establishes the extent to which the centrality dependent
enhancement of baryon production persists.

The particle-type dependence of $v_2$ and $R_{CP}$ provides a
stringent test for models of heavy-ion collisions.  In particular
the larger values of $\Lambda+\overline{\Lambda}$ $v_2$ compared
to their smaller suppression manifested in $R_{CP}$ suggests that
for $p_T < 4.0$~GeV/c a particle production mechanism beyond the
framework of energy loss and fragmentation exists in central Au+Au
collisions. The particle- and $p_T$-dependence of $v_2$, and
$R_{CP}$ are consistent, however, with expectations based on the
hadronization of a bulk partonic matter by coalescence or
recombination. As such, the constituent-quark-number scaled $v_2$
reflects the anisotropy established in a partonic stage and
provides strong evidence for the existence of a quark-gluon plasma
in Au+Au collisions at RHIC.

}

\begin {document}
\graphicspath{images/}

\makeintropages         

\chapter{Introduction to Relativistic Heavy-Ion Collisions}
\indent

By colliding heavy nuclei at relativistic energies scientists are
able to test the nature of nuclear matter at high temperature and
density, to produce conditions similar to those thought prevalent
in the early universe, and to search for previously unstudied
states of nuclear matter. In this chapter, we discuss the
essential components of the theory thought to govern heavy-ion
collisions, and we introduce the analysis topics that will be
presented in this thesis.

\section{QCD--Asymptotic Freedom and Confinement}

\begin{figure}[htpb]
\centering\mbox{
\includegraphics[width=0.60\textwidth]{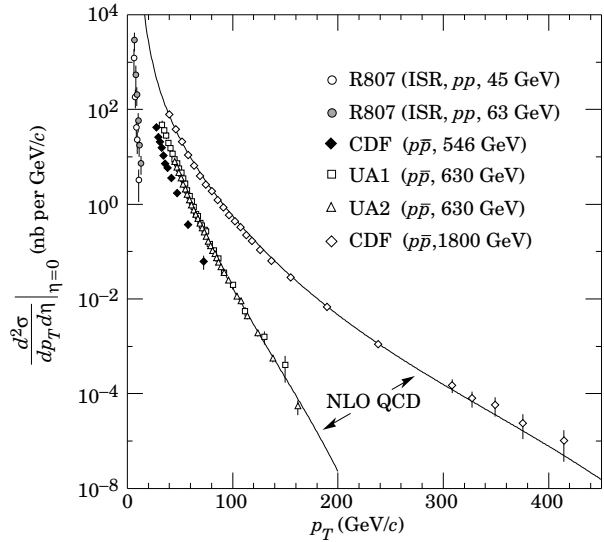}}
\caption[Jet production in proton (anti-)proton collisions]
{Differential cross-sections for single jet production at
pseudo-rapidity $\eta = 0$ as a function of the jet transverse
momentum $p_T$ in proton (anti-)proton collisions. Jets are
somewhat collimated sprays of particles produced when quarks or
gluons collide, transfer (and carry away) a lot of momentum, and
then fragment into a spray of hadrons. The curves represent pQCD
calculations for the collisions at center-of-mass energy
$\sqrt{s}=630$ and 1800 GeV.} \label{fig:jets}
\end{figure}

Matter is made of leptons, quarks, and force mediators. Quarks,
the building blocks of nucleons (and all hadronic matter), carry a
property analogous to electric charge called \textit{color}. The
theory that describes the forces between colored objects and that
is thought to be the correct theory for strong interaction
is called quantum chromodynamics (QCD). In QCD, just as the
electromagnetic force is carried by photons, the color force (or
strong force) is carried by gluons. However, whereas photons carry
no electric charge, gluons do carry color charge so they can
interact directly with each other, and whereas the electrodynamic
coupling constant $\alpha = \frac{1}{137}$, the strong coupling
constant $\alpha_s$ can be larger than one. As a consequence of
the direct gluon-gluon coupling the \textit{effective} coupling
constant for the strong force becomes smaller at shorter
distances. This effect is known as \textit{asymptotic freedom}.
Asymptotic freedom means the force between quarks is stronger at
larger distances so quarks seem to remain confined to a small
($\sim$1 fm$^3$) region in colorless groups of two (mesons) or
three (baryons). Because the effective strong coupling is only
small at short distances, perturbation theory can only be used
with QCD for interactions involving large momentum transfers
(\textit{i.e. hard processes}). Although perturbative QCD (pQCD)
is in very good agreement with experimental observations involving
hard processes (see Figure~\ref{fig:jets} for example~\cite{pdg}),
it cannot be used to calculate QCD predictions for the processes
that dominate the universe at present: soft processes

Explicit QCD Lagrangian calculations of the force between quarks
can only be made in the limits of weak and strong coupling. To
understand the behavior of colored objects where pQCD is not a
valid approximation, physicists rely on numerical path integrals
of the QCD Lagrangian on a discretized lattice in four-dimensional
Euclidean space-time. It is the formulation of Lattice QCD with a
strong coupling approximation that first demonstrated how quarks
are confined~\cite{lattice_wilson}.

In principle, the lattice formulation of QCD can be used to
perform numerical calculations for all physical regimes.  In
practice, however, there are regimes where approximations used to
simplify the calculations fail and the computations become
technically very challenging.

\section{Deconfined Quark Matter}

\begin{figure}[htpb]
\centering\mbox{
\includegraphics[width=0.60\textwidth]{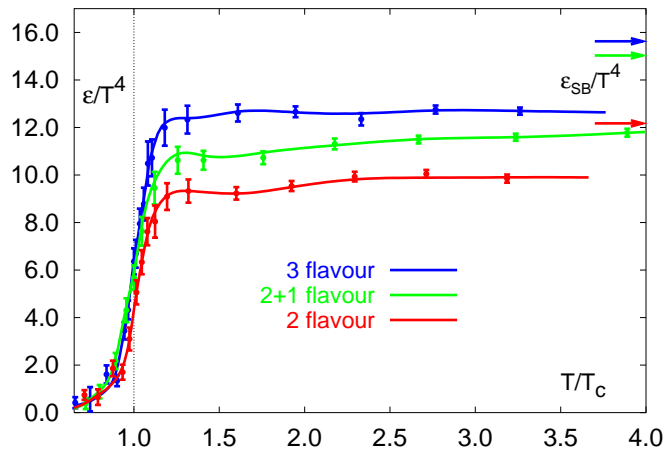}}
\caption[Lattice calculations of energy density] {The energy
density in QCD from lattice calculations. When the temperature $T$
reaches the critical temperature $T_c$, the number of degrees of
freedom rapidly rises indicating that quarks and gluons become
relevant degrees of freedom. The arrows represent the
Stefan-Boltzmann values for asymptotically high temperature.}
\label{fig:lattice}
\end{figure}

In the strong coupling regime the energy required to separate two
quarks increases linearly with the distance between them. As a
result, we have never observed deconfined quarks: a deconfined
quark is taken as one that can move in a volume much larger than
the volume of a proton. Recent advances in the formulation of
thermodynamical lattice QCD at finite temperature and density
however, suggests that when sufficiently high temperature and
density are reached, quarks become effectively deconfined.
Figure~\ref{fig:lattice}~\cite{Karsch:2001vs} shows that the ratio
of the energy density scaled by $T^4$ (where $T$ is the system
temperature) $\epsilon/T^4$ quickly increases at a critical
temperature $T_C$. The magnitude of $\epsilon/T^4$ reflects the
number of degrees of freedom in the thermodynamic system. The rise
corresponds to a transition in the system to a state where the
quarks and gluons have become relevant degrees of freedom.

The idea of a new state of matter where deconfined quarks and
gluons are the relevant degrees of freedom is not new. In 1973,
shortly after asymptotic freedom was shown to arise from QCD
theory~\cite{Gross:1973id,Politzer:1973fx}, deconfined quark
matter was postulated as the true state of nuclear matter at high
energy density at the center of neutron
stars~\cite{Collins:1975ky}:
\begin{quote}
A neutron has radius of about 0.5--1 fm, and so has a density of
about $8 \times 10^{14}$~gm/cm$^3$, whereas the central density of
a neutron star can be as much as $10^{16}-10^{17}$~gm/cm$^3$.  In
this case, one must expect the hadrons to overlap, and their
individuality to be confused.  Therefore, we suggest that there is
a phase change, and that nuclear matter at such high densities is
a quark soup.
\end{quote}
Later, in the fall of 1974, at a workshop on heavy-ion collisions,
T.D. Lee discussed the need for a physics program to study quark
matter~\cite{Lee:1975kn}:
\begin{quote}
Hitherto, in high-energy physics we have concentrated on
experiments in which we distribute a higher and higher amount of
energy into a region with smaller and smaller dimensions. In order
to study the question of ``vacuum,'' we must turn to a different
direction; we should investigate some ``bulk'' phenomena by
distributing high energy over a relatively large volume.
\end{quote}

\begin{figure}[htpb]
\centering\mbox{
\includegraphics[width=0.60\textwidth]{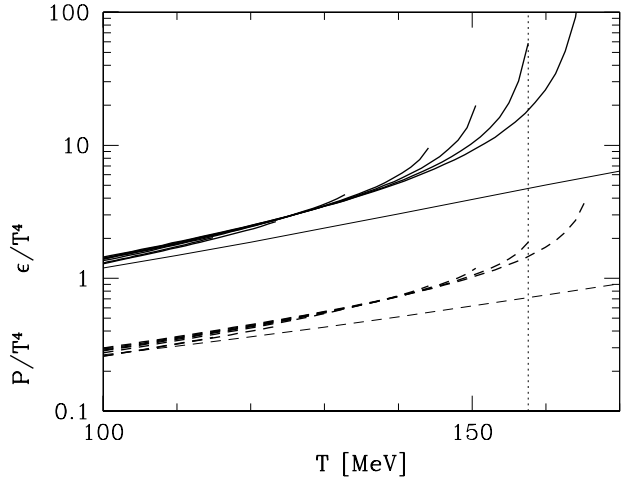}}
\caption[Statistical hadronization model] {The energy density
$\epsilon$ and pressure $P$, scaled by $T^4$ from a statistical
model~\cite{Rafelski:2002ga}. The various lines show results for
different hadron mass spectra. The results show an increase in the
energy density degrees of freedom at a critical temperature near
$T = 158$~MeV.} \label{fig:stat_model}
\end{figure}
Not all conceptualizations of the cross-over from hadronic degrees
of freedom to a new form of matter relied on QCD or the knowledge
of quarks. In 1951, Pomeranchuk postulated an upper limit to the
temperature of hadronic matter based on the finite size of
hadrons~\cite{Pomeranchuk:1951}.  In the late sixties, Hagedorn's
approach involving a self-similar hadronic resonance composition
pointed to a similar limit~\cite{Hagedorn:1965st}. We now believe
these limits reflect a transition to a state of matter with quarks
and gluons as deconfined constituents. Figure~\ref{fig:stat_model}
shows the scaled energy density $\epsilon/T^4$ and scaled pressure
$P/T^4$ derived from a statistical model of a hadronic
gas~\cite{Rafelski:2002ga}.

\section{Goals of Heavy-Ion Physics}

The creation and study of bulk matter made of deconfined quarks
and gluons (\textit{i.e.} a quark-gluon plasma or QGP) was one of
the prime motivations for building the Relativistic Heavy-Ion
Collider (RHIC). The interaction of high-energy, colliding beams
of heavy nuclei generates matter of extreme density and
temperature. The temperatures and densities reached are expected
to be similar to those thought to have prevailed in the very early
universe, prior to the formation of protons and neutrons. The
observation and study of matter in these conditions will be
relevant to the nuclear physics community, the astrophysics
community and the high-energy physics community. One also expects
this research to have a significant impact on many in the general
public since the nature of our universe at the earliest stages and
the transitions that produced the matter we are familiar with
today are interesting to most naturally curious or inquisitive
people.

By colliding large nuclei at high energy a window is opened onto
an asymptotic regime of QCD. The exploration of this region of the
QCD phase diagram is an exciting scientific endeavor. Many
questions will be addressed in heavy-ion research programs: %
How well does the system thermalize? %
In the early universe, how did matter hadronize? %
Is there a first order phase transition, second order phase transition or smooth cross-over? %
How is fragmentation affected by the dense system created in the collisions? %
What is the role of chiral symmetry breaking in the transition from %
deconfined partons to hadrons? %
The measurements presented here provide insight into how well the
matter created at RHIC thermalizes and how it subsequently
hadronizes.

Learning about dense nuclear matter is also important to the
astrophysics community. Heavy-ion physics can potentially provide
insight into the structure of neutron stars (\textit{i.e.} their
mass-radius relationship, their thermal evolution, their upper
mass limit). In addition, reaching a better understanding of dense
nuclear matter will help determine whether a new class of stars,
quark stars, are likely or unlikely to exist in our universe.

Perhaps the most exciting discoveries made will be those that are
least expected.  The heavy-ion collisions at RHIC constitute an
exploration into the unknown and one should be ready to be
surprised.  We do not know, for example, what, if any, exotic
states may be produced in the hadronization of bulk quark matter.
Candidates include multi-quark states, exotic atoms, and large
droplets of strange-quark matter.

\section{Experimental Observations}

\begin{figure}[htpb]
\centering\mbox{
\includegraphics[width=0.60\textwidth]{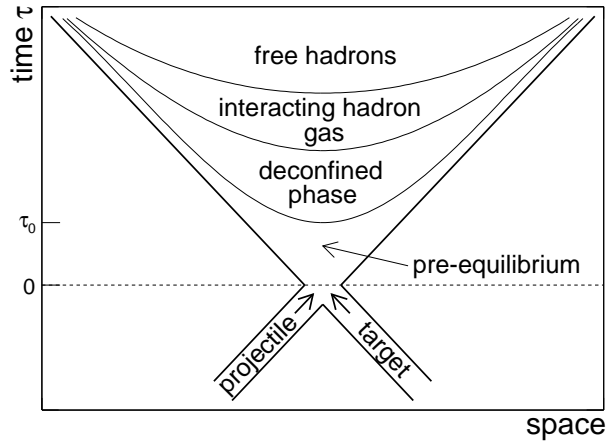}}
\caption[Collision evolution diagram] {A sketch of the expected
evolution of a relativistic heavy-ion collision.}
\label{fig:evolution}
\end{figure}
Figure~\ref{fig:evolution} depicts the space-time evolution of a
heavy-ion collision. Four possible stages of the evolution are
shown: a pre-equilibrium stage, an equilibrated-deconfined-parton
stage, an interacting-hadron-gas stage and finally a free-hadrons
stage.  The experiments at RHIC detect hadrons in the free-hadron
stage of the collision evolution. Probing the early stage of the
collision evolution with particles measured in this final stage is
a significant challenge. In this thesis we will present
measurements thought to be sensitive to the early part of the
collision evolution and to a possible deconfined-parton phase.

\subsection{Initial Conditions}

\begin{figure}[htpb]
\resizebox{.50\textwidth}{!}{\includegraphics{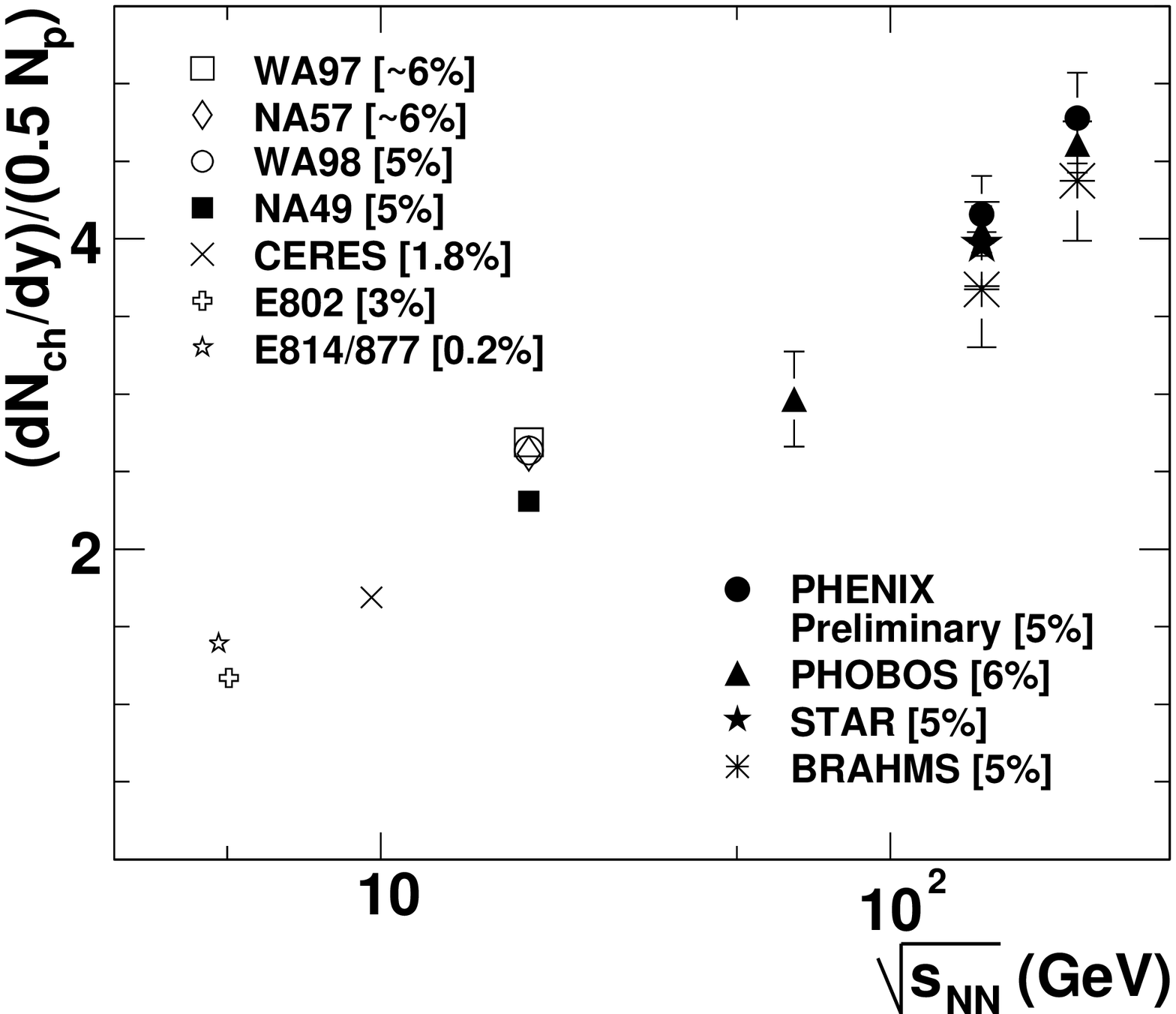}}
\resizebox{.50\textwidth}{!}{\includegraphics{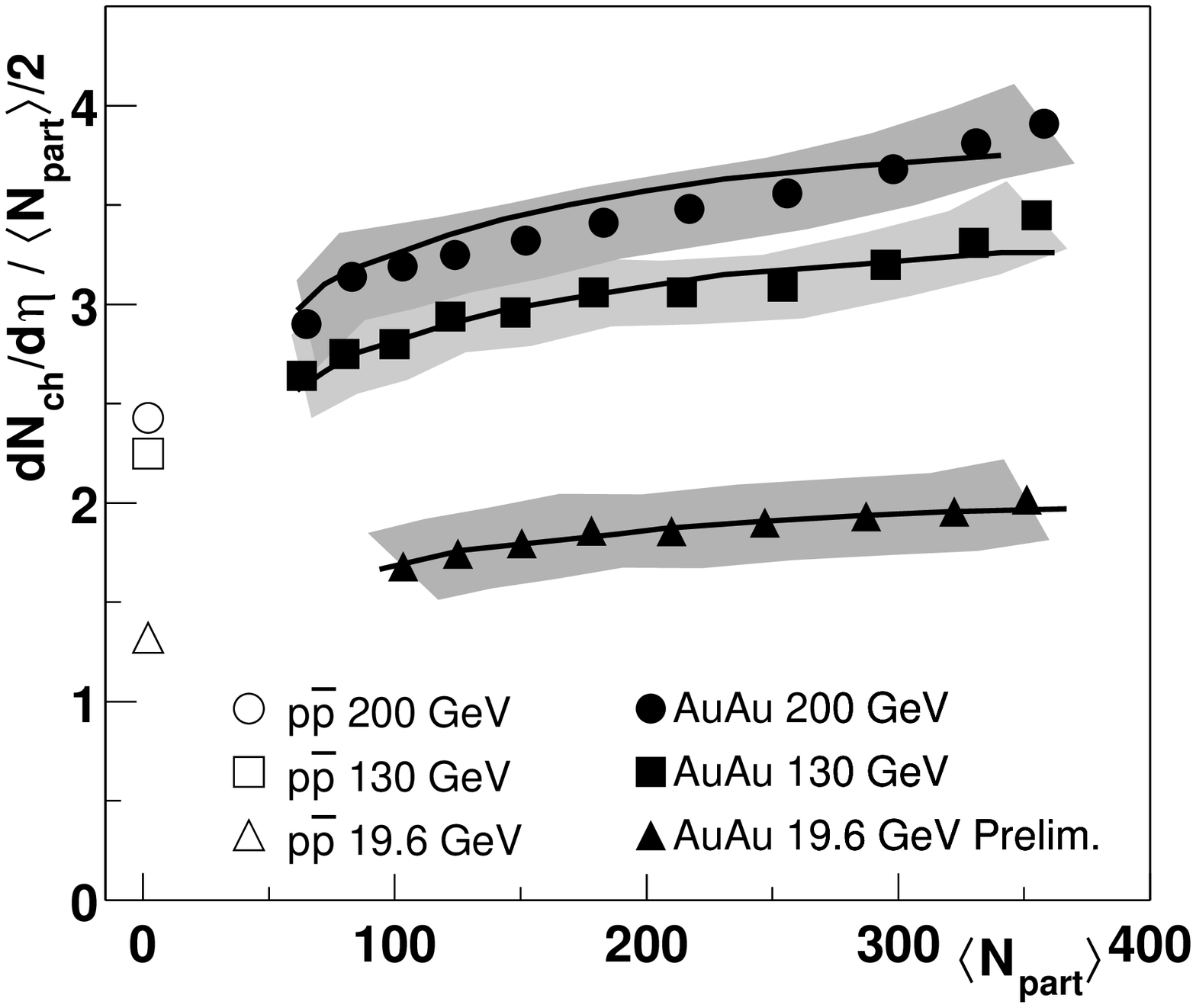}}
\caption[Rapidity density] {Left: Charged particle rapidity
density at mid-rapidity scaled by the number of participating
nucleon pairs $0.5 \times$ N$_{\mathrm{part}}$ versus
$\sqrt{s_{_{NN}}}$~\cite{Bazilevsky:2002fz}. Right: Scaled
pseudo-rapidity density for Au+Au and p+p collisions at
$\sqrt{s_{_{NN}}}=$ 19.6, 130, and 200~GeV versus
N$_{\mathrm{part}}$~\cite{Back:2002ft}.} \label{fig:dndy}
\end{figure}

It is not known a-priori that an equilibrated-deconfined-parton
phase can be created by colliding heavy ions in the laboratory.
The large energy densities reached in central collisions
(\textit{i.e.} head-on collisions) however, significantly surpass
estimates of the energy densities needed to reach the
deconfinement phase transition.  The initial energy density
$\epsilon$ of the produced medium can be determined using the
Bjorken estimate~\cite{Bjorken:1983qr}
\begin{equation}
\epsilon = \left(\frac{dN_{h}}{dy} \right)_{y=0}\frac{w_h}{\pi
R_{A}^{2}\tau_{0}},
\label{eq:bjorken}
\end{equation}
where $(\frac{dN_{h}}{dy})_{y=0}$ is the number of hadrons per
unit rapidity produced at mid-rapidity, $w_h$ is the average
energy of the hadrons, $R_A$ is the nuclear radius, and $\tau_0$
is the formation time of the medium.  The formation time is not
known but is generally taken to be approximately one fm/c. The
density of normal nuclear matter is approximately
$0.16$~GeV/fm$^3$. Lattice calculations predict that the phase
transition to deconfined quarks and gluons occurs near 1.0
GeV/fm$^3$. The Bjorken estimate for the initial energy density in
central Pb+Pb collisions with $\sqrt{s_{_{NN}}}=17$~GeV at the
CERN-SPS experiment is 3.5 GeV/fm$^3$~\cite{Satz:2002ku}. The
estimate from RHIC for central Au+Au collisions at
$\sqrt{s_{_{NN}}}=130$~GeV is 4.6 GeV/fm$^3$~\cite{Zajc:2001va}.
For the top RHIC energy ($\sqrt{s_{_{NN}}}=200$~GeV), $\epsilon
\sim 5.0$~GeV/fm$^3$~\cite{D'Enterria:2003rr}. These estimates of
$\epsilon$ far exceed the energy density thought necessary to
generate deconfined partonic matter. Given these large densities,
collective behavior due to multiple interactions is expected. An
important question to ask then is; are the interactions copious
enough and rapid enough to thermalize the dynamic and expanding
matter created in the laboratory? Answering this question will be
a challenge to the experiments at RHIC. Figure~\ref{fig:dndy}
(left) shows how the rapidity density per participating nucleon
pair increases as a function of
$\sqrt{s_{_{NN}}}$~\cite{Bazilevsky:2002fz}.

\begin{figure}[htpb]
\centering\mbox{
\includegraphics[width=0.60\textwidth]{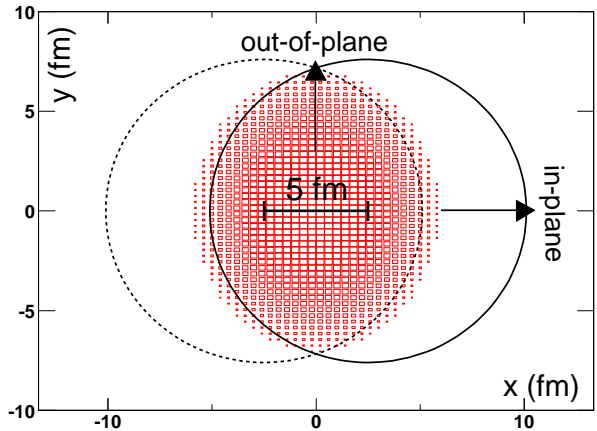}}
\caption[Au+Au collision overlap density] {The overlap density for
Au nuclei colliding off-axis. The beam directions are in and out
of the of the page. The two large circles represent the outline of
the incoming or outgoing Au nuclei. The impact parameter b---the
distance between the center of the two colliding nuclei in the
azimuthal plane---is 5 fm. The reaction plane is by the beam axis
and the vector connecting the centers of the two nuclei.}
\label{fig:overlap}
\end{figure}
In addition to the center-of-mass energy $\sqrt{s_{_{NN}}}$, the
initial conditions of heavy-ion collisions also depend on the
centrality of the collision. An off-axis nucleus-nucleus collision
will have a smaller number of participating nucleons
(N$_{\mathrm{part}}$), a smaller system size, and a smaller
initial energy density. Figure~\ref{fig:dndy} (right) shows the
rapidity density per participating nucleon pair versus
N$_{\mathrm{part}}$~\cite{Back:2002ft}. We also note that for
nuclei colliding off-axis, the overlap region will be asymmetric.
In Figure~\ref{fig:overlap} we plot the overlap density for Au
nuclei colliding with impact parameter b = 5 fm. A Woods-Saxon
distribution is used for the density profile of the Au nuclei.

Most observables in heavy-ion collisions are integrated over the
azimuthal angle and, as such, they are insensitive to the
azimuthal asymmetry of the initial source. In this thesis we
discuss measurements sensitive to the conversion of the initial
spatial anisotropy to a final momentum-space anisotropy. The
spatial anisotropy can be quantified by estimating the
eccentricity $\varepsilon$ of the initial source,
\begin{equation}
\varepsilon = \frac{\langle y^2-x^2\rangle}{\langle
y^2+x^2\rangle}.
\end{equation}
Extracting the mean eccentricity of the initial source $\langle
\varepsilon \rangle$ for a given centrality interval is helpful
for understanding the event-by-event anisotropy in the final state
momentum distributions. Analytic of initial eccentricities can be
found in Appendix~\ref{app:geo}.

\subsection{Event-by-event Momentum-space Anisotropy}\label{sec:v2}

\begin{figure}[htpb]
\centering\mbox{
\includegraphics[width=0.70\textwidth]{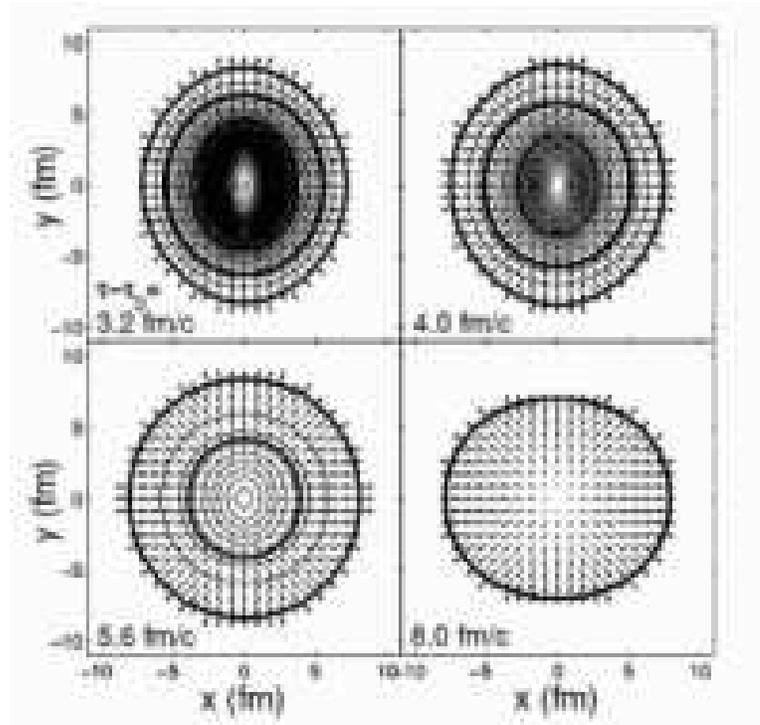}}
\caption[Source shape evolution] {The evolution of the source
shape is shown from a model where a heavy-ion collision is treated
as a hydrodynamic system~\cite{Kolb:2000sd}. The initial shape is
extended out-of-plane. By 8 fm/c after the formation time
($\tau-\tau_0$), the shape has deformed to an in-plane extended
source. In this model, the anisotropy in momentum-space measured
by $v_2$ is dominated by the early stages.} \label{fig:hydro_evol}
\end{figure}

Anisotropy in the distribution of a particle in momentum-space is
thought to be sensitive to the early stage of the collision
system. The anisotropy of the source will be largest immediately
after the collision occurs. As the system evolves, the spatial
anisotropy is converted by multiple interactions into a
momentum-space anisotropy. With time, the interactions will cause
the spatial distribution to become more isotropic. For this
reason, it's believed that the final azimuthal momentum-space
anisotropy is primarily built up in the initial moments of the
system's evolution. Figure~\ref{fig:hydro_evol} shows the
evolution of the source shape calculated from a model where the
collision system is described by hydrodynamic
equations~\cite{Kolb:2000sd}.

The azimuthal anisotropy of the transverse momentum distribution
for a particle can be described by expanding the azimuthal
component of the particle's momentum distribution in a Fourier
series,
\begin{equation}
\frac{d^3n}{p_Tdp_Tdyd\phi} = \frac{d^2n}{p_Tdp_Tdy}\left[1 +
2\sum_{\alpha}{v_{\alpha} \cos\left(\alpha[\phi-\Psi_{RP}
]\right)} \right].
\end{equation}
The harmonic coefficients, $v_{\alpha}$, are anisotropy
parameters, $p_T$, $y$, and $\phi$ are the respective transverse
momentum, rapidity, and azimuthal angle for the particle, and
$\Psi_{RP}$ is the reaction plane
angle~\cite{Poskanzer:1998yz}\footnote{The reaction plane is
defined by the beam axis and the vector connecting the centers of
the two colliding nuclei. For high energy collisions, in the
laboratory reference frame the Au nuclei are Lorentz-contracted
along the beam axis. As such, the vector connecting the colliding
nuclei is nearly perpendicular to the beam axis and the reaction
plane can be characterized by its azimuthal angle. }. The second
coefficient $v_2$ (customarily called \textit{elliptic flow})
measures the elliptic component of the anisotropy. Due to the
shape of the source created in an off-axis collision, $v_2$ is the
largest and most studied of the anisotropy parameters.
\begin{figure}[htpb] \centering\mbox{
\includegraphics[width=0.50\textwidth]{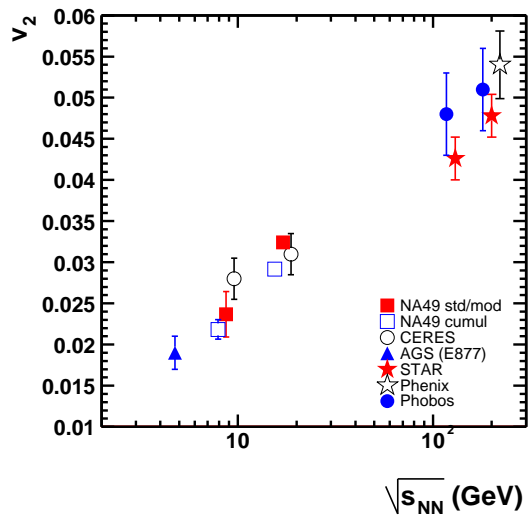}}
\caption[Elliptic flow excitation function] {The integrated
anisotropy parameter $v_2$ near mid-rapidity for mid-central
events (roughly 12--34\% central) plotted versus collision
energy~\cite{Alt:2003ab}.} \label{fig:v2_roots}
\end{figure}
Figure~\ref{fig:v2_roots} shows the energy dependence of $v_2$ for
charged particles near mid-rapidity.  For this energy range $v_2$
is positive and rising monotonically with the nucleon-nucleon
center of mass energy. At lower energies $v_2$ is
negative~\cite{Adamova:2002qx}.

\begin{figure}[htpb]
\resizebox{.5\textwidth}{!}{\includegraphics{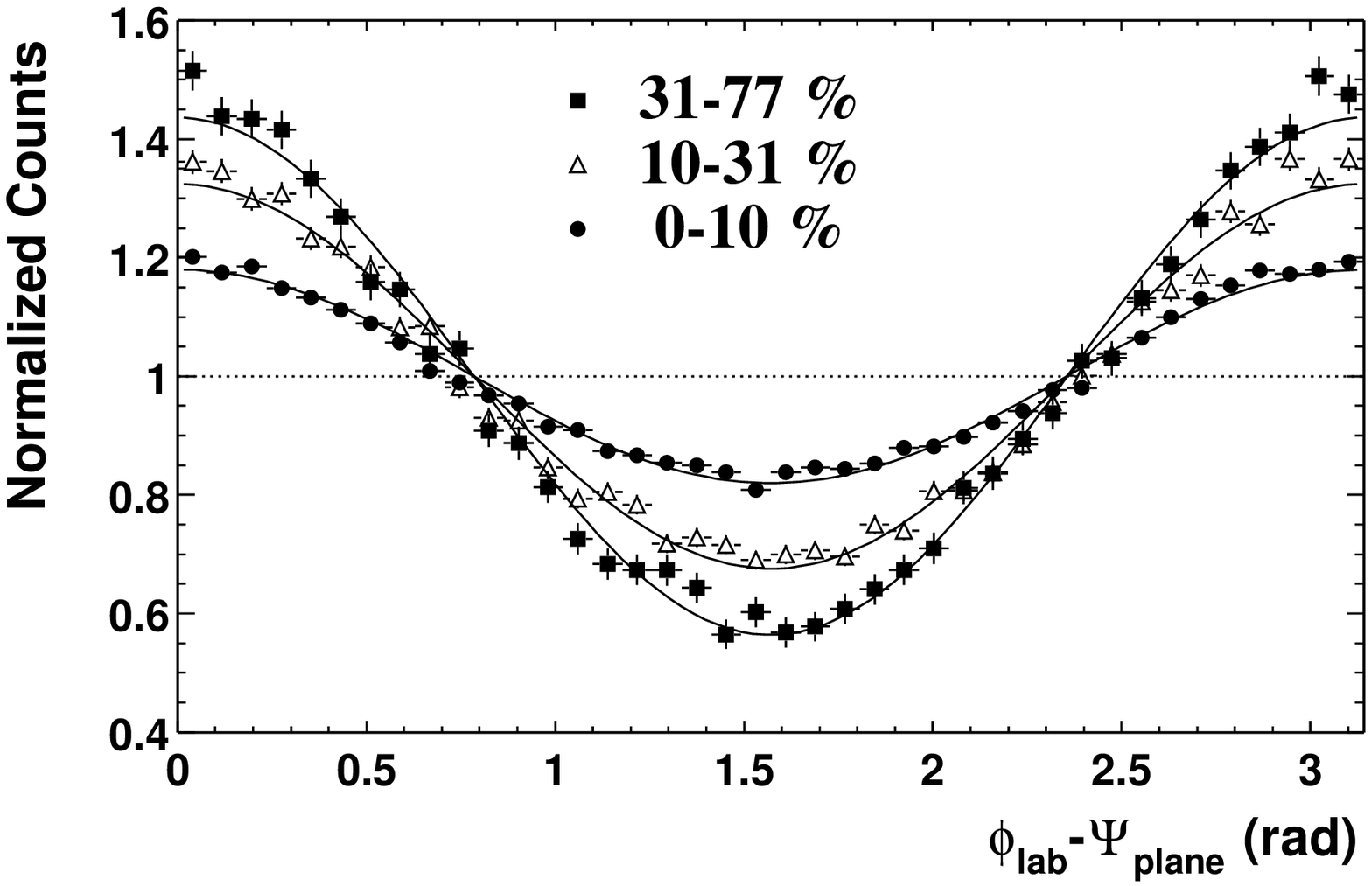}}
\resizebox{.5\textwidth}{!}{\includegraphics{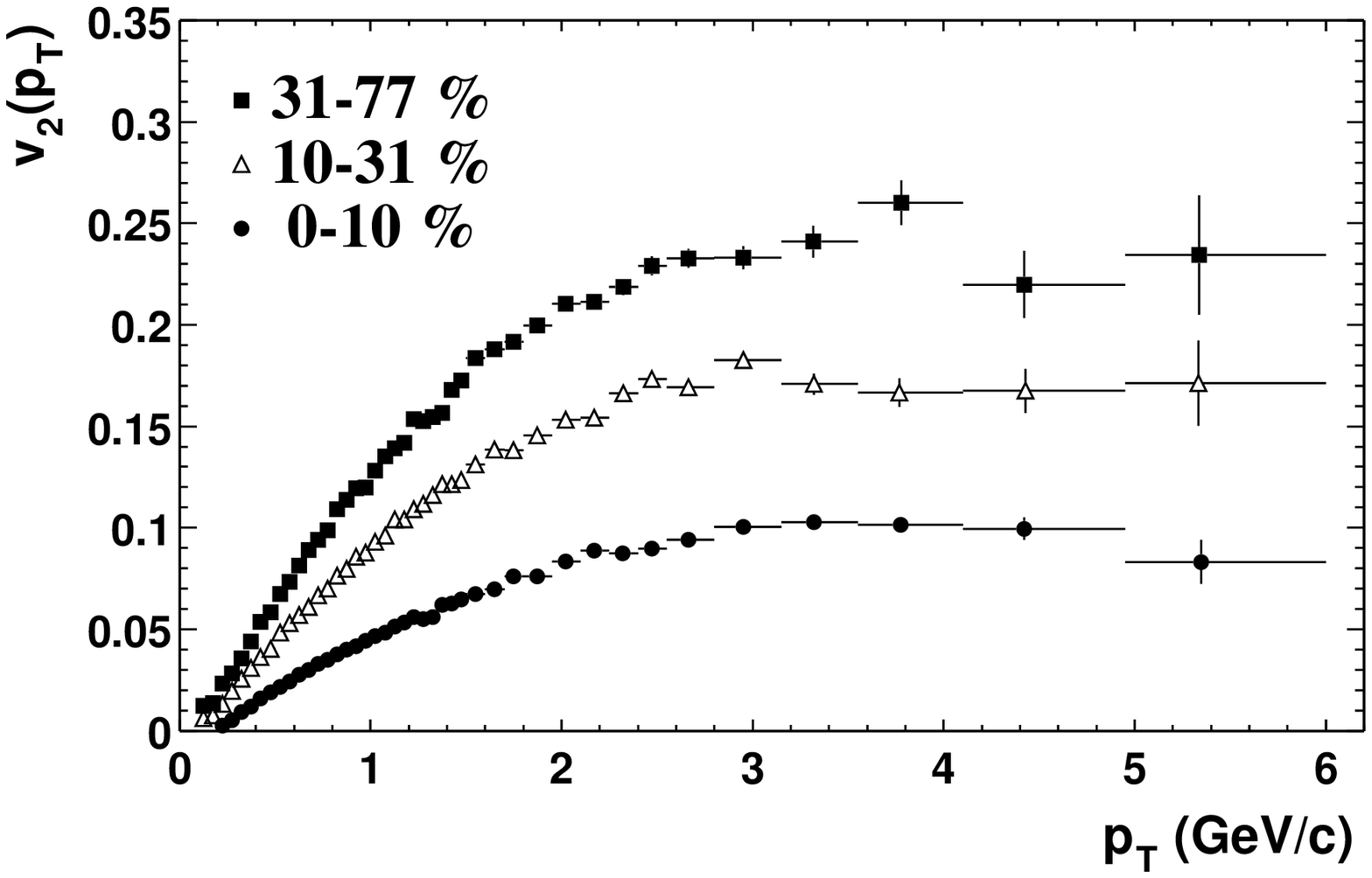}}
\caption[Charged hadron $v_2$] {Left: The distribution of charged
particles with $2.0 < p_T < 6.0$~GeV/c in the azimuthal plane from
Au+Au collisions at $\sqrt{s_{_{NN}}}=130$~GeV. The 0-10\%,
10-31\%, and 31-77\% represent different classes of centrality
where 0-10\% is the most central. Right: The differential $v_2$
for in three centrality intervals~\cite{Adler:2002ct}.}
\label{fig:v2_hadrons}
\end{figure}

Multiple interactions are necessary to develop a momentum-space
anisotropy from a coordinate-space anisotropy. If each
nucleon-nucleon collision is independent, the final momentum
distribution will represent a superposition of random collisions
and will therefore be isotropic. The azimuthal momentum-space
distribution of charged hadrons with $2.0 < p_T < 6.0$~GeV/c, for
three centrality intervals, is shown in
Figure~\ref{fig:v2_hadrons} (left). Figure~\ref{fig:v2_hadrons}
(right) shows how $v_2$ for charged hadrons changes with $p_T$
(differential $v_2$). The magnitude of $v_2$ is smallest in
central events because the initial eccentricity $\varepsilon$ is
smaller.

The large saturated values of $v_2$ at high $p_T$ are a surprising
result from RHIC. Although hydrodynamic models predict a monotonic
increase of differential $v_2$, it is believed that hydrodynamic
models must fail at higher values of $p_T$ where their assumptions
become invalid.  The measurement of a large $v_2$ at high $p_T$
gives rise to the question: ``how does the initial spatial
anisotropy manifest itself in the distribution of high $p_T$
particles?''  One explanation is that high energy partons lose
energy as they pass through the matter created in the collisions.
Since the source is asymmetric, the amount of energy loss will
depend on the direction the parton travels. As such, energy loss
can lead to a momentum-space anisotropy that reflects the initial
spatial anisotropy of the source. We will discuss energy loss and
the suppression of high $p_T$ particle production in
section~\ref{sec:raa}.

\begin{figure}[htpb]
\centering\mbox{
\includegraphics[width=0.70\textwidth]{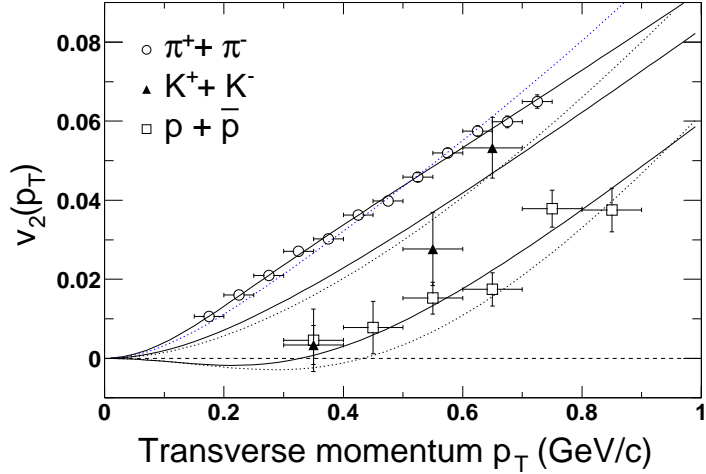}}
\caption[Identified particle elliptic flow] {The differential
elliptic flow for identified particles at mid-rapidity from Au+Au
collisions at $\sqrt{s_{_{NN}}}=130$~GeV~\cite{Adler:2001nb}. The
curves represent fits to hydrodynamic inspired parameterizations.}
\label{fig:v2_identified}
\end{figure}

Figure~\ref{fig:v2_identified} shows the differential $v_2$ at mid
rapidity for identified particles at low $p_T$ ($p_T < 1$~GeV/c)
where particles can be identified by their energy loss in the
detector gas~\cite{Adler:2001nb}. The hydrodynamic models predict
a mass-ordering for elliptic flow with less massive particles
having larger elliptic flow for all values of $p_T$.
The large $v_2$ and its mass-ordering at low $p_T$ are consistent
with the hydrodynamic limit for the conversion of spatial
anisotropy to momentum anisotropy (where local thermal equilibrium
has been
assumed)~\cite{hydro:Huovinen:2001cy,hydro:Ollitrault:1992bk,hydro:Sorge:1998mk,hydro:Teaney:2000cw}.
At intermediate $p_T$ ($1.5 < p_T < 4$~GeV/c), however, while
hydrodynamic models predict a monotonic increase, the charged
hadron $v_{2}$ saturates at a value approximately independent of
$p_T$. After the first year of RHIC data taking, the particle-type
dependence of $v_2$ in the high momentum region remained an open
question. In this thesis we present measurements of $v_2$ for
$K_S^0$ and $\Lambda+\overline{\Lambda}$ at mid-rapidity from
Au+Au collisions at $\sqrt{s_{_{NN}}}=200$~GeV/c that extend up to
$p_T \sim 6$~GeV/c.

\subsection{Nuclear Modification of Particle Production}\label{sec:raa}


Like $v_2$, high $p_T$ hadron production---presumably through
scatterings of partons involving large momentum transfer---is also
thought to probe the early stage of heavy-ion collisions.
High-energy partons passing through dense matter are predicted to
lose energy by induced gluon
radiation~\cite{dedx:Gyulassy:1990ye,dedx:Baier:2000mf,dedx:Gyulassy:2003mc}.
Since the total energy loss depends on the color charge density of
the medium, nuclear modification of the high $p_T$ particle yields
can probe the dense, perhaps deconfined-partonic matter created by
the collision.

Partonic energy loss or \textit{jet-quenching} can be studied by
measuring the modification of particle production in nuclear
collisions. A nuclear modification factor can be formed by taking
the ratio of the particle yields in nucleus-nucleus collisions and
the particle yields in proton-proton collisions. The ratio is then
scaled by $T_{AA}=\langle \mathrm{N_{binary}}
\rangle/\sigma^{NN}_{inel}$ to account for the trivial increase in
the yield with the system size:
\begin{equation}
R_{AA}(p_T) =
\frac{d^2n^{AA}/dp_Td\eta}{T_{AA}d^2\sigma^{NN}/dp_Td\eta},
\end{equation}
where $\eta$ is the pseudo-rapidity and $\mathrm{N_{binary}}$ is
the number of binary nucleon-nucleon collisions. In the absence of
nuclear effects, at high $p_T$, $R_{AA}$ is expected to be unity.
In the low $p_T$ region the yield is not expected to scale with
$\mathrm{N_{binary}}$. The $p_T$-scale where the high $p_T$ regime
begins is an experimental observable that our measurements will
address.

\begin{figure}[htpb]
\centering\mbox{
\includegraphics[width=0.45\textwidth]{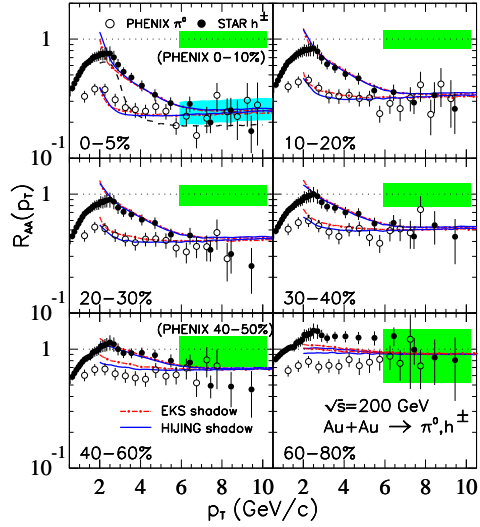}}
\caption[Charged hadron and $\pi^0$ $R_{AA}$] {The nuclear
modification parameter $R_{AA}$ for charged
hadrons~\cite{highpt:Adams:2003} and for neutral
pions~\cite{highptpi0:Adler:2003qi} in Au+Au collisions at
$\sqrt{s_{_{NN}}}=200$~GeV. The panels show different Au+Au
collision centrality intervals with the most central in the top
left. The curves show model calculations based on partonic energy
loss and other nuclear effects~\cite{dedx:Wang:2003aw}.}
\label{fig:hadron_raa}
\end{figure}

Figure~\ref{fig:hadron_raa} shows $R_{AA}$ for charged hadrons and
neutral pions $\pi^0$ from Au+Au collisions at
$\sqrt{s_{_{NN}}}=200$~GeV. The high $p_T$ yields in central Au+Au
collisions are suppressed with respect to $\mathrm{N_{binary}}$
scaling. The suppression is largest for central collisions while
the yields in peripheral collisions are consistent with
expectations from $\mathrm{N_{binary}}$ scaling. The suppression
is approximately independent of $p_T$ for $p_T > 3$~GeV/c for
$\pi^0$ and for $p_T > 5$~GeV/c for charged hadrons.

\begin{figure}[htpb]
\centering\mbox{
\includegraphics[width=0.60\textwidth]{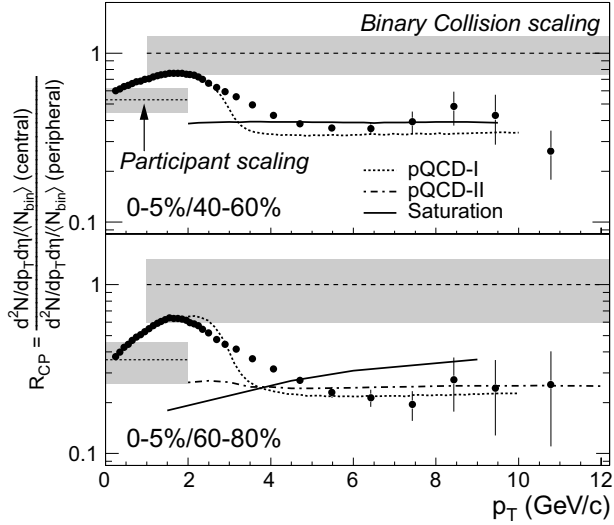}}
\caption[Charged hadron $R_{CP}$] {Nuclear modification of charged
particle production measured at $\sqrt{s_{_{NN}}}=200$~GeV from
central Au+Au events (0--5\% of the collision cross-section)
compared to peripheral Au+Au events (40--60\% and 60--80\% of the
collision cross-section)~\cite{highpt:Adams:2003}.}
\label{fig:hadron_rcp}
\end{figure}
Like $R_{AA}$, the ratio of the yields in central and the yields
in peripheral collisions ($R_{CP}$) also can measure nuclear
modifications to particle production:
\begin{equation}
R_{CP}(p_T) = \frac{\left [dn/\left (\mathrm{N_{binary}}dp_T
\right ) \right ]^{\mathrm{central}}}{\left [dn/\left
(\mathrm{N_{binary}}dp_T \right ) \right ]^{\mathrm{peripheral}}}.
\end{equation}
When $R_{AA}$ for peripheral events follows $\mathrm{N_{binary}}$
scaling, $R_{CP} \approx R_{AA}^{central}$. The ratio $R_{CP}$
typically has smaller systematic uncertainties than $R_{AA}$ and
does not require the measurement of a p+p reference spectrum. The
charged hadron $R_{CP}$ in Figure~\ref{fig:hadron_rcp} shows a
suppression of particle yields in central events compared to
scaled peripheral events. The suppression of charged hadrons is
roughly constant for $p_T > 5$~GeV/c. The dependence on
particle-type of the suppression and the $p_T$-scale for its onset
remained an open question after the first year of RHIC collisions.
In this thesis we present the measurement of $R_{CP}$ for $K_S^0$
and $\Lambda+\overline{\Lambda}$ from Au+Au collisions at
$\sqrt{s_{_{NN}}}=200$~GeV up to $p_T \sim 6$~GeV/c.

As mentioned in section~\ref{sec:v2}, energy loss can also
manifest itself in $v_2$. By suppressing the yield of large $p_T$
particles more in the out-of-plane direction than the in-plane
direction\footnote{The in-plane and out-of-plane directions are
perpendicular to the beam axis. The in-plane direction lies along
the vector connecting the colliding nuclei.}, energy loss can
cause an anisotropy in the final momentum distribution. The
particle-type dependence of $v_2$ and $R_{CP}$ will be a powerful
test of the energy loss hypothesis. If energy loss governs the
development of $v_2$ {\em and} $R_{CP}$ then we expect either no
particle-type dependence or we expect the particle with the larger
$v_2$ to also have a larger suppression.

\subsection{Other Observations}

\begin{figure}[htpb]
\resizebox{.5\textwidth}{!}{
\begin{rotate}{\begin{rotate}{\begin{rotate}{
\includegraphics{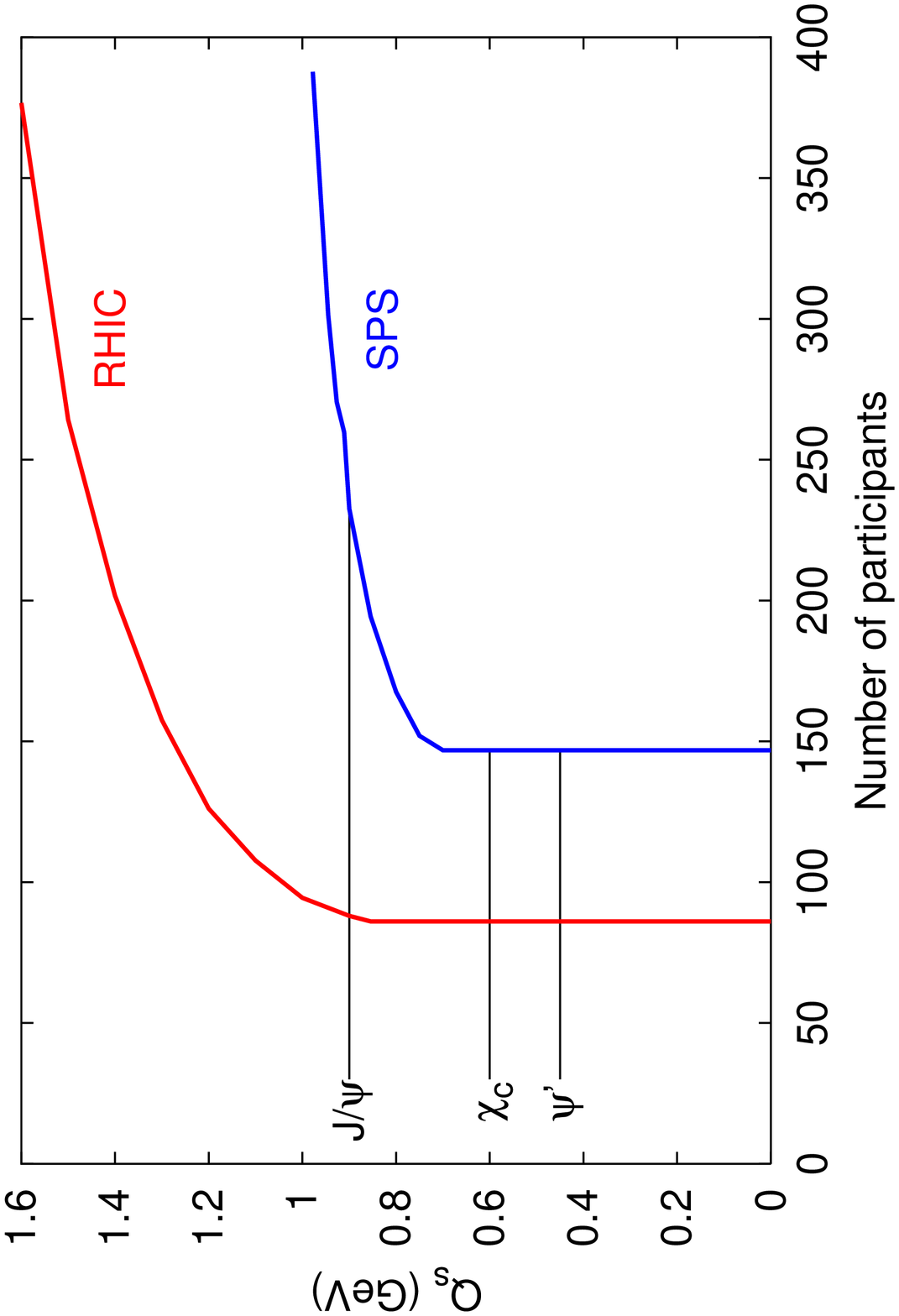}
}\end{rotate}}\end{rotate}}\end{rotate} }
\resizebox{.5\textwidth}{!}{\includegraphics{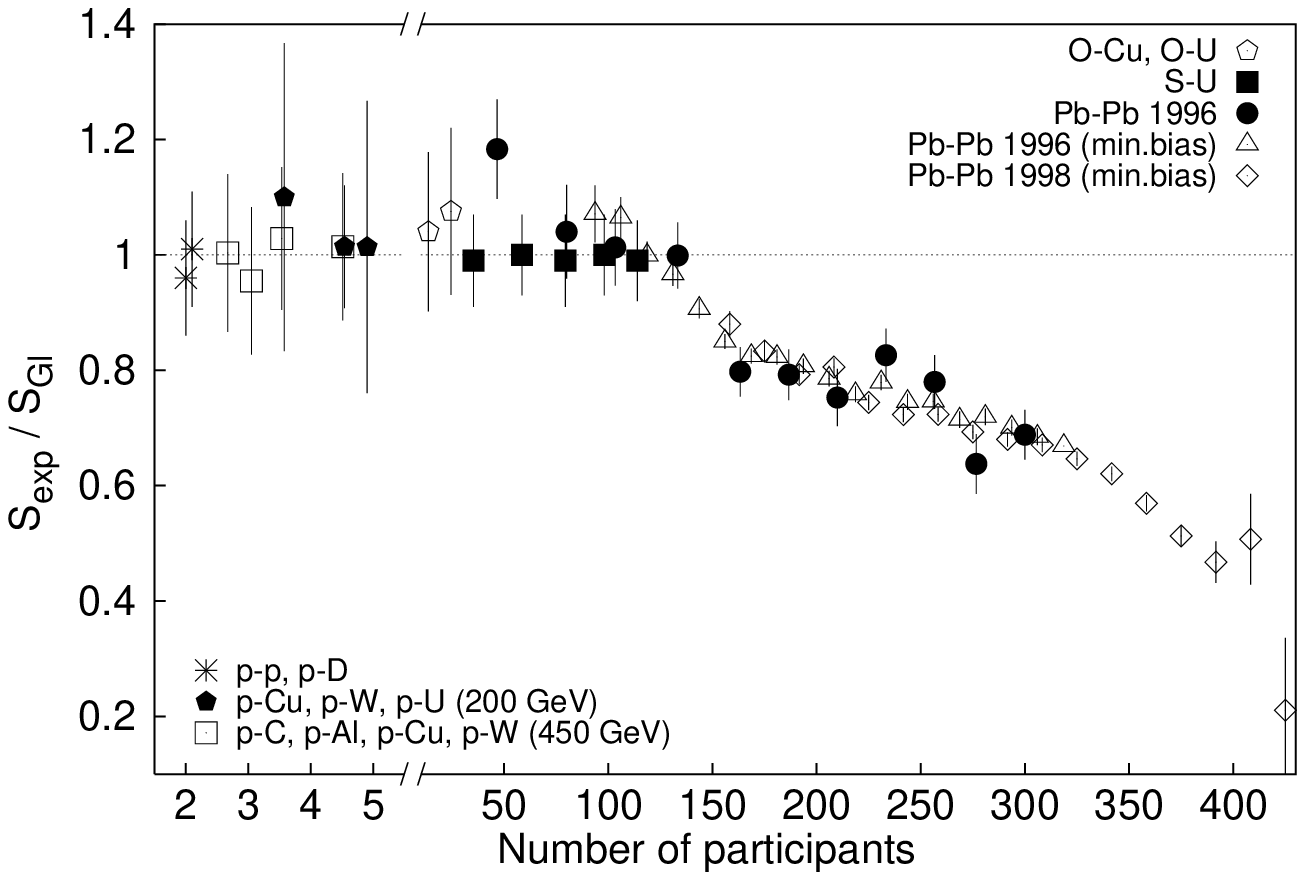}}
\caption[J/$\psi$ suppression] {Left: Centrality dependence of the
percolation scale Q$_S$ where charmonium suppression is thought to
set in for RHIC energy and SPS energy~\cite{Perc:Digal:2002bm}.
Right: The ratio of the measured and expected J/$\psi$ yield
showing a step like suppression pattern as the number of
participants in the collisions system
increases~\cite{Satz:2002ku}.} \label{fig:jpsi}
\end{figure}
\begin{figure}[htpb]
\centering\mbox{
\includegraphics[width=0.80\textwidth]{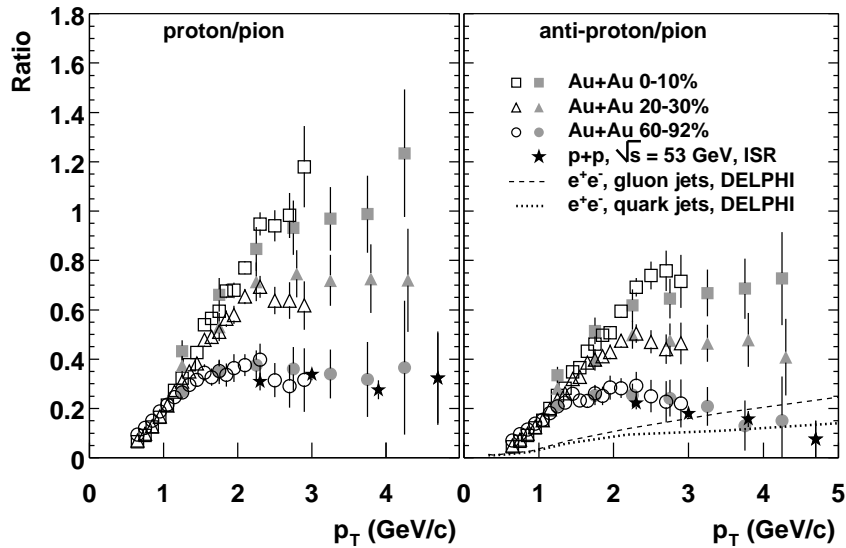}}
\caption[Baryon enhancement] {The proton/pion and anti-proton/pion
ratio~\cite{proton:Adler:2003kg}.} \label{fig:baryon}
\end{figure}
\begin{figure}[htpb]
\centering\mbox{
\includegraphics[width=0.80\textwidth]{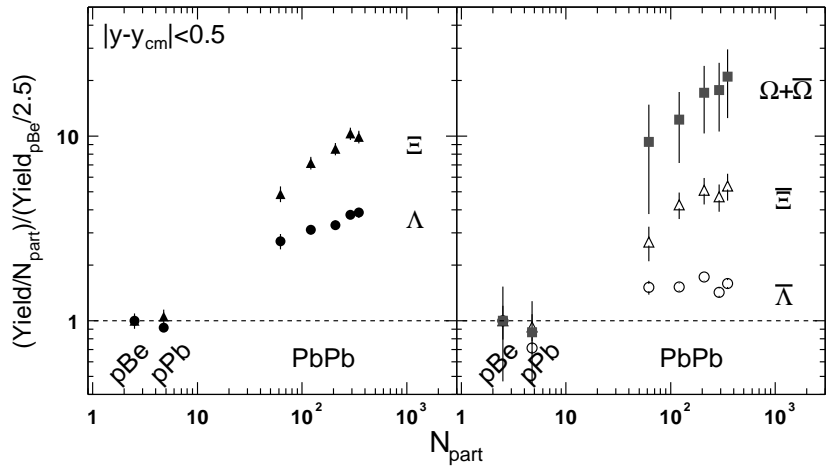}}
\caption[Strangeness enhancement] {Strange particle yields scaled
by $\mathrm{N_{part}}$ normalized by p-Be
collisions~\cite{NA57:Fanebust:2002xb}.} \label{fig:strange}
\end{figure}

Other important observations made in heavy-ion collisions include
the suppression of J$/\psi$ production, strangeness enhancement,
the coincidence of particle ratios with statistical model
predictions, the enhancement of baryon production, and the
reduction of the net baryon number. Figure~\ref{fig:jpsi} (right)
shows the ratio of the expected and measured J$/\psi$ yields
versus N$_{\mathrm{part}}$. The step like behavior was interpreted
as being caused by the dissolution of successive charmonium states
in a new form of matter created in heavy-ion collisions. First the
$\psi^{\prime}$ and $\chi_C$ dissolve and then the J$/\psi$
dissolves. Figure~\ref{fig:jpsi} (left) shows the percolation
scale where the dissolution of different charmonium states should
set in versus N$_{\mathrm{part}}$~\cite{Perc:Digal:2002bm}.
Figure~\ref{fig:baryon} shows the enhancement of the (anti-)proton
to pion ratio at intermediate $p_T$ in central Au+Au collisions
relative to $e^++e^-$, p+p, or peripheral Au+Au collisions. The
enhancement of baryon production will be studied further in this
thesis. Figure~\ref{fig:strange} shows the enhancement of strange
particle production in heavy-ion collisions relative to p+Be
collisions. The enhancement increases with the strange quark
content; \textit{i.e.} $\Omega^-$(sss) $> \Xi^-$(dss) $>
\Lambda$(uds)~\cite{NA57:Fanebust:2002xb}.

\section{Thesis Outline}

In this thesis, we present the first measurement of $R_{CP}$ for
$K_{S}^{0}$ and $\Lambda+\overline{\Lambda}$ for Au+Au collisions
at $\sqrt{s_{_{NN}}}=200$~GeV and the first measurements of $v_2$
for $K_{S}^{0}$ and $\Lambda+\overline{\Lambda}$ for Au+Au
collisions at $\sqrt{s_{_{NN}}}=130$ and 200~GeV. Our emphasis is
on probing the early stage of heavy-ion collisions, mapping out
the transition between $p_T$ regions (\textit{i.e.} soft,
intermediate, hard, etc.), and understanding how hadronization
modifies the observables we measure. In mapping out the $p_T$
regions we hope to learn what processes dominate particle
production within each region. Studying the variation in yields
with centrality ($R_{CP}$) and azimuthal angle ($v_2$) for
different particle species will help us understand the
hadronization mechanisms in heavy-ion collisions. This information
will be helpful for characterizing the matter created in heavy-ion
collisions.

In Chapter 2 we will discuss the facilities used to study
heavy-ion collisions. The Relativistic Heavy-Ion Collider (RHIC)
will be described, an introduction to particle tracking detectors
will be given, and the Solenoidal Tracker at RHIC (STAR) detector
system will be reviewed. Chapter 3 contains details of the
analysis methods. In Chapter 4 we present the results of the
analysis and in Chapter 5 we discuss these results, draw
conclusions, and present an outlook for future work. In the
appendices we include a description of the coordinates system in
the transverse plane, calculations of the nuclear overlap density
for Au+Au collisions, definitions for the kinematic variables used
in this thesis and a list of STAR collaborators and institutions.


\chapter{Experimental Set-up}
    \section{The Relativistic Heavy-Ion Collider }
\indent \label{sec:rhic}

The Relativistic Heavy-Ion collider (RHIC) at Brookhaven National
Lab (BNL) is designed to collide counter-rotating heavy-ion beans
at energies up to 100 GeV/u.  RHIC is the first facility to
collide heavy-ion beams. The center-of-mass energy for these
collisions is roughly a factor of ten times greater than the
highest energies reached with the previous fixed target heavy-ion
experiments. Parameters for existing and future relativistic
heavy-ion facilities are given in Table~\ref{tab:colliders}. RHIC
consists of two concentric rings of super-conducting magnets
(cooled to below 4.6 degrees Kelvin) that focus and guide the
beams and a radio frequency ($rf$) system that captures,
accelerates and stores the beams. The ring's diameters are
approximately 1.22 km.
\begin{table}[hbt]
\centering\begin{tabular}{lcccccccc} \hline \hline
                         & AGS & AGS & SPS & SPS & SPS & RHIC & RHIC & LHC \\
\hline
Start year               & 1986 & 1992 & 1986 & 1994 & 1999 & 2000 & 2001 & 2006 \\
A$_{max}$                & $^{28}$Si & $^{197}$Au & $^{32}$S & $^{208}$Pb & $^{208}$Pb & $^{197}$Au & $^{197}$Au & $^{208}$Pb \\
E$_P^{max}$ [AGev]       & 14.6 & 11 & 200 & 158 & 40 & $0.91${\sffamily\footnotesize E}$4$ & $2.1${\sffamily\footnotesize E}$4$ & $1.9${\sffamily\footnotesize E}$7$ \\
$\sqrt{s_{_{NN}}}$ [GeV] & 5.4 & 4.7 & 19.2 & 17.2 & 8.75 & 130 & 200 & 6000 \\
$\sqrt{s_{_{AA}}}$ [GeV] & 151 & 934 & 614 & $3.6${\sffamily\footnotesize E}$3$ & $1.8${\sffamily\footnotesize E}$3$ & $2.6${\sffamily\footnotesize E}$4$ & $4${\sffamily\footnotesize E}$4$ & $1.2${\sffamily\footnotesize E}$6$ \\
$\Delta y/2$             & 1.72 & 1.58 & 2.96 & 2.91 & 2.22 & 4.94 & 5.37 & 8.77 \\
\hline \hline
\end{tabular}
\caption[Heavy-ion colliders] {RHIC compared to existing and
future facilities; A$_{max}$ is the maximum species mass number,
E$_{P}^{max}$ is the maximum (equivalent) fixed-target beam energy
per nucleon, $\sqrt{s_{_{NN}}}$ is the maximum center of mass
energy per nucleon, $\sqrt{s_{_{AA}}}$ is the total center of mass
energy, and $\Delta y/2$ is the rapidity gap from the beam to
mid-rapidity~\cite{raf:let:book}.}\label{tab:colliders}
\end{table}

\begin{figure}[htpb]
\centering\mbox{
\includegraphics[width=0.9\textwidth]{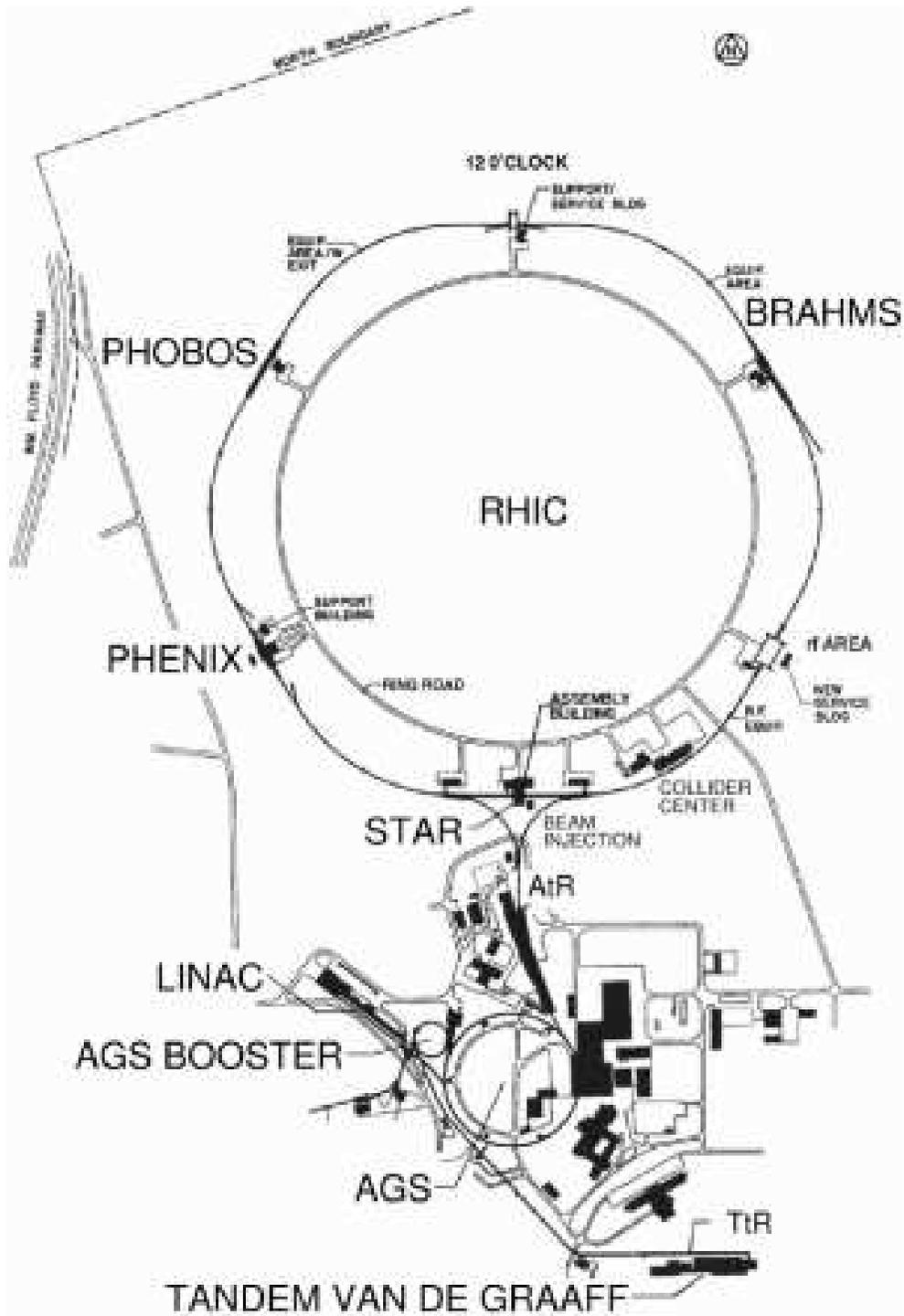}}
\caption[The BNL--RHIC facility] {A diagram of the Brookhaven
National Laboratory collider complex including the accelerators
that bring the nuclear ions up to RHIC injection energy (10.8
GeV/u for $^{197}$Au$^{79}$).} \label{fig:rhic_injector}
\end{figure}
Figure~\ref{fig:rhic_injector} shows the BNL accelerator complex
including the accelerators used to bring the gold ions up to RHIC
injection energy.  In the first of the Tandem Van de Graaff
accelerators, gold ions in a charge state $Q=-1e$ accelerate to 15
MeV. The ions then pass through a stripping foil (located between
the Van de Graaffs) where electrons are knocked off so that their
most probable charge state becomes $Q=+12e$. With their charge
changed from negative to positive, the ions gain another 1 MeV/u
of energy as they accelerate through the second Van de Graff, back
to ground potential. On exiting the Tandem, the ions pass through
a second stripping foil bringing their most probable charge to
$Q=+32e$. They are then injected into the Booster synchrotron and
accelerated to 95 MeV/u. A stripper foil in the transfer line
between the booster and the Alternating Gradient synchrotron (AGS)
increases their charge state to $Q=+77e$. In the AGS the ions are
accelerated to 10.8 GeV/u. They are extracted from the AGS and
passed through one final stripper foil where the remaining K-shell
electrons are removed ($Q=+79e$). Finally, they are injected into
RHIC where they are accelerated to top energy and can be stored
for up to 10 hours. Table~\ref{tab:rhic_specs} lists important
parameters for RHIC.

\begin{table}[hbt]
\centering\begin{tabular}{lc} \hline \hline
Top Au+Au $\sqrt{s_{_{NN}}}$          & 200 GeV\\
Ave. luminosity $\mathcal{L}$ (10 hour store)  & $\sim 2 \times 10^{26}$ cm$^{-2}$s$^{-1}$ \\
Bunches per ring                         & 60 \\
Gold ions per bunch                      & $10^9$ \\
Crossing points                          & 6 \\
Beam lifetime (store length)             & $\sim 10$ hours \\
RHIC circumference                       & 3833.845 m \\
\hline \hline
\end{tabular}
\caption[RHIC parameters] {Nominal RHIC parameters for Au+Au
collisions.} \label{tab:rhic_specs}
\end{table}

%

\section{RHIC Experimental Program} \indent

To date, RHIC has generated collisions between gold nuclei at
$\sqrt{s_{_{NN}}} =$ 22, 56, 130, and 200 GeV, between protons at
$\sqrt{s_{_{NN}}} = 200$ GeV, and between gold and deuterium
nuclei at $\sqrt{s_{_{NN}}} = 200$ GeV.
Table~\ref{tab:rhic_performance} shows the luminosity achieved at
the end of RHIC Run-2 (Au+Au collisions at $\sqrt{s_{_{NN}}} =
200$~GeV). The STAR experiment recorded integrated luminosities
$\mathcal{L} \sim 2.8 \:b^{-1}$ and $\mathcal{L} \sim 80 \:\mu
b^{-1}$ for RHIC Run-1 (Au+Au collisions at $\sqrt{s_{_{NN}}} =
130$~GeV) and RHIC Run-2 respectively. Most of the integrated
luminosity comes late in the runs, after the collider is tuned.
\begin{table}[hbt]
\centering\begin{tabular}{ccccc} \hline \hline
Bunches & Ions/Bunch & $\mathcal{L}_{peak}$ & $\mathcal{L}_{ave}$ (store) & Integrated \\
 {} &  {} &  {[cm$^{-2}$s$^{-1}$]} &  {[cm$^{-2}$s$^{-1}$]} &  {$\mathcal{L}$ [($\mu$b)$^{-1}$]} \\
\hline
55 & $6 \times 10^8$ & $3.7 \times 10^{26}$ & $3.7 \times 10^{26}$ & $\sim 80$ \\
 \hline \hline
\end{tabular}
\caption[RHIC Run-2 performance] {The performance of RHIC during
the 2001 Au+Au run (Run-2).} \label{tab:rhic_performance}
\end{table}

There are four experimental collaborations at RHIC; the PHOBOS
collaboration with 107 members from 8 institutions, the BRAHMS
collaboration with 51 members from 14 institutions, the PHENIX
collaboration with 328 members from 52 institutions, and the STAR
collaboration with 293 members from 39
institutions~\footnote{These numbers are taken from each
collaborations author list as of July 2003 and do not represent
the total number of people working on the experiment. There are,
for example, over 450 scientist and engineers working on the
PHENIX experiment.}.

\begin{table}[htpb]
\centering\begin{tabular}{|p{66mm}|p{66mm}|} \hline
\epsfig{file=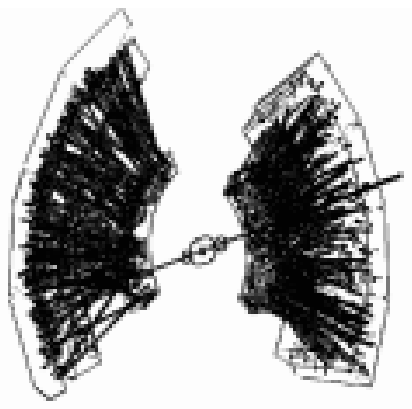, width=60mm} &
\epsfig{file=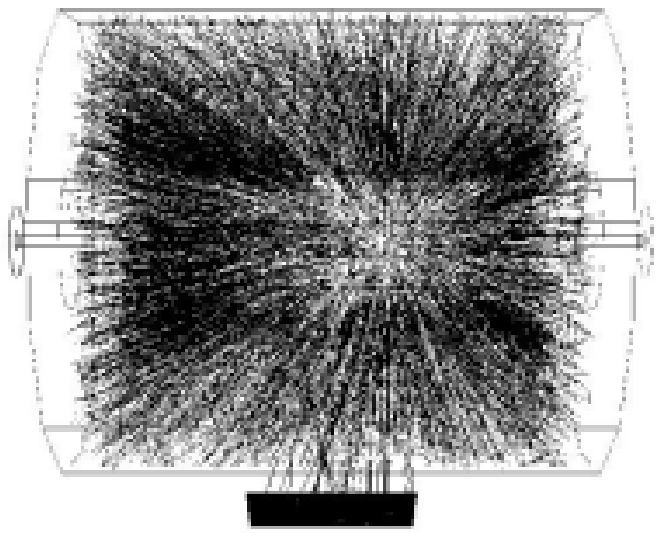, width=60mm} \\
\multicolumn{1}{|c|}{\textbf{PHENIX event display}} &
\multicolumn{1}{c|}{\textbf{STAR event display}} \\
\hline
Two muon spectrometers cover the pseudo-rapidity region $1.1 <
|\eta| < 2.4$ and azimuth angle $0 < \phi < 2\pi$. A central
spectrometer with two arms and tracking sub-systems (each
subtending $\pi/2$ radians) covers $|\eta| < 0.35$. With a smaller
acceptance and faster detectors the emphasis is on triggering on
rarer probes, hadron identification and electron identification.
& A large acceptance solenoidal tracking detector with particle
identification covers the full azimuth ($|\phi|<\pi$), $|\eta| <
2.0$ and $2.5< |\eta| < 4.0$. Subsystems include a central TPC,
two forward TPCs, a silicon vertex tracker and a barrel
electromagnetic calorimeter. The emphasis is on global event
characterization, resonance identification, fluctuations and
event-by-event variables.
\\
\hline
\epsfig{file=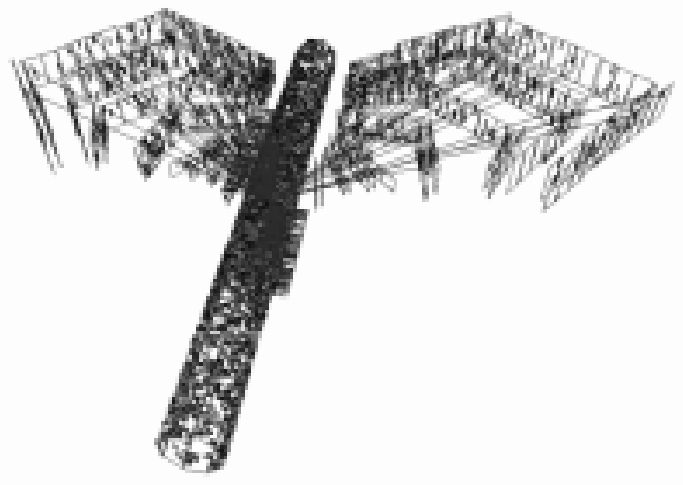, width=60mm} &
\epsfig{file=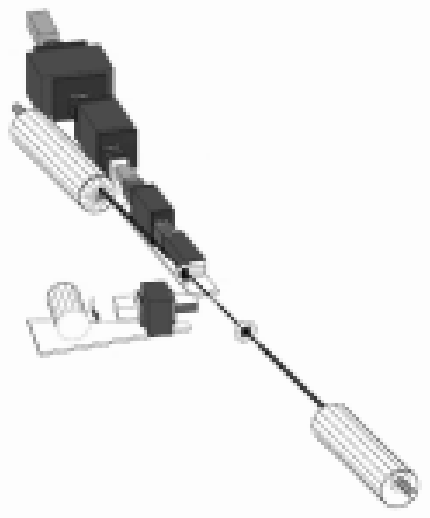, width=55mm} \\
\multicolumn{1}{|c|}{\textbf{PHOBOS event display}} & \multicolumn{1}{c|}{\textbf{BRAHMS detector}} \\
\hline
Measurements of charged particles are made across a full
solid-angle with a multiplicity detector. Two small acceptance
spectrometer arms allow for particle identification at
mid-rapidity. Multiplicity measurements across a broad range of
$\eta$ and $p_T$ are emphasized.
& Designed to provide good particle identification across a broad
rapidity and $p_T$ range ($0<y<4$; $0.2 < p_T < 3.0$~GeV/c) with
two small solid-angle spectrometers. Measuring particle production
at forward angles is emphasized.
\\
\hline
\end{tabular}
\caption[RHIC experiments] {Summary of RHIC experiments.}
\label{tab:rhic_detectors}
\end{table}

In this thesis we present an analysis of Au+Au collisions recorded
by the STAR detector during the summer of 2000 and the winter of
2001. Approximately $5 \times 10^5$ and $5 \times 10^6$ usable
events were recorded at $\sqrt{s_{_{NN}}} = 130$~GeV and
$\sqrt{s_{_{NN}}} = 200$~GeV respectively.

\clearpage

    \section{Particle Tracking Detectors}
\indent

The primary detector used for the analysis presented in this
thesis is the STAR Time Projection Chamber (TPC).  The TPC is
designed to do particle tracking which facilitates the
identification of secondary vertices from weak decays (e.g.
$K^{0}_{S} \rightarrow \pi^{+} + \pi^{-}$). In the following we
give an introduction to high-energy particle tracking
technology~\cite{Tracking:Safarik:2003ia}.

\subsection{History of Particle Tracking}

Since the beginning of particle physics, when J.J.~Thomson
realized that the \textit{cathode rays} he was studying were not
``rays'' but streams of subatomic charged particles instead, our
understanding of the subatomic realm and the mechanics that
governs it has depended strongly on our ability to detect the
\textit{tracks} of charged particles. Thomson was able to surmise
the existence of electrons, measure their charge to mass ratio and
even measure their velocity as they were emitted by a hot filament
because he could see their trajectory as they passed through
crossed electric and magnetic fields.  During the years since
Thomson's experiments in 1897 many techniques have been developed
to detect or \textit{visualize} charged particle tracks---nuclear
emulsions, cloud chambers, bubble chambers, spark chambers,
streamer chambers, other gas detectors, solid-state detectors and
so on.  All of these techniques rely on very fast \textit{charged}
particles ionizing atoms as they pass through matter.  The
ionization left along the paths of the high-energy particles can
then act as catalysts for reactions that leave an observable
trace---such as, a bubble, a spark, condensation, or a charge
cascade.

Many of these particle detecting techniques have---because of
their inherent limitations---been abandoned for the most part in
favor of \textit{gas} or \textit{solid state} detectors.
Experiments at the newest colliders like RHIC or the LHC rely
almost exclusively on these two techniques because they lend
themselves well to triggering, high event rates and the
digitization of huge amounts of data.  Bubble chambers however,
are of particular historic importance and produced a wealth of
information from their inception in 1952 to well into the 1970s.
Their importance to high-energy physics was acknowledged with a
Nobel prize in 1960 awarded to Donald Glaser.  Glaser struck upon
the idea of a bubble chamber when he saw the tracks created by
bubbles in beer.

\subsection{Bubble Chambers---Three Decades of Physics and Two Nobel Prizes}

\textbf{Bubble chambers} initially used liquid in a super-heated
state to detect the ionization left along the tracks of
high-energy charged particles passing through the liquid.  In 1952
Glaser used diethyl-ether heated to $\sim 100^{\circ}$C above its
boiling point to build the first bubble chamber.  The super-heated
liquid, when struck by cosmic rays, began boiling violently and a
photograph made using a fast camera showed tracks left behind by
the high-energy charged particles created by a cosmic ray.  It is
presumed that after a high-energy particle causes the initial
ionization along its path, heat generated by recombination is
responsible for the boiling and bubble formation in the liquid.

Improvements to this technique, including the use of pistons to
create a sudden pressure drop to induce bubble formation, led to
greater precision and larger chambers that could be placed in a
magnetic field.  Perhaps the most notable improvement to the
bubble chamber came when Luis Alvarez substituted hydrogen for the
ether used by Glaser.  Alvarez's chamber produced much clearer
tracks and this technological advance was of such value that he
won the Nobel prize for physics in 1968 for his work, the second
Nobel prize awarded for work related to the development of the
bubble chamber.

In bubble chambers, the fluids in the chamber act both as the
target and as the detector so different fluids---some cryogenic
and some room temperature---were eventually used to suit the
purpose of the experiment.  Cryogenic liquids consisted of the
simplest nuclei like, H$_2$, D$_2$, He, Ne, Ar and Xe while the
room temperature ``heavy liquids'' like propane (C$_3$H$_8$) and
Freon (CF$_2$Cl$_2$ or CF$_3$Br) offered short interaction
lengths.  The typical size of a bubble in a bubble chamber is
$\sim$~10~$\mu$m and the bubble density can be used to determine
$\beta \equiv v/c$ for the passing particle.

The advantages bubble chambers offered kept them in widespread use
for three decades, from the 1950's to well into the 1970's.  They
had good spatial resolution (10~--~150~$\mu$m), a large sensitive
volume, $4\pi$ geometrical acceptance; and they permitted the use
of a variety of materials as targets.  Eventually however, as
physics began requiring more complex triggers and as large-volume
high-precision detectors demanding electronic data recording came
in use, the bubble chambers disadvantages rendered them obsolete.
The analysis of photographs was a tedious task requiring expensive
projectors for scanning the images and the whole process was only
modestly scalable so that only limited statistics could be
achieved. Bubble chambers were also complicated to operate,
required cryogenics, and were a safety concern.  In addition,
bubble chambers weren't compatible with particle colliders---the
now dominant high-energy accelerator, they provided no triggering
for low cross-sections and they had a relatively long sensitive
time ($\sim$~1~ms) which necessitates a lower beam luminosity.

\subsection{Streamer chambers---a Precursor to Modern Gas Detectors}

The \textbf{streamer chamber} developed by G.E.~Chikovani in
1963---an improvement on the spark chamber---overcame
\textit{some} of the limitations of the bubble chamber and was the
predecessor of the \textit{gaseous detectors} of today.  Like the
bubble chamber however, the streamer chamber also relied on
photographic film to record the tracks of streamers, placing a
limit on the statistics available for analysis.

The \textbf{spark chamber} uses a large potential across two
parallel planes of electrodes to induce electrical breakdown---a
spark---in gas between the electrodes.  The ionization left from
the passing of a high-energy particle acts as the catalyst for the
spark.  The spark chamber however, can only measure the position
of the track in the direction parallel to the electric field to
within the spacing of the electrodes.  The \textit{streamer
chamber} overcomes this limitation by applying a high-voltage
pulse for a short duration ($\sim$~15~ns). The strong electric
field ($\sim$~20~kV/cm) from the high-voltage pulse induces an
incomplete spark discharge.  These electron avalanches or
streamers form all along the particles path and the radiation of
the gas in the streamer plasma can be recorded optically. Streamer
chambers were built with sensitive volumes of several cubic meters
that recorded particle tracks in any direction with equal
efficiency.  The density of the streamers can be used for particle
identification up to particle momenta of $\sim$~1~GeV/c.

The two major advantages of the streamer chamber over the bubble
chamber are its ability to be triggered by external devices and
its very short sensitive time ($\sim$~1~$\mu$s).  Eventually
however, its use of photographic film, its limited spatial
resolution ($\geq~300~\mu$m) and its relatively long dead time
($\sim$~300~ms) turned out to favor the gaseous detectors that
would rely on electronic, not optical, recording techniques.

\subsection{Today's Tracking Detectors}

Almost all tracking detectors, other than the solid state and
gaseous detectors using electronic recording techniques have been
abandoned. The modern-era detectors have shorter sensitive times
and shorter dead times so that the beam intensity of the particle
accelerator can be increased and greater statistics can be
recorded.  These newer detectors also tend to be easier to operate
and have greater spatial resolution.

\subsection{Gas Detectors}

Most gas detectors---multi-wire proportional chambers (MWPC),
drift chambers, straw tubes, cathode strip or pad chambers, time
projection chambers (TPC) and micro-strip gas chambers
(MSGC)---use the \textit{proportional counting mode} of operation.
In this mode, the electrons from the primary electron-ion pairs
created by the high-energy charged particle, are directed in an
electrostatic field toward a very high field region
(10--100~kV/cm) surrounding an anode wire of small radius. In this
region, the fast electrons gain enough energy to create secondary
electron-ion pairs. Each new electron produced by ionization, in
turn, creates more electrons-ion pairs; the development of this
avalanche or cascade is called \textit{gas multiplication}. Most
of the electrons in the avalanche are created very close to the
wire so they are collected within a few nanoseconds. The heavier
ions---also predominantly produced near the wire---move more
slowly across a larger potential difference.  As they do so, they
induce a signal that can be detected with an amplifier and used
for position and energy loss (dE/dx) measurements.

The mode of operation of a gaseous detector is determined by the
response of the ions and gas to the field strength surrounding the
anode. In the \textit{proportional mode} of operation the field
strength is great enough to induce gas amplification---typically
$10^4$--$10^6$ times---but is not so great that it leads to
complete breakdown or non-negligible space charge effects:caused
by the build up of longer-lived positive ions. In the proportional
mode of operation the signal is \textit{proportional} to the
number of primary electron-ion pairs which is in turn proportional
to the energy lost by the traversing particle. The measured dE/dx
can then be used for particle identification (PID).

A \textbf{multi-wire proportional chamber} consists of planes of
independent wires---typically spaced 1--2 mm apart---set between
two planes of cathodes at a distance of 3--4 times the wire
spacing.  A negative voltage is applied to the cathodes and the
wires are held at ground.  Each wire then acts as a proportional
counter for primary electrons-ion pairs left along a particles
track. The distance from cathode to anode is typically about 1 cm
and the wire diameter should be about 20--50 $\mu$m. The spatial
resolution is given by $d/\sqrt{12}~=~300 - 600~\mu$m, where d is
the wire pitch or spacing.  For this invention G.~Charpak was
awarded the 1992 Nobel prize in physics.

A \textbf{drift chamber} is a multi-wire proportional chamber with
a large wire pitch---from several centimeters up to 50 cm but more
typically 5 cm.  Track position is determined by measuring the
time electrons need to reach the anode wires.  The speed of the
electron drift depends on the gas used and the pressure in the
chamber and is typically $\sim$~5~cm/$\mu$s so that a timing
resolution of 1~ns gives a spatial resolution of $\sim$~50~$\mu$m.
Different geometries and configurations can be used in order to
create constant fields pointing toward the anode wires.  A
\textbf{straw tube} is a drift chamber composed of an individual
straw shaped cathode (diameter of $\sim$~5~mm) with a single anode
wire in the center.  A ``continuous'' tracker can be constructed
by packing many layers of straw tubes together.  Straw tube
detectors tolerate high loads because they don't use a common gas
volume and can achieve a resolution of about 150~$\mu$m with
coarse time measurements.

A \textbf{time projection chamber} (TPC) is a drastic variation on
a simple drift chamber.  A TPC consist of a large
three-dimensional gas filled vessel with readout detectors on a
wall at the end of a drift volume. The readout detectors are
usually \textit{cathode pad chambers}. A strong electric field
across the TPC produced by a cathode on the wall opposite the
readout planes creates the drift field. When a charged particle
creates electron-ion pairs within the drift volume the strong
drift field prevents them from recombining. The much lighter
electrons move quickly toward the readout chambers. The drift
field is chosen so that it is not strong enough to create
secondary electron-ion pairs: typically hundreds of volts per
centimeter. The readout chamber is separated from the drift volume
by a \textit{gating grid}.  The gating grid is a plane of wires
that electrostatically separates the amplification region from the
drift region.  The gating grid prevents the ions created in the
amplification region from getting back into the drift region and
allows for triggering of the detector; when an interesting event
occurs the gating grid wires are set to voltages that allow
electrons to pass through.

The TPC readout chambers typically consist of an anode wire plane
between a ground wire plane and a cathode pad plane. The signal
induced on the anode wires is typically detected via image charges
on several nearby pads. The position of the electron-ion cascade
in the anode wire direction can be determined precisely by fitting
a modified Gaussian to the signals on several consecutive pads.
This measurement gives two transverse coordinates and the drift
time gives the third coordinate, making the TPC a fully
three-dimensional detector.  Unlike other historic
three-dimensional detectors however, such as the bubble or
streamer chambers, the TPC is read completely electronically.  The
TPC also has the advantage that it has no pulsed very
high-voltages and is fast compared to historic detectors: its
speed is determined by the maximum drift time which, for large
chambers is $\sim$~100~$\mu$s.

Inhomogeneities in the drift field and effects due to magnetic
fields however, can distort the drift path of the electrons and
further degrade the resolution.  The electron clouds also diffuse
at a rate of hundreds of $\mu$m/$\sqrt{cm}$ due to elastic
rescattering in the gas as they drift toward the readout chamber.
The TPC requires careful tuning of the drift field and a high
degree of gas purity.  Many parameters like drift length, track
angle, or the number of primary ions affect the spatial resolution
but a typical value is $\sim$~500~$\mu$m.

\subsection{Solid State Detectors}

Solid state detectors---\textit{silicon micro-strip detectors,
silicon pixel detectors and silicon drift detectors}---offer very
good resolutions of $\pm$~10~--~100~$\mu$m and are now in common
use. Every detector planned at the Large Hadron Collider at CERN
(LHC) will use trackers based on silicon devices.  Silicon
detectors require only 3.6~eV of energy from the traversing
particle to create an electron-hole pair.  That is roughly one
order-of-magnitude less than gas detectors require ($\sim$~30~eV).
This, along with silicon's higher density, means that the number
of electron-hole pairs created by a minimum ionizing particle
(MIP) in silicon is much greater than the number of electron-hole
pairs created over the same distance in gas.  In 1~$\mu$m of
silicon a MIP produces $\sim$~100 charge pairs. To produce that
much charge in gas would require several centimeters. As a result,
unlike gaseous detectors, silicon detectors don't require signal
amplification inside the detector; in a typical silicon detector a
MIP will produce 20~--~30 thousand electrons.

Typically, silicon detectors are built using $\sim$~300~$\mu$m
thick, high resistivity, n-doped silicon plates with a thin
p-doped layer on one side.  A reverse bias voltage---positive on
the n-side and negative on the p-side---is applied to deplete the
silicon of free charge carriers and to create an electric field
that will cause the electrons and holes to drift to opposite
surfaces where readout structures are organized.  The highly
developed state of silicon technology allows for the production of
many different readout structures.

The readout structures for both silicon strip and silicon pixel
detectors are layers of aluminum applied to the surface of the
silicon.  \textbf{Silicon strip detectors} use a solid layer of
aluminum on the n-doped side of the silicon and a sequence of
aluminum strips on the p-doped side.  The strips typically have a
pitch (d) of $\sim$~50~$\mu$m.  The resolution for this pitch is
$\sim$~15~$ \mu$m or $d/\sqrt{12}$.  The charge collected on the
strip is electronically integrated and read out as an analogue or
digital signal.  If strips are placed on both sides of the
silicon---a double sided silicon strip detector---two coordinates
can be measured simultaneously.

The \textbf{silicon pixel detector} uses pixels instead of strips
and has the advantage that it is a true two dimensional
micro-detector. Amplifier circuitry however, needs to be connected
to each pixel which typically has a surface area of only $\sim 50
\times 400 \mu$m$^2$. This is done using specially designed
readout chips that are bump-bonded to the detector silicon.
Silicon pixel detectors are fast, have very low noise (small
capacitance) and have excellent pattern recognition for high
particle densities.  They require however, a large number of
readout channels (100 million is not uncommon), are very fragile
and offer many technological challenges in development.

\textbf{Silicon drift detectors} have two-dimensional capabilities
but by their design avoid the large number of channels required by
silicon pixel detectors.  Silicon drift detectors use a silicon
wafer, with an array of anodes arranged at one edge and cathodes
at the other.  An electric field drifts the primary
electrons---from the track of a passing particle---through the
silicon, toward the array of anodes.  A typical drift speed is
$\sim$~15~mm/$\mu$s.  The anode position along the edge of the
wafer and the drift time give two coordinates for the position of
the track.  The third coordinate is given by the position of the
wafer and, like a spark chamber, is only known to within the
thickness of the wafer.  The signal on the anodes can be read out
at $\sim$~40~MHz---a very high frequency---but the time for the
electrons to drift to the anodes ($\sim$~5~$\mu$s), makes it a
relatively slow detector.  In addition, these detectors require
very precise climate control because of the dependence of the
drift time on temperature.  The high resolution of all these solid
state detectors however, makes them ideal for constructing vertex
chambers that are particularly useful for detecting heavy-flavor
particles.

    \section{The STAR Detector System}
\indent

\begin{figure}[phbt]
\centering\mbox{
\includegraphics[width=0.85\textwidth]{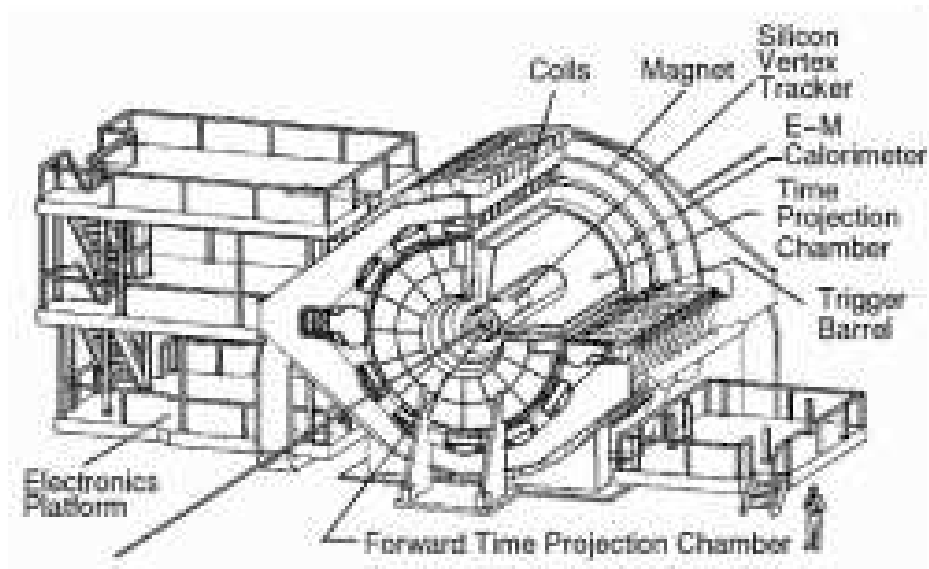}}
\caption[The STAR experiment]{View of the STAR detector system.
\label{fig:stardetector}}
\end{figure}
The STAR detector~\cite{stardet} (Figure~\ref{fig:stardetector})
is an azimuthally symmetric, large acceptance, solenoidal detector
designed to measure many observables simultaneously. The detector
consists of several subsystems and a large Time Projection Chamber
(TPC) located in a 0.5 Tesla solenoidal analyzing magnet.

\begin{figure}[htbp]
\centering\mbox{
\includegraphics[width=1.00\textwidth]{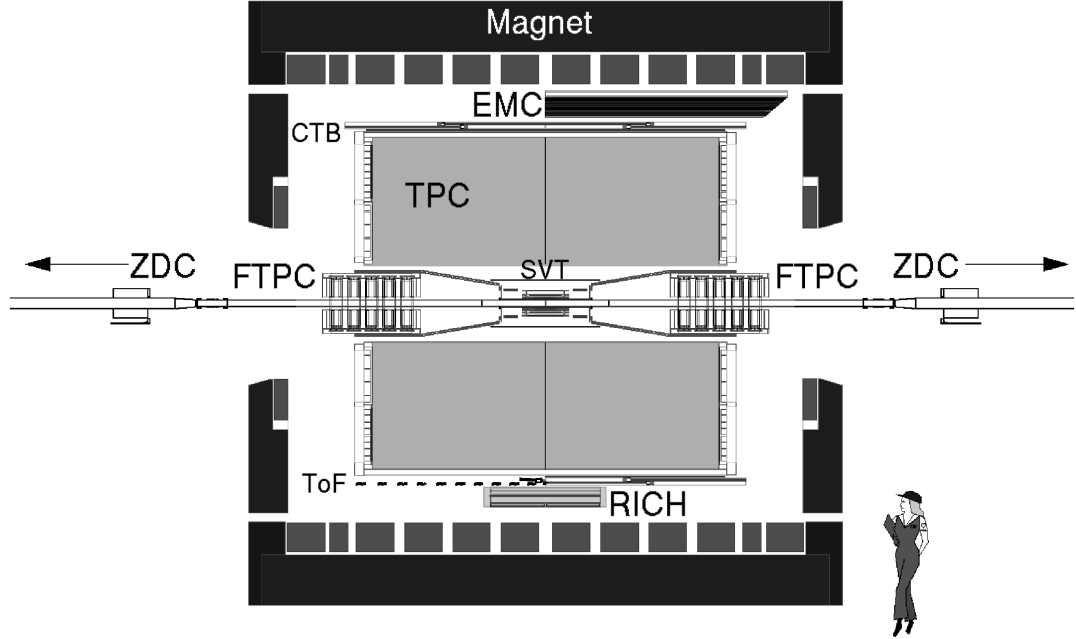}}
\caption[STAR 2001 layout] {Cutaway of the STAR detector in its
2001 configuration; including a partial installation of the
electromagnetic calorimeter (EMC), the temporary ring-imaging
Cherenkov detector (RICH), and a time-of-flight detector (ToF)
prototype. } \label{fig:star2001}
\end{figure}
The layout of the STAR detector system as it was for Run-2 is
shown in Figure~\ref{fig:star2001}. The active subsystems included
two RHIC-standard zero-degree calorimeters (ZDCs) that detect
spectator neutrons, a central trigger barrel (CTB) that measures
event multiplicity, a ring-imaging Cherenkov and time-of-flight
detector that extend particle identification to higher $p_T$, 10
percent of the full barrel electromagnetic calorimeter to measure
photons, electrons and the transverse energy of events, and four
tracking detectors. The tracking detectors are the main TPC, two
forward TPCs, and the silicon vertex tracker (SVT).

The TPC is STAR's primary detector~\cite{tpc} and can track up to
$\sim 4 \times 10^3$ particles per event. For collisions in its
center, the TPC covers the pseudo-rapidity region $|\eta| < 1.8$.
It can measure particle $p_T$ within the approximate range $0.07 <
p_T < 30$~GeV/c. The momentum resolution $\delta p/p$ depends on
$\eta$ and $p_T$ but for most tracks $\delta p/p \sim 0.02$. The
full azimuthal coverage of the STAR detector ($-\pi < \phi < \pi$)
makes it ideal for detecting weak decay vertices, reconstructing
resonances and measuring $v_2$ and other variables requiring
event-by-event characterization.

\begin{figure}[htbp]
\centering\mbox{
\includegraphics[width=1.00\textwidth]{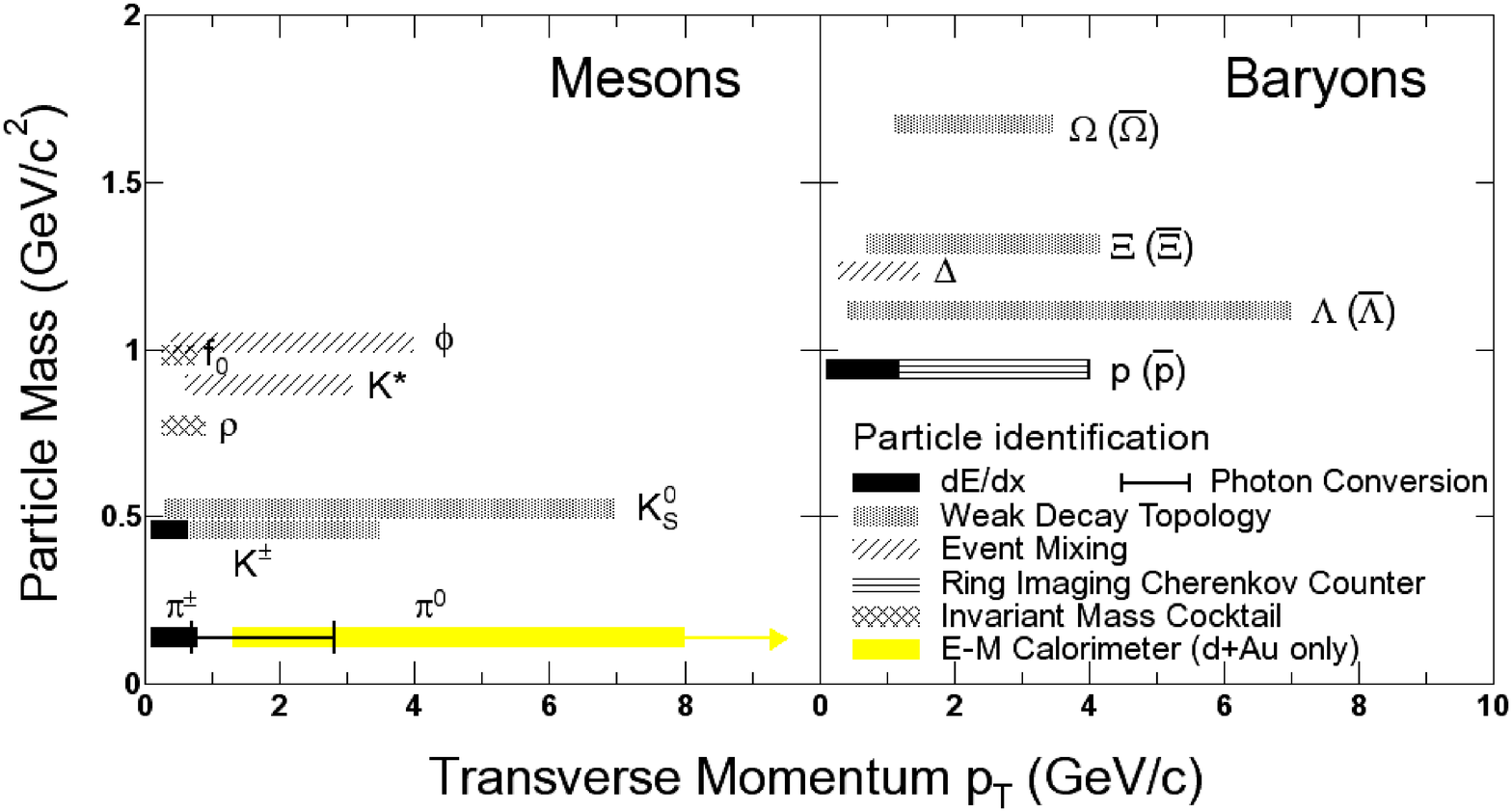}}
\caption[STAR identified particles] {The $p_T$ reach of STAR's
particle identification capabilities with the 2001 detector setup.
} \label{fig:pid}
\end{figure}
Figure~\ref{fig:pid} illustrates the STAR detector's particle
identification capabilities during Run-2. These capabilities will
be further enhanced with detector upgrades, larger data samples,
and more advanced triggering to select rare events. Most of the
measurements illustrated in Figure~\ref{fig:pid} are limited in
$p_T$ coverage by the statistics available. Using the topology of
their weak decays in the TPC, the $K_S^0$ and
$\Lambda(\overline{\Lambda})$ were identified across the largest
$p_T$ range ($0.3 < p_T < 7.0$~GeV/c). The kinematic reach of
these and other topologically identified particle measurements
(\textit{i.e.} $\Xi(\overline{\Xi})$, and
$\Omega(\overline{\Omega})$) will reach their limit when the
momentum of the daughter tracks becomes too high to be accurately
measured in the TPC. As the momentum resolution worsens the
invariant mass calculation will become less accurate. As a result,
the width of the mass peak will broaden. In addition, low $p_T$
particles mis-measured as high $p_T$ particles will start to
dominate the less prominent high $p_T$ signal (feed-down). The
$p_T$ scale where the analysis fails has not been extensively
studied but should depend on the specific particles decay
topology. We naively expect the $K_S^0$ identification to fail
first, around $p_T \sim 15$~GeV/c, where the high $p_T$ signal
will be dominated by low $p_T$ feed-down. For comparison, the
$\pi^0$ identification in the EMC is limited by the detector
technology to roughly $1.5 < p_T < 20$~GeV/c.

With detector upgrades and increased data samples, STAR has the
potential to measure the yield of heavy-flavor mesons and baryons
(particularly for D mesons), charmonium production (J/$\psi$), and
direct photon production. Given its extensive array of particle
identification and event characterization capabilities, the STAR
detector is particularly well suited for characterizing the matter
created in heavy ion collisions.

\subsection{The STAR Trigger Detectors}

The bunch crossing rate at RHIC is $\sim 10$ MHz while the
read-out rate for the STAR TPC is $\sim 100$ Hz. When the
interaction rates approach the bunch crossing rates, the STAR
trigger must reduce the event rate by five orders of magnitude.
The STAR trigger needs to reject background, such as beam-gas
interactions (expected rate $\sim 100$ Hz), select events that
best further our physics goals, and issue triggers to the other
detectors. Furthermore, the future success of STAR may depend on
the ability to trigger on rare events.

With recent upgrades, the STAR detector system has four fast
detectors that can be used as trigger detectors: the central
trigger barrel (CTB), the zero-degree calorimeters (ZDC), a
multi-wire counter (MWC), and the barrel electromagnetic
calorimeter (EMC). In addition, a beam-beam counter (BBC), a
forward $\pi^0$ detector (FPD), and an endcap electromagnetic
calorimeter (EEMC) will become available for
triggering~\cite{trigger}.

The CTB measures the charged particle multiplicity. With 240
scintillator slats each covering $\pi$/30 in $\phi$ and 0.5 in
$\eta$, the whole CTB covers $-1.0 < \eta < 1.0$ and $-\pi < \phi
< \pi$ at a radius of four meters. Its multiplicity resolution is
$\sim 3$\% for multiplicities $> 1000$.

Each RHIC experiment has two ZDC's to monitor beam interactions.
The ZDC's detect the neutrons freed from the Au ions when a
collision occurs (spectator neutrons). The STAR ZDC's are located
$\pm 18.25$ m from the nominal interaction region and subtend an
angle $\theta < 0.002$ radians. Each ZDC consists of three modules
with a series of tungsten plates and layers of wavelength shifting
fibers that route Cherenkov light to a photo-multiplier tube. The
timing of the ZDC signals is also used to locate the longitudinal
position of the interaction vertex.

\begin{figure}[htpb]
\resizebox{.5\textwidth}{!}{\includegraphics{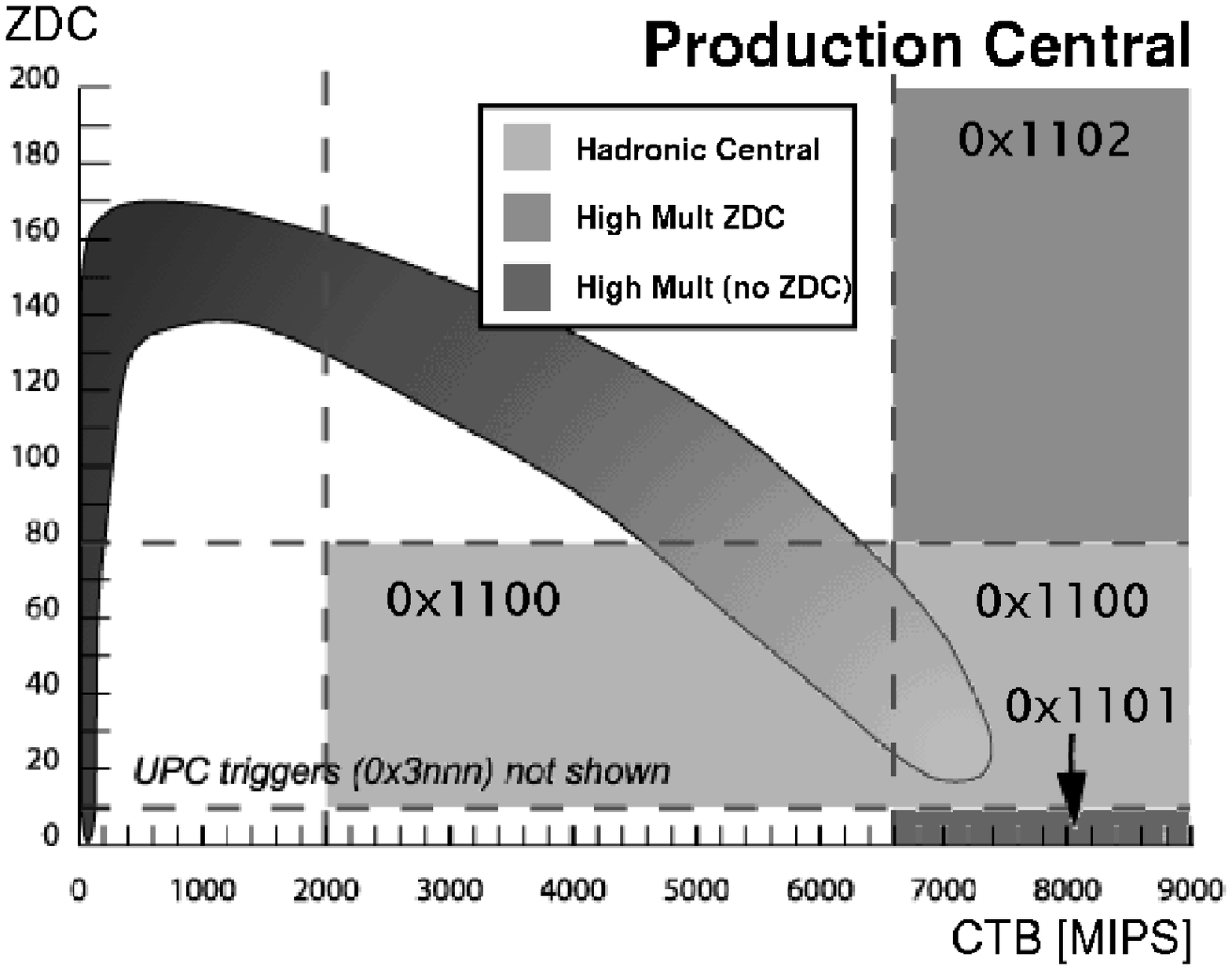}}
\resizebox{.5\textwidth}{!}{\includegraphics{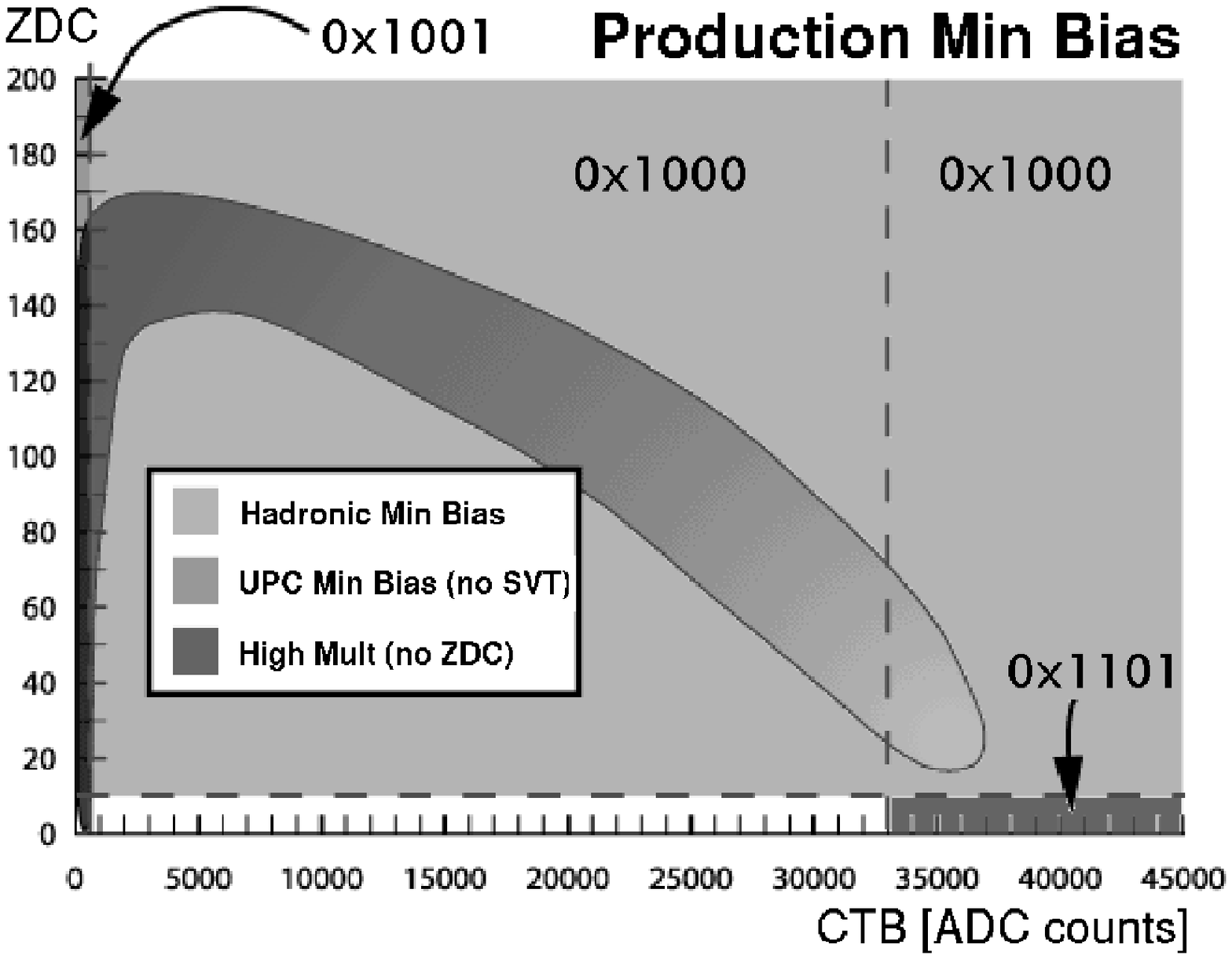}}
\caption[CTB versus ZDC] {A diagram illustrating the STAR trigger
scheme for central (left) and minimum-bias (right) triggers. A
central event will have a low ZDC count and a high CTB count. }
\label{fig:ctbzdc}
\end{figure}

\begin{table}[hbt]
\centering\begin{tabular}{l|l} \hline \hline
{Trigger}               & {Conditions}  \\
\hline
{Hadronic Minbias} & \small{[ZDCe $\ge 5$ \& ZDCw $\ge 5$] \& CTB $\ge 15$ mips} \\
{Hadronic Central}  & \small{[ZDCe $\ge 5$ \& ZDCw $\ge 5$] \& ZDCsum $< 85$} \\
{} & {\& [Vertex Cut] \& CTB $\ge 2000$ mips}\\
\hline \hline
\end{tabular}
\caption[Trigger conditions]{ The two trigger settings used in
this thesis from the 2001 Au+Au data taking.} \label{tab:trigger}
\end{table}
During the 2000 and 2001 Au+Au runs the CTB and ZDC were used to
study minimum-bias, peripheral and central Au+Au collisions.
Table~\ref{tab:trigger} lists the ZDC and CTB conditions for the
two trigger settings used in this analysis; hadronic minimum-bias
and hadronic central. Figure~\ref{fig:ctbzdc} illustrates the
selection scheme for these triggers in the ZDC verses CTB plane.

\subsection{The STAR Time Projection Chamber}
\label{sec:tpc}

\begin{figure}[ht]
\rotatebox{0}{ \centerline{\hbox{
\centering\psfig{figure=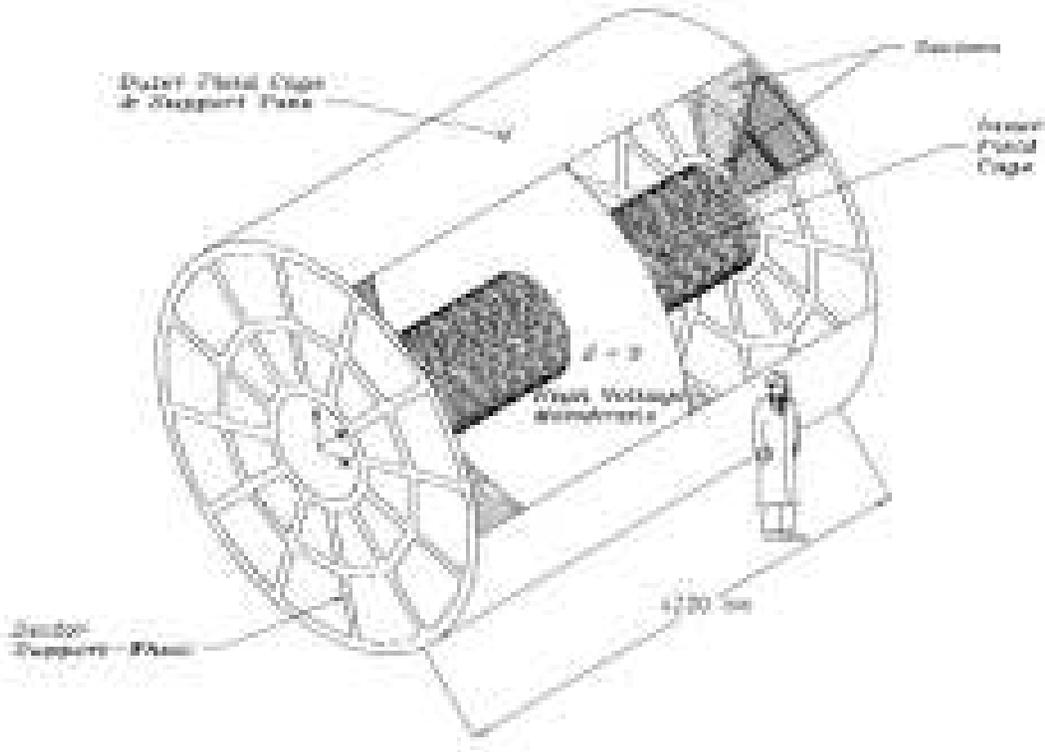,height=4.5in,clip=t}
}}} \caption[Sectioned view of the STAR TPC] {Sectioned view of
the STAR TPC showing the inner/outer field cage, the central
membrane, and inner/outer sectors. \label{fig:tpc_structure}}
\end{figure}
The STAR TPC (Figure~\ref{fig:tpc_structure}) surrounds the
beam-beam interaction region. The inner and outer radii of its
drift volume are 50 cm and 100 cm respectively.  The drift length
from the central membrane to either of the ground planes is 209.3
cm. The central membrane is typically held at 28 kV. A chain of
183 resistors and equipotential rings along the inner and outer
field cage create a uniform drift field from the central membrane
to the ground planes where the anode wires and pad planes are
organized into 12 sectors.

\begin{figure}[htbp]
\centering\mbox{
\includegraphics[width=0.70\textwidth]{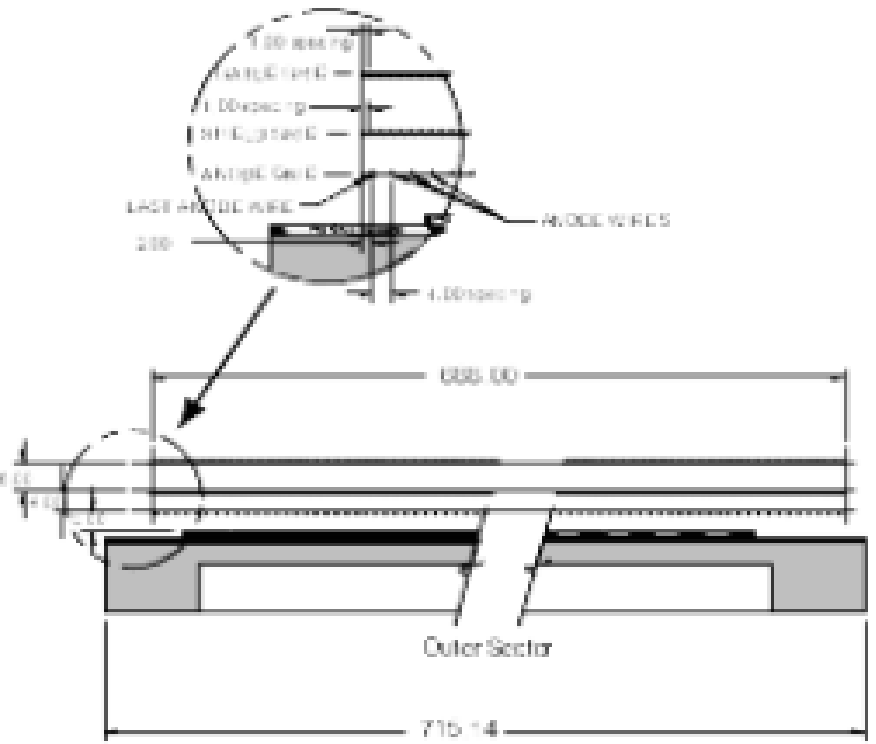}}
\caption[TPC outer wire cutaway] {Cutaway of the pad plane of a
TPC outer sub-sector.} \label{fig:padplane}
\end{figure}

Figure~\ref{fig:padplane} shows a cutaway view of the readout pad
planes of an outer sub-sector. The first of three wire planes is
used as a \textit{gating grid}. The anode wires are located
between a shielding wire plane and the cathode pad plane. In the
open configuration the voltage on the gating grid wires is set so
that ions pass through freely. When it is closed the field lines
terminate on the gating grid wires and the electrons and ions
cannot pass. When the TPC is not being read-out the gating grid is
closed and prevents ions from drifting back into the TPC drift
volume where they can interfere with the uniformity of the drift
field.

The second wire plane shields the TPC drift region from the strong
fields around the anode wires.  As electrons drift past the gating
grid and the shield plane they accelerate towards the anode wires
and initiate a charge amplifying cascade.  The $x$-$y$ position of
the electron-ion pair left in the TPC by a high energy particle is
determined by the position of the cathode pads that detect the
cascade. The $z$ position is determined by the time bucket and the
drift velocity. With 136,608 pad planes and 512 time buckets, the
TPC has over 70 million three-dimensional pixels. In addition, we
use the signal from three adjacent pads to better determine the
cluster centroid and so, the resolution in the pad row direction
is significantly smaller than the pad size. The resolution depends
on the position and orientation of a track relative to the pads,
but it is typically 0.5--1.0 mm.

\begin{figure}[htbp]
\centering\mbox{
\includegraphics[width=1.00\textwidth]{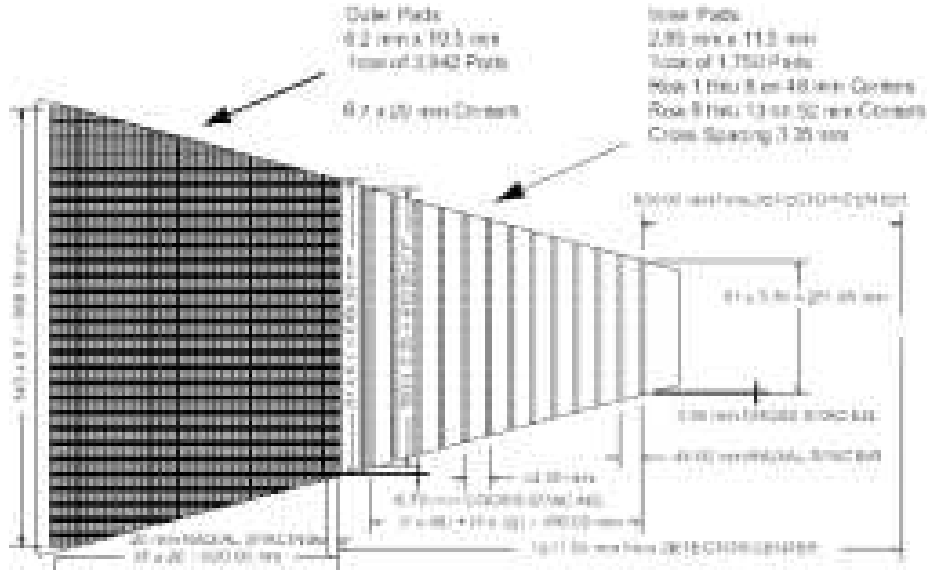}}
\caption[TPC pad plane] { A diagram of a TPC pad plane sector. }
\label{fig:sector}
\end{figure}
Figure~\ref{fig:sector} shows one sector of the TPC pad plane. The
inner sub-sector is designed to handle the higher track density
near the collision vertex. Table~\ref{tab:geometry} lists the
dimensions of the inner and outer sub-sectors. Because of the size
of the front-end electronics, the inner-pad coverage cannot be
made continuous.
\begin{table}[hbt]
\centering\begin{tabular}{l|ll} \hline \hline
{} & {Inner Sub-sector} & {Outer Sub-sector}  \\
\hline
Pad size (mm)             & $2.85 \times 11.50$ & $6.20 \times 19.50$ \\
Pad isolation gap (mm)    & 0.5 & 0.5 \\
Pad rows                  & 13 & 32 \\
Number of pads            & 1750 & 3942 \\
Anode to pad spacing (mm) & 2.0 & 4.0 \\
Anode voltage (V)         & $1170$ & $1390$ \\
Anode gain                & $3770$ & $1230$ \\
\hline \hline
\end{tabular}
\caption[Inner/outer sub-sector geometry]{ Geometry of the inner
and outer sub-sectors.} \label{tab:geometry}
\end{table}

\begin{figure}[htbp]
\rotatebox{0}{ \centerline{\hbox{
\centering\psfig{figure=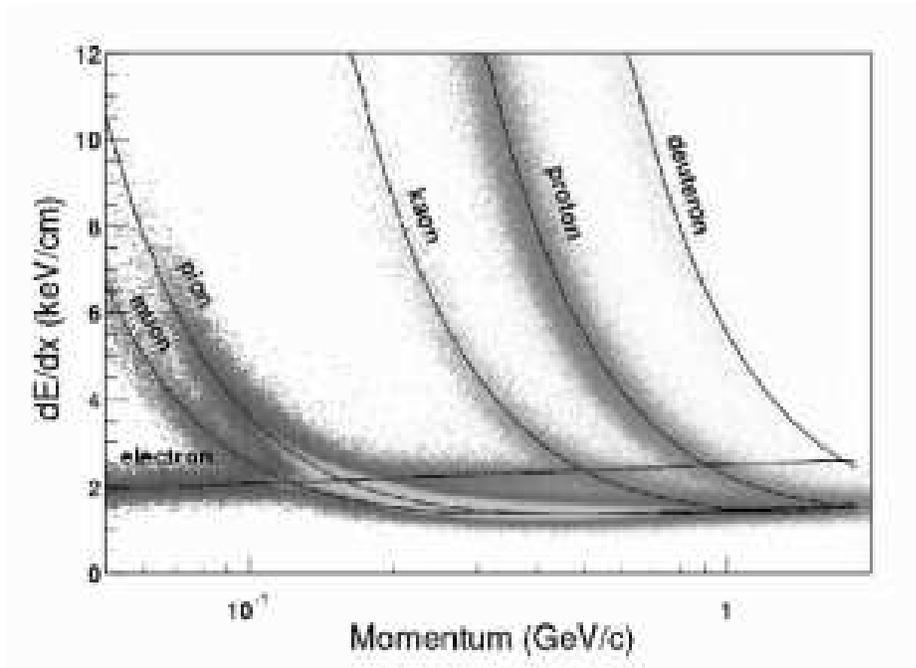,height=3.5in,clip=t} }}}
\caption[STAR TPC dE/dx]{The energy loss of charged particles
traversing the TPC gas. \label{fig:dedx}}
\end{figure}

In addition to tracking charged particles, the TPC is also able to
identify particles by their mass. High energy charged particles
lose energy as they traverse the TPC gas. The average energy loss
depends on their velocity, not their momentum $p$. At a given $p$
below 0.8~GeV/c pions, kaons and protons suffer significantly
different average energy losses. As such, in this region,
measurements of the energy deposited along a particles trajectory
can be used to identify the particle. Figure~\ref{fig:dedx} shows
the energy loss (dE/dx) measured from tracks in the STAR TPC where
the bands correspond to particles with different masses.

\subsection{STAR TPC Gas System} \indent

The TPC gas system~\cite{gas} supplies the TPC with either one of
two gas mixtures---P10 (Ar 90\% + CH$_4$ 10\%) while the detector
is operating or C$_2$H$_6$ 50\% + He 50\% for purging the TPC when
it is not in use. The TPC gas mixture must satisfy multiple
requirements. It is the medium where the particles being tracked
induce ionization, the medium those electron-ion pairs drift
through, and the medium where the electron multiplication takes
place. The convenience and safety of the gas is also considered.

\begin{table}[hbt]
\centering\begin{tabular}{l|l|l} \hline \hline
\multicolumn{3}{c}{Drift Characteristics}  \\
\hline
{Drift Velocity (Maximum)}& {5.45 cm/$\mu$s} & {at 130 V/cm}    \\
{Longitudinal Diffusion}  & {320 $\mu$m/$sqrt(cm)$} & {0.5 Tesla Field}    \\
{Transverse Diffusion}    & {185 $\mu$m/$sqrt(cm)$} & {0.5 Tesla Field} \\
\hline
\multicolumn{3}{c}{Ionization Characteristics} \\
\hline
{Charge Created}          & {227 electrons}    & {from a 5.9 keV X-ray (Fe$^{55}$)}    \\
\hline
\multicolumn{3}{c}{Gain (N/N$_0$) Characteristics} \\
\hline
{Inner Sector Gain}        & {$\approx 3770$} & {for V$_{anode} = 1170$ V}   \\
{Outer Sector Gain}        & {$\approx 1230$} & {for V$_{anode} = 1390$ V}   \\
\hline \hline
\end{tabular}
\caption[TPC gas characteristics]{ Characteristics of the TPC gas
mixture, P10.} \label{tab:tpcgas}
\end{table}

The electron drift velocity in P10 is relatively fast and it peaks
and saturates at a relatively low electric field (130 V/cm).
Operating with a drift field in the saturated region minimizes
variations in the drift velocity. Some of the important
characteristics of P10 are listed in table~\ref{tab:tpcgas}.

The drift velocity and the gas gain are both sensitive to the
pressure and purity of the gas.  The TPC gas pressure varies with
atmospheric pressure so both of these are monitored. Ionization
induced by lasers at fixed locations in the TPC are used to
measure the drift velocity. The gas gain is monitored by a gain
chamber and by observing the energy loss of tracks in the TPC.
Table~\ref{tab:tpcgas2} lists characteristics of the TPC gas
system.

\begin{table}[hbt]
\centering\begin{tabular}{l|l} \hline \hline
\multicolumn{2}{c}{System Characteristics}  \\
\hline
{TPC Volume}           & {$50000$ liters} \\
{Internal TPC Pressure} & {$2.0 \pm 0.03$ mbar} \\
{Recirculation Flow}    & {$36000$ liters/hour} \\
{Oxygen Content}        & {$< 25$ ppm}  \\
{Water Content}         & {$< 20$ ppm} \\
\hline \hline
\end{tabular}
\caption[Gas system parameters]{ Gas system parameters.}
\label{tab:tpcgas2}
\end{table}

\subsection{TPC Gas Gain Monitor} \indent

The UCLA nuclear physics group has been responsible for the
construction and installation of a chamber designed to keep a
minute-by-minute record of the gain of the STAR TPC gas. At
present, we base our corrections for changes in the gas
gain---thought to be primarily due to variation in the gas
pressure---on measurements of the TPC gas pressure.  An attempt is
also made to correct for residual variations in the gain by
estimating the ionization energy loss in the TPC gas (dE/dx) for
tracks which qualify as proton candidates. This step requires
averaging together data taken over several hours. The final
adjusted gain is then used to make a better measurement of the
dE/dx of tracks as they traverse the TPC. As described in
Section~\ref{sec:tpc}, this critical measurement is correlated
with a particles momentum to provide a means of particle
identification (PID).


\begin{figure}[phbt]
\centering\mbox{
\includegraphics[width=0.95\textwidth]{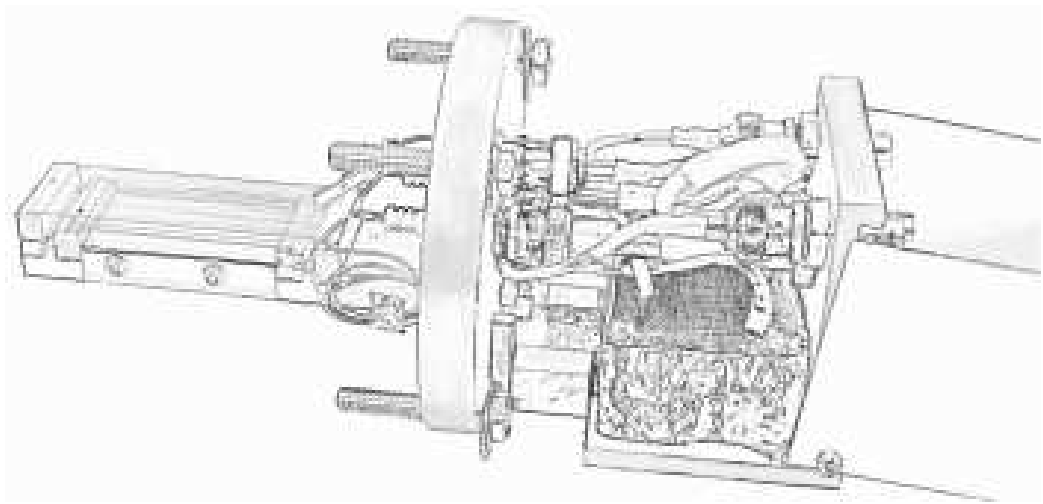}}
\caption[Gain monitor chamber] { The gain monitor chamber with the
wire planes and the shielded electronics exposed.  The shielding
for the wire planes and the Fe$^{55}$ source are not shown.  The
aluminum outer housing is folded back and locked into place when
the chamber is in use so that the high-voltage wires are not
exposed.  } \label{photo}
\end{figure}

With the construction and installation of the new gain monitor
chamber complete, we are able to measure the TPC gas gain directly
and as such may be able to improve the measurement of dE/dx and
PID at STAR.  A source of radiation with a known energy is used to
ionize the TPC gas flowing through the chamber. The deposited
charge accelerates toward the chamber's high voltage anode wires
and a charge amplifying cascade ensues. The pulse generated by the
cascade depends on the energy of the incident radiation and the
gas gain. We use an Fe$^{55}$ source emitting 5.9 keV photons. The
signal from the anode wires is amplified and conditioned with an
Amptek A225 pre-amplifier and shaping amplifier and an Amptek A206
voltage amplifier and low-level discriminator.  The magnitudes of
these pulses are analyzed with an Amptek (PMCA600A) multi-channel
analyzer (MCA).

We fit the spectrum of pulse heights from the MCA to an
exponential function for the background noise and two Gaussian
functions; one for the 5.9 keV peak and another for the secondary
photon escape peak at 2.7 keV. The variation in the position of
the primary peak is used to monitor the relative magnitude of the
gas gain.  We are able to keep the noise level low in the spectrum
and have found that the resolution of the Gaussian peak is
approximately 13\%.


The MCA is read and controlled by a PC (BEATRICE.STAR.BNL.GOV)
located in the data acquisition (DAQ) room of the STAR hall.  No
wires can connect the `outside world' to the electronics platform
where the MCA is located or to the detector where the chamber is
located. Instead, the PC controls the MCA via optical fibers and a
SITECH 2506 fiber-optic modem.  The chamber is attached to a pipe
flange on one of four exhaust manifolds on the face of the TPC.
Its wire planes extend into the pipe where the P-10 exhaust flows
from the TPC. The chamber (shown in Figure~\ref{photo}) is built
to replicate the behavior of the TPC pad planes. Its
geometry---including the diameter of the wire used---matches the
geometry of the outer sub-sector pads in the TPC. The chamber is
electrically isolated from the pipe and the wire planes are
shielded from stray fields by a wire mesh cage surrounding them.
The Fe$^{55}$ source is mounted inside the wire mesh cage several
centimeters above the wire planes.  When in operation the anode
wire plane is held at +1390 V.  The anode plane is located between
a grounded pad plane and a wire ground plane.  The chambers face
is constructed out of non-conducting material so that the bolts
used to attach the chamber to the exhaust manifold are isolated
from the rest of the chamber.  This is necessary because the
exhaust manifold does not share the same ground as the TPC and the
monitor chamber.

\begin{figure}[htbp]
\centering\mbox{
\includegraphics[width=1.0\textwidth]{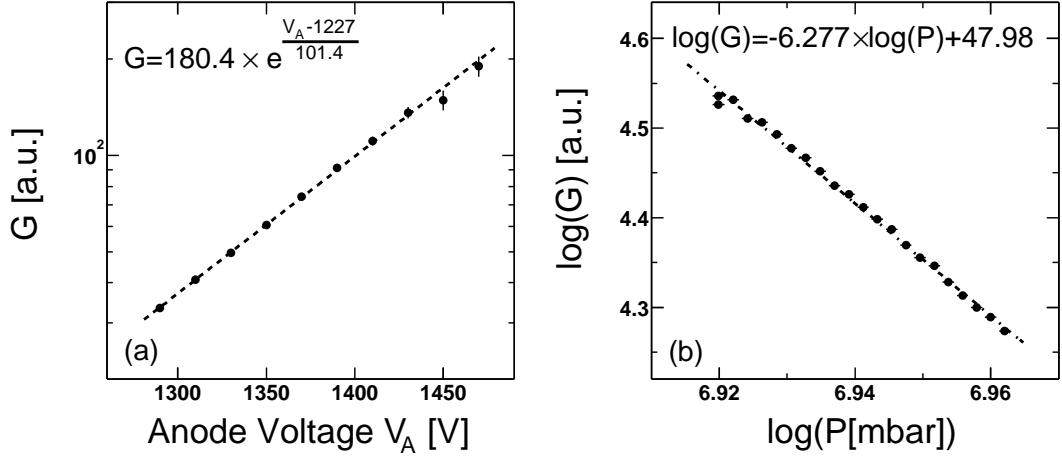}}
\caption[Gain monitor testing and calibrations]{ Left: Gas gain
versus anode wire voltage. Right: Pressure dependence of the gas
gain in P10.} \label{calibrations}
\end{figure}

The chamber was tested at UCLA before installation. The gas gain
variation with respect to the anode wire voltage is shown in
Figure~\ref{calibrations} (a). Although we do not anticipate that
the gain monitor will ever be used with any voltage other than
+1390 V, we ran this test to compare to other gain monitors and to
understand how variations in the supply voltage could affect the
measured gain.  Figure~\ref{calibrations} (b) shows the dependence
of the gas gain on the gas pressure. The expected relationship of
gain to pressure is given by:
\begin{align}
\frac{dG}{G} &\approx -A \frac{dP}{P},\\
\log(G) &\approx -A \log(P) + B, \label{eq:gvsp}
\end{align}
where $G$ and $P$ are the gain and pressure respectively. The
coefficient $A$ is expected to be 6.7 $\pm 1.5$~\cite{blum} and
$B$ is an arbitrary constant. In our calibrations we find $A=6.277
\pm 0.003$ where the error is statistical only.

\begin{figure}[hbt]
\centering\mbox{
\includegraphics[width=1.0\textwidth]{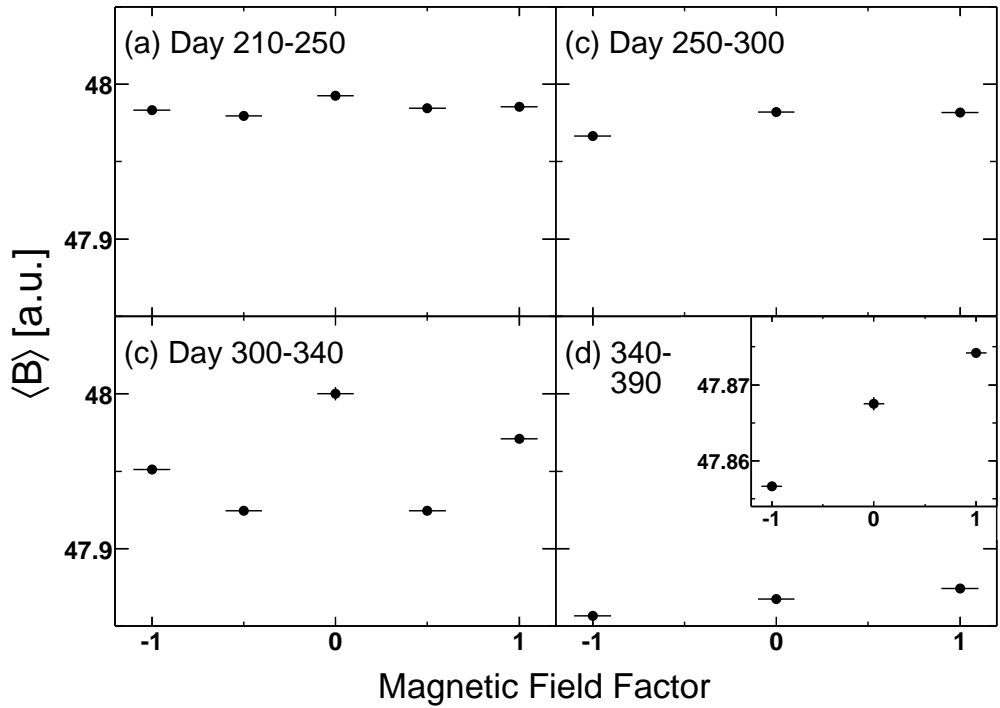}}
\caption[Gain versus magnetic field]{ The variation of the gain
with the magnetic field---after accounting for the pressure
dependence---is shown for four different time periods. The
dependence is not strong in the upper-left plot (earlier in the
Au+Au Run-2) but subsequent measurements show a strong field
dependence. In the inset of panel (d) the vertical axis is
adjusted to emphasize the field dependence. } \label{magField}
\end{figure}

The gain monitor recorded data during the entire 2001--2002 data
taking.  The software proved to be very robust and required little
or no maintenance or intervention from the detector operators or
shift crew.  Several troubling features in the data however,
became apparent during the data taking.  It was noted early in the
Run-2 that the gain measurement varied systematically with the
magnetic field setting (shown in Figure~\ref{magField}). Possible
causes include variations in the response of the electronics with
the magnetic field setting, a change in the charge amplification
caused by the orientation of the wire planes relative to the
magnetic field, or gas leaks exacerbated by the magnetic field.
For the 2003 d+Au collisions the gain monitor was realigned to
match the orientation of the TPC pad planes.

\begin{figure}[hbtp]
\centering\mbox{
\includegraphics[width=1.0\textwidth]{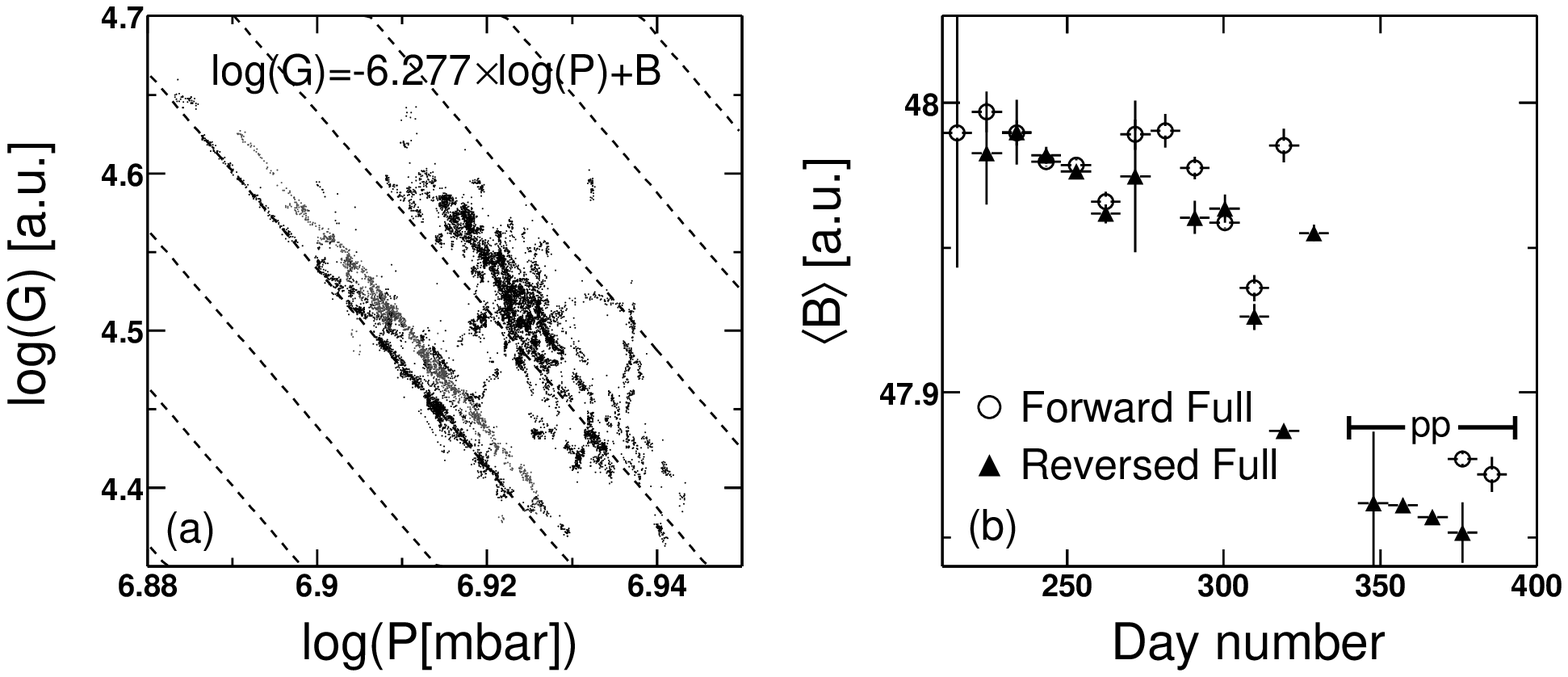}}
\caption[Gain versus pressure and time]{ Left: Scatter plot of
$\log(G)$ versus $\log(P)$. The contours represent the expected
behavior for $\log(G)=-6.277\log(P)+B$. Right: Variation in gain
with time: the mean value of the ``constant'' $B$ is shown after
the expected pressure dependence has been parameterized.}
\label{logGlogP}
\end{figure}

We also find that the relative magnitude of the gas gain decreased
with time; as seen in Figure~\ref{logGlogP} (b).  This time
dependence was observed in other calibration data and has been
accounted for in the dE/dx calibrations.  The importance of the
gain monitor chamber however, can be seen in Figure~\ref{logGlogP}
(a). Scatter is seen in the plot of $\log(G)$ versus $\log(P)$.
This may indicate that there are still variations in the gain not
taken into account with the current calibration method---a method
that is insensitive to gain variations on a short time scale.

Further study of the gain monitor and the gain monitor data is
needed before we use it for calibrations.  The electronics were
returned to UCLA where they were tested in a 0.3 Tesla magnetic
field for variations in pulse height with field direction. No
effect was observed.  We have also built and installed an adapter
that rotates the gain monitor so that the wire planes are
perpendicular to the magnetic field. In the 2001--2002 data taking
the gain monitor data was recorded in the online database, but was
not propagated to the off-line database where it can be easily
used for calibrations. This will be changed for future data taking
and we anticipate that the gain monitor will be used for
calibrations as well as diagnostics.

\clearpage

\chapter{Analysis Methods}

The technique for finding $K_S^0$, $\Lambda$, or
$\overline{\Lambda}$ candidates with the STAR detector and
calculating their $p_T$ and $y$ distributions is well
established~\cite{Lam:Adler:2002pb,huilong}\footnote{The author
thanks H. Long for his assistance in making the measurements
presented in this thesis. His work with weak-decay-vertex finding
has become a cornerstone of the STAR collaborations scientific
program.}. Up to now, measurements of identified particle $v_2$
have relied on pure particle identification (90\%
purity\footnote{Purity is defined as the raw yield of the particle
at a given dE/dx value, divided by the sum of all other particle
yields with the same dE/dx.}) via dE/dx measurements in the TPC
gas~\cite{Adler:2001nb}. We've adapted the $v_2$ analysis method
to calculate $v_2$ for particles identified only on a statistical
basis. With this method it is possible to calculate $v_2$ for
identified particles independent of the particle sample's
signal-to-background ratio.

In this chapter, we describe the selection criteria for events,
tracks and the $K_S^0$, $\Lambda$, or $\overline{\Lambda}$
candidates. Details of the $R_{CP}$ and spectra
measurements---including an analysis of systematic errors---are
given. Finally, the analysis methods for measuring $v_{2}$ of
$K_S^0$, $\Lambda$, and $\overline{\Lambda}$ are presented along
with a discussion of the systematic errors associated with the
$v_2$ analysis.

\section{Event and Track Selection}
\indent

\begin{table}[hbt]
\centering\begin{tabular}{l|cc|cc} \hline \hline
Data set & \multicolumn{2}{c|}{Minimum-bias} & \multicolumn{2}{c}{Central} \\
\hline
 & Recorded & Used & Recorded & Used \\
\hline
Run-1 ($\sqrt{s_{_{NN}}}=130$ GeV) & $0.8 \times 10^6$ & $0.20 \times 10^6$ &  $0.8 \times 10^6$ & $0.18 \times 10^6$ \\
Run-2 ($\sqrt{s_{_{NN}}}=200$ GeV) & $4.0 \times 10^6$ & $1.6 \times 10^6$ & $3.5 \times 10^6$ &  $1.5 \times 10^6$ \\
\hline \hline
\end{tabular}
\caption[Recorded and usable events] {The number of Au+Au events
used in our analysis.} \label{tab:events}
\end{table}
%
To date, for the STAR experiment, the number of events that are
useful for our analysis has been $\sim 40 \%$ of the total
recorded. Table~\ref{tab:events} lists the number of events
recorded and used for Run-1 and Run-2. Events for which no primary
vertex is found are discarded. For Run-1, events with $z$-vertex
further than 75 cm from the TPC center were discarded. For Run-2,
improvements in the accelerator allowed STAR to select only
collisions within 25 cm of the TPC center. Still more events are
discarded to remove trigger biases.
A large sample of minimum-bias data was taken with a tight
$z$-vertex cut applied in the level zero trigger that biased the
sample.  These events require more careful analysis and are not
included in this analysis.

\begin{figure}[htpb]
\centering\mbox{
\includegraphics[width=0.80\textwidth]{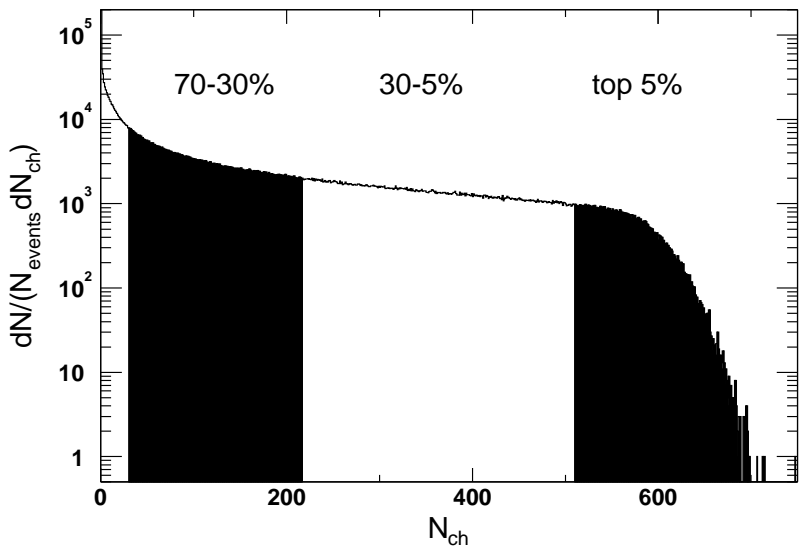}}
\caption[Charged particle multiplicity] {The TPC charged particle
multiplicity distribution for $\sqrt{s_{_{NN}}}=200$~GeV.
N$_{\mathrm{ch}}$ is the number of primary tracks in $|\eta| <
0.5$. The fractions of the total cross-section used for the
analysis of $v_2$ are shown. } \label{fig:evolution}
\end{figure}
The multiplicity as measured by the TPC---not the CTB---is used to
define STAR's centrality intervals.  The TPC reference
multiplicity for Run-2 is the total number of \textit{primary
tracks} in the TPC with 10 or more fit points, having
$|\eta|<0.5$, and a distance of closest approach (DCA) to the
primary vertex less than 3 cm. A primary track is defined by a
helix fit to the TPC points \textit{and} to the primary vertex;
the global track fits do not include the primary vertex. For Run-1
primary tracks within $|\eta|<0.75$ were used to define the
multiplicity.


\begin{table}[hbt]
\centering\begin{tabular}{lcc} \hline \hline
{} & Event plane & $K_S^0$ and $\Lambda(\overline{\Lambda})$ \\
\hline
Track set & Primary & Global \\
\hline
DCA to primary vertex (cm)    & $< 3$    & \textit{na} \\
Number of hits                & $> 15$   &  $> 15$     \\
Number of hits/possible hits & $> 0.52$ & \textit{na} \\
$|\eta|$                      & $< 1.5$  & \textit{na} \\
Momentum (GeV/c)              & $0.1 < p_T < 2.0$ & $p_T > 0.05$ \\
\hline \hline
\end{tabular}
\caption[Track selection criteria] {Track selection criteria for
Au+Au collisions at $\sqrt{s_{_{NN}}} =
200$~GeV.}\label{tab:track_cuts}
\end{table}
Table~\ref{tab:track_cuts} list the selection criteria for tracks
used in the analysis of $\sqrt{s_{_{NN}}} = 200$~GeV data. For the
$K_S^0$, $\Lambda$, or $\overline{\Lambda}$ reconstruction when
the dE/dx of a track can be used to identify the particle type, an
additional dE/dx cut is made. For $0.2 < p_T < 2.0$~GeV/c, pion
candidates are required to have a dE/dx value within $6 <
\sigma_{\pi} < 5$ and proton candidates are required to have a
dE/dx value within $2.85 < \sigma_{p} < 10$. These cuts are very
loose and only act to exclude tracks which are obviously not of
the correct type. The most effective selection criteria in the
identification of $K_S^0$, $\Lambda$, or $\overline{\Lambda}$
particles are the decay topology cuts.

\section{Decay Vertex Topology: Yield Measurements}
\indent

\begin{figure}[htbp]
\centering\mbox{
\includegraphics[width=1.0\textwidth]{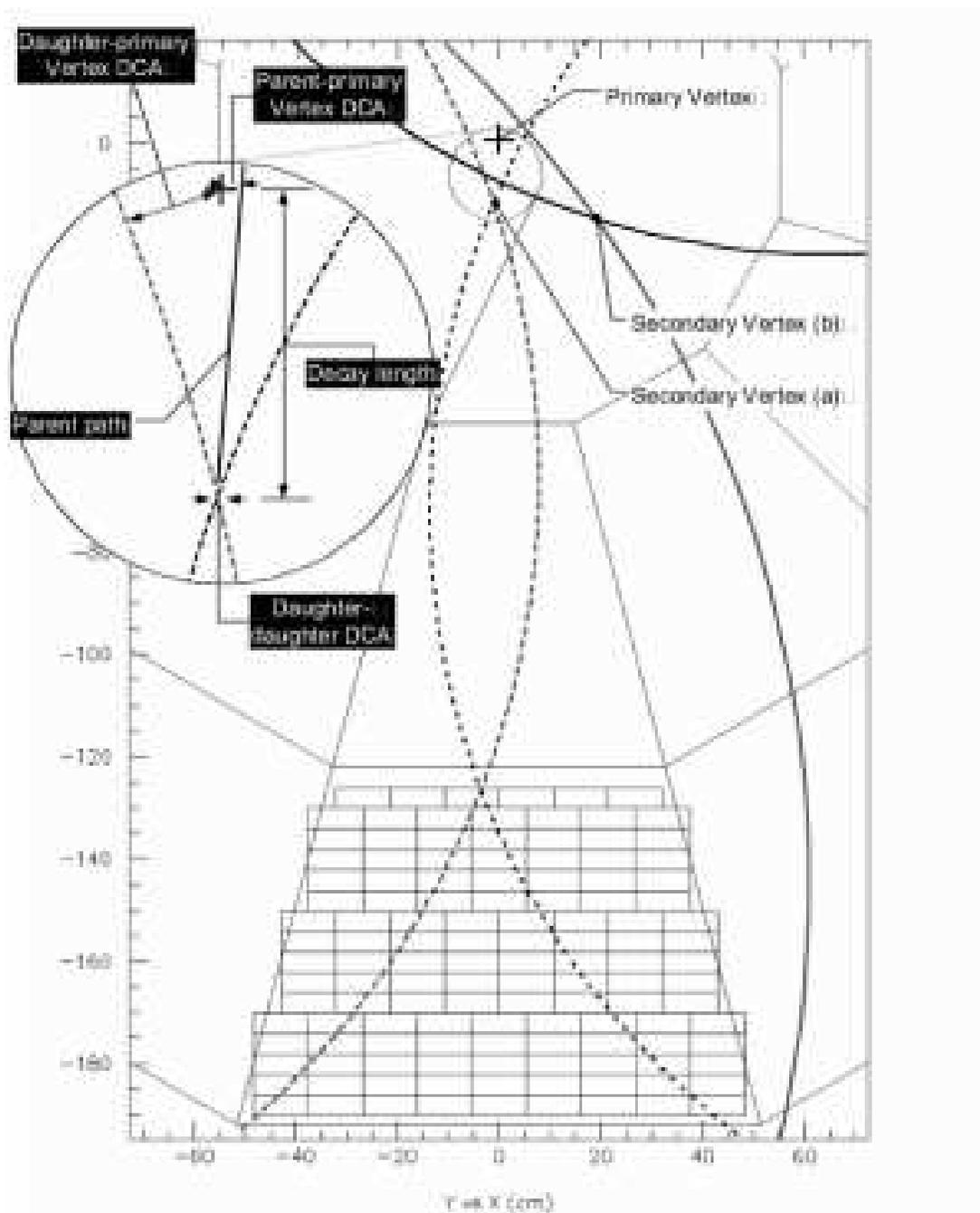}}
\caption[Neutral particle weak decay]{ A sketch of the geometry of
a neutral particle decay in the TPC. Case "a" shows charged
daughter tracks curving towards each other after they are created
in the decay while in case "b" they are curving away from each
other.} \label{fig:v0decay}
\end{figure}

We identify the $K_S^0$, $\Lambda$ and $\overline{\Lambda}$
candidates from the charged daughter tracks produced in the weak
decays: $K_S^0 \rightarrow \pi^+\pi^-$, $\Lambda \rightarrow
p\pi^-$ and $\overline{\Lambda} \rightarrow \overline{p}\pi^+$. To
select the $K_S^0$, $\Lambda$, or $\overline{\Lambda}$ candidates,
we calculate the distance of closest approach (DCA) between all
combinations of selected global tracks within an event. We define
the four-momenta of the daughter particles by assuming they
originated from the points on the two helices where the DCA
occurs, and by choosing a mass hypothesis appropriate for the
weak-decay channel. We use the four-momentum of the two daughter
particles to calculate the invariant mass and kinematic properties
of the candidate.


\begin{table}[hbt]
\centering\begin{tabular}{lccc} \hline \hline
{Candidate $p_T$ (GeV/c)} & {$<1.6$} & {1.6--3.0} & {$>3.0$}  \\
\hline
{Daughter--daughter DCA }       & {$<0.80$}    & {$<0.80$}    & {$<0.80$}    \\
{Daughter--$V_{prim}$ DCA } & {$>3.00$}    & {$>3.00$}    & {$>0.50$}    \\
{Decay Length }                 & {4.0--25.0} & {4.0--40.0} & {5.0--60.0} \\
{$V^0$--$V_{prim}$ DCA }    & {$<0.80$}    & {$<0.80$}    & {$<0.80$}    \\
\hline \hline
\end{tabular}
\caption[Candidate selection criteria ($R_{CP}$)]{ Selection
criteria for the final analysis of the $K_{S}^{0}$ $R_{CP}$. Units
are centimeters except where indicated.} \label{tab:v0cuts2}
\end{table}

Further selection criteria (\textit{i.e.} cuts) are applied to the
orientation of the two tracks---with respect to each other and
with respect to the primary vertex---to increase the probability
that the track combination is associated with a real decay.
Figure~\ref{fig:v0decay} illustrates the geometry of a
neutral-particle decay vertex ($V^0$). For $K_S^0$, $\Lambda$ or
$\overline{\Lambda}$ decays in a magnetic field, two equally
probable cases occur: the daughter tracks curve towards each other
or the daughter tracks curve away from each other. The geometric
variables used to select $K_S^0$, $\Lambda$ or
$\overline{\Lambda}$ decays are shown in the figure.
Table~\ref{tab:v0cuts2} shows the $K_S^0$ selection criteria used
for the spectra and $R_{CP}$ analysis. We choose the vertex
geometry cuts to minimize the statistical and systematic
uncertainty in the measured $R_{CP}$.

\subsection{Invariant Mass Distributions} \indent

The $K_S^0$, $\Lambda$, or $\overline{\Lambda}$ particles are not
identified on a particle-by-particle basis but their uncorrected
yields are extracted from the peak at their known masses in the
invariant mass distributions. The yield is estimated by fitting a
smooth function to the combinatorial background outside the peak
region. We determined that the background is dominated by
combinatorial counts by rotating all positive tracks 180 degrees
in the transverse plane and reconstructing the $K^0_S$ and
$\Lambda(\overline{\Lambda})$ decay vertices.
This procedure destroys all real vertices within our acceptance so
that we can describe the combinatorial contribution to the
invariant mass distributions.

\begin{figure}[htbp]
\centering\mbox{
\includegraphics[width=1.00\textwidth]{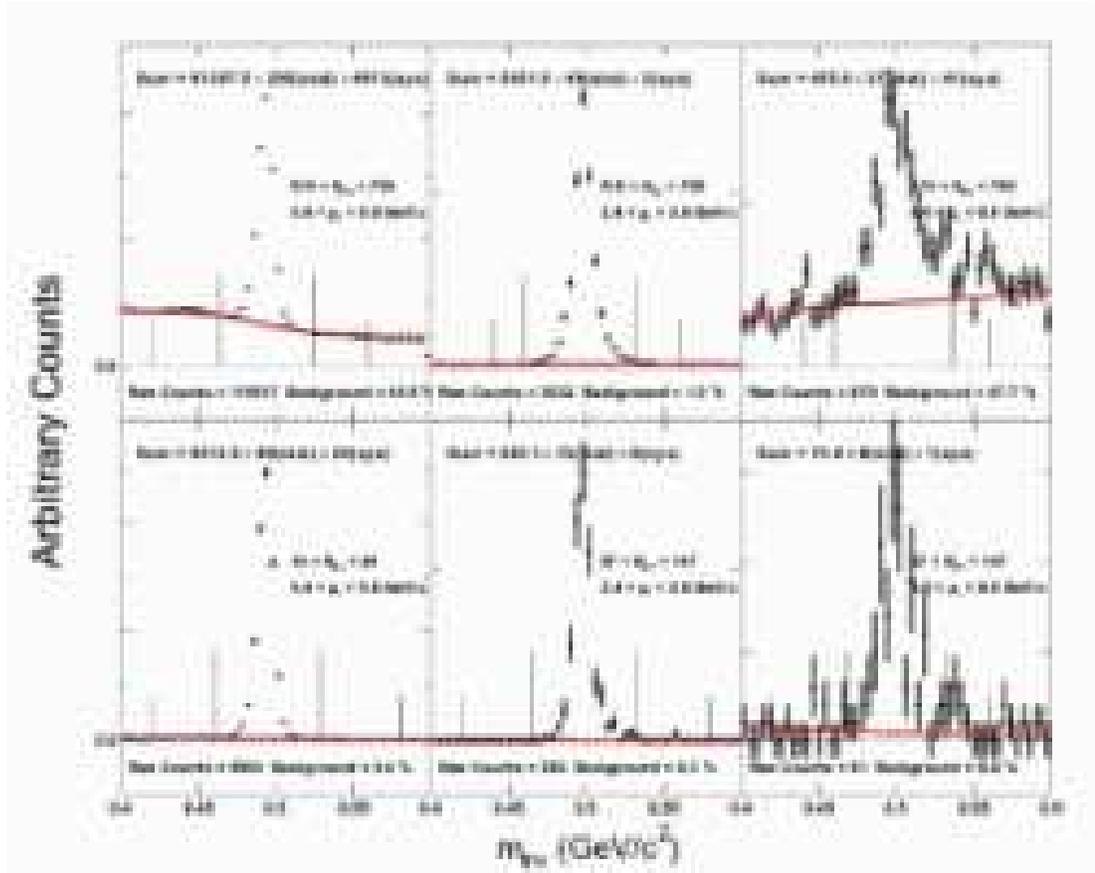}}
\caption[Invariant mass]{ Invariant mass distributions for
$\pi^+\pi^-$ from 0--5\% central collisions (top) and 40--60\%
central collisions (bottom) in three $p_T$ intervals.}
\label{fig:minv}
\end{figure}

The observed masses, $496 \pm 8$~MeV/$c^2$ for $\pi^+\pi^-$ and
$1116 \pm 4$~MeV/$c^2$ for $p\pi$, are roughly consistent with
accepted values~\cite{pdg} and the widths are determined by the
momentum resolution of the detector. For $p_T < 1.5$~GeV/c,
however, the $K_S^0$ peak is shifted to a lower mass. At $p_T =
0.2$~GeV/c the peak is shifted by the greatest amount, 10~MeV/c.
This shift is, for the most part, replicated by simulations and is
attributed to energy loss suffered by the daughter particles in
the detector material. Figure~\ref{fig:minv} shows invariant mass
distributions for the $K_S^0$ $R_{CP}$ analysis. When the same
selection criteria are used for all $p_T$ and centrality, the
combinatorial background is larger for lower $p_T$ and for more
central events. At higher $p_T$, where the size of our data sample
is limited, we place less stringent requirements on the
candidates.  As a result, the combinatorial background is quite
large in central events for $p_T > 3.0$.

\subsection{Detector, Tracking and Reconstruction Efficiency}
\indent

Simulations are used to calculate the efficiency of the detector
and the tracking software~\cite{huilong}.  The TPC response to
Monte-Carlo generated $K_S^0$, $\Lambda$, or $\overline{\Lambda}$
decays is simulated. The simulated clusters (the pixel level TPC
response) are then embedded into real events and these events are
passed into the $K_S^0$, $\Lambda$, or $\overline{\Lambda}$
reconstruction chain. Reconstructed candidates are then associated
with the embedded particles so that the efficiency of the detector
and the reconstruction chain can be estimated.

\begin{figure}[htbp]
\centering\mbox{
\includegraphics[width=0.99\textwidth]{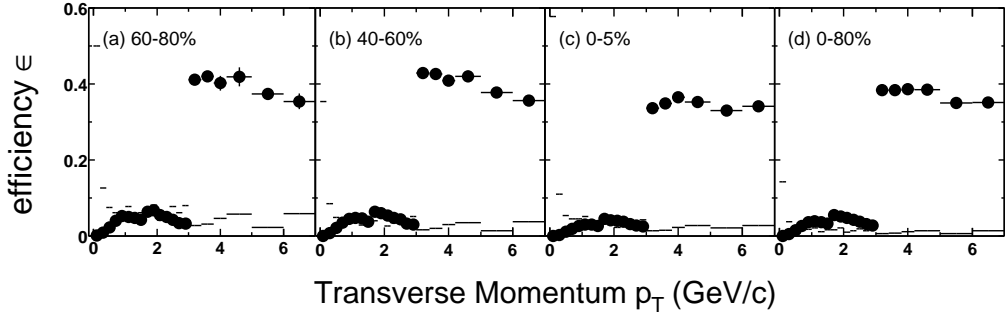}}
\caption[$K_S^0$ reconstruction efficiency]{ The $K_S^0$
efficiency for the selection criteria listed in
Table~\ref{tab:v0cuts2}. The discontinuities in the efficiency at
$p_T=1.6$ and 3.0~GeV/c reflect changes in the selection criteria
(see Table~\ref{tab:v0cuts2}). The line histograms show the
relative magnitude of the statistical error for the efficiency.}
\label{fig:eff}
\end{figure}

Figure~\ref{fig:eff} shows the efficiency correction factor for
$K_S^0$ in the three centrality intervals used in the calculation
of $R_{CP}$. The number of simulated $K_S^0$ particles we embed at
a given $p_T$ is varied to approximate the true $p_T$ dependence
of the yield.  The \textit{slope} of the $p_T$ spectrum is
characterized by the inverse slope parameter T of an exponential
fit. By matching the slopes in the simulations to the real slopes,
the bin-sharing for the simulated particles replicates the
bin-sharing for real particles. In this way, the efficiency
correction also acts as a feed-down correction. With a slope on
the embedded particle yield, it is necessary to embed within
limited $p_T$ ranges to generate high statistics at large $p_T$.
To match the feed-down in real data, simulated data cannot be used
near the edge of the embedded $p_T$ range.
Table~\ref{tab:embedding} lists the slope parameters, the $p_T$
ranges, and the number of events used for the embedding.

\begin{table}[hbt]
\centering\begin{tabular}{cccc} \hline \hline
{Embedded $p_T$ range (GeV/c)} & {T (MeV/c$^2$)} & {$p_T$ coverage (GeV/c)} & {Events}  \\
\hline
{0.0--1.4}  & {$300$} & {0.0--1.2} & {$39$ k} \\
{1.0--2.2}  & {$315$} & {1.2--2.0} & {$45$ k} \\
{1.8--2.8}  & {$330$} & {2.0--2.6} & {$49$ k} \\
{2.4--3.4}  & {$330$} & {2.6--3.4} & {$36$ k} \\
{3.0--5.2}  & {$450$} & {3.4--5.0} & {$15$ k} \\
{4.6--10.0} & {$500$} & {5.0--8.0} & {$19$ k} \\
 \hline \hline
\end{tabular}
\caption[Embedded data]{ Embedded data for $K_S^0$ analysis.}
\label{tab:embedding}
\end{table}


\subsection{Systematic Uncertainties} \indent

Systematic errors for the spectra and $R_{CP}$ are introduced from
uncertainties in the detection efficiency, the reconstruction
efficiency, the background subtractions and from mis-measurements
of the candidates $p_T$.  The momentum resolution $\delta p_T/p_T$
is estimated from simulations. The other systematic uncertainties
are studied by varying the $K_S^0$, $\Lambda$, and
$\overline{\Lambda}$ selection criteria. Changing the selection
criteria varies the number of background counts and tests how well
the distributions in the simulated data match the real data.  When
all the relative distributions are accurately simulated, changing
the cuts will not change the efficiency corrected particle yields.

\begin{figure}[htbp]
\centering\mbox{
\includegraphics[width=0.80\textwidth]{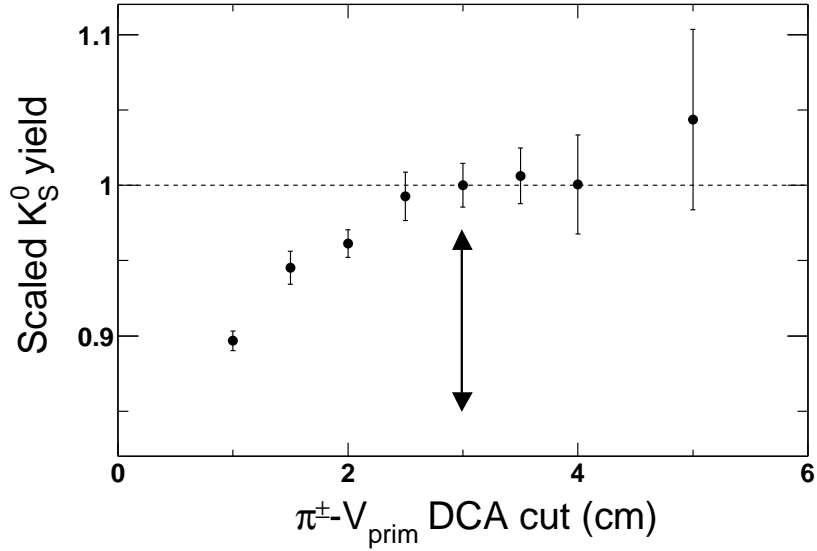}}
\caption[Systematic yield variation]{ The $K_S^0$ yield for
different $\pi^{\pm}-V_{prim}$ DCA selection criteria, relative to
a 3.0 cm cut.} \label{fig:sysdca}
\end{figure}
We found that the $K_S^0$ yield depends strongly on the
$\pi^{\pm}-V_{prim}$ DCA cut. This variation is most likely caused
by either space-charge or E$\times$B distortions which arise as
ionization drifts toward the TPC pad planes.
Figure~\ref{fig:sysdca} shows the measured $K_S^0$ yield for
several $\pi^{\pm}-V_{prim}$ DCA values.  The yields are scaled by
the yield from the $\pi^{\pm}-V_{prim}$ DCA $> 3.0$ cm analysis.
No $p_T$ dependence is seen in the variation of the scaled yield,
so the $p_T$ integrated yield is used in Figure~\ref{fig:sysdca}.
For 3.0 cm and above the yield is independent of the
$\pi^{\pm}-V_{prim}$ DCA, so we used a 3 cm cut for our final
analysis.

\begin{figure}[htbp]
\centering\mbox{
\includegraphics[width=0.90\textwidth]{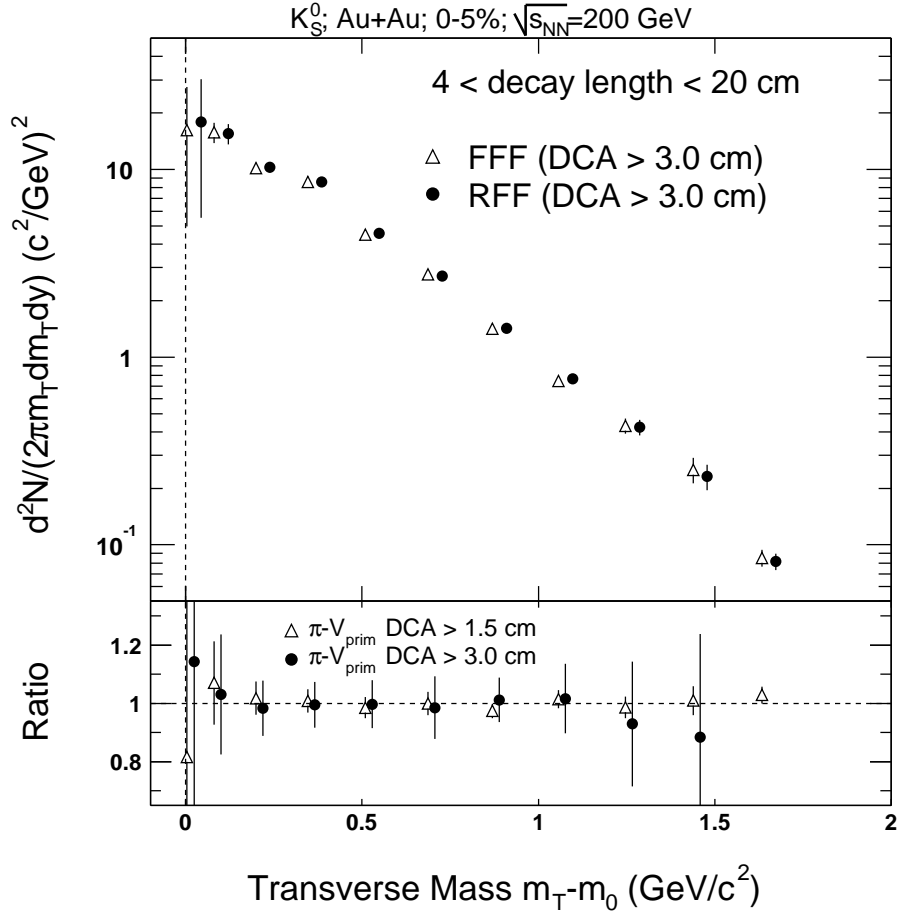}}
\caption[Magnetic field dependence]{ $K_S^0$ yield for forward and
reversed field settings.} \label{fig:sysfield}
\end{figure}
Figure~\ref{fig:sysfield} shows the $K_S^0$ spectra for forward
and reversed field settings.  Early in the analysis an
approximately 8\% systematic deviation was seen between the yields
from events with the two field settings.  By selecting only
$K_S^0$ candidates with a decay length from 4--20 cm the deviation
is removed. Final estimates for the systematic errors on $R_{CP}$
are dominated by variation in the yields with different selection
criteria. Table~\ref{syserr_rcp} lists systematic uncertainties
for $K_S^0$ and $\Lambda+\overline{\Lambda}$ $R_{CP}$.

\begin{table}[hbt]
\centering\begin{tabular}{l|r@{.}lr@{.}lr@{.}l|r@{.}lr@{.}lr@{.}l}
\hline \hline
~ & \multicolumn{6}{c}{$K_S^0$} \vline & \multicolumn{6}{c}{$\Lambda + \overline{\Lambda}$} \\
\hline
$p_T$ (GeV/c) & 1&0 & 2&5 & 4&0 & 1&0 & 2&5 & 4&0 \\
\hline $R_{CP}$ (bg)
    & $\pm0$&$04$ & $\pm0$&$02$ & $\pm0$&$08$ & $\pm0$&$02$ & $\pm0$&$04$ & $\pm0$&$06$ \\
$R_{CP}$ (eff)
    & $\pm0$&$10$ & $\pm0$&$10$ & $\pm0$&$10$ & $\pm0$&$10$ & $\pm0$&$10$ & $\pm0$&$10$ \\
\hline $\delta p_T/p_T$
    & $\pm0$&$012$ & $\pm0$&$027$ & $\pm0$&$030$ & $\pm0$&$016$ & $\pm0$&$027$ & $\pm0$&$037$ \\
\hline \hline
\end{tabular}
\caption[Systematic errors ($R_{CP}$)]{The systematic errors from
background (bg), and the efficiency calculation (eff) are listed
for $R_{CP}$ (0--5\%/40--60\%) at three $p_T$ values along with
the $p_T$ resolution ($\delta p_T/p_T$). The values listed are
relative errors and do not include the overall normalization
uncertainty in $R_{CP}$ from the calculation of
$\mathrm{N_{bin}}$. } \label{syserr_rcp}
\end{table}

\section{Reconstructing the Reaction Plane}
\indent

The real reaction plane is not known, but the \textbf{event
plane}, an experimental estimator of the true reaction plane, can
be calculated from the azimuthal distribution of primary
tracks~\cite{Poskanzer:1998yz}.
The selection criteria for the primary tracks used to calculate
the event plane are given in Table~\ref{tab:track_cuts}.
We require the ratio of the number of space points to the expected
maximum number of space points for each track to be greater than
0.52 to prevent split tracks from being counted twice.
For the analysis using $\sqrt{s_{_{NN}}} = 130$~GeV data the
events are required to have a primary vertex within 75~cm
longitudinally of the TPC center ($z$-vertex). During the Au+Au
running with $\sqrt{s_{_{NN}}} = 200$~GeV the $z$-vertex
distribution was narrower so those events are required to have
$|z$-vertex$| < 25$~cm.
These cuts do not bias our analysis.

For experiments at RHIC energies---unlike those at lower
energies---the reaction plane is assumed to be transverse to the
beam axis.  As such, since it is not necessary to rotate the flow
coordinate system in the polar direction, only the transverse
direction is considered. The error introduced by this assumption
will go as the square of the polar flow angle, $\theta_{Flow} \sim
\langle p_{x} \rangle /p_{beam} \ll 1$, and is negligible.


With a perfect detector the azimuthal distribution of the event
plane would be isotropic.  In a realistic environment, however,
limitations to a detectors acceptance lead to a bias in the
estimation of the reaction plane. \textbf{Acceptance corrections}
are introduced to account for both the limited coverage and the
imperfect efficiency of a real detector. With the STAR detector,
detector biases are removed from the event plane distribution by
applying weights to the tracks used in its calculation. The
\textbf{$\mathbf{\phi}$-weights} are generated by inverting the
normalized $\phi$ distribution for tracks from many events. When
other weights are included in the event plane calculation
(\textit{e.g.} $p_T$ weights) they are also applied to the
$\phi$-weights. In this way, after the $\phi$-weights are applied
the azimuthal distribution for tracks is---by
construction---isotropic.

The acceptance of the STAR detector system depends on the
kinematic variables of a particle ($\eta$, $\phi$ and $p_T$), the
longitudinal position of the collision vertex in the TPC, the
multiplicity of the event and the hour-by-hour state of the
detector. During Run-2 electronics failures resulted in the
temporary removal of read-out (RDO) boards from the data
acquisition chain. The masking and unmasking of RDO boards changed
the detectors acceptance with time. To ensure that all these
variations are accounted for properly, the $\phi$-weights are
calculated separately for positive or negative $\eta$, for
positive or negative z-vertex position, for magnetic field
polarity, for nine different centralities and for four different
detector states.

\begin{figure}[htb]
\centering\mbox{
\includegraphics[width=1.0\textwidth]{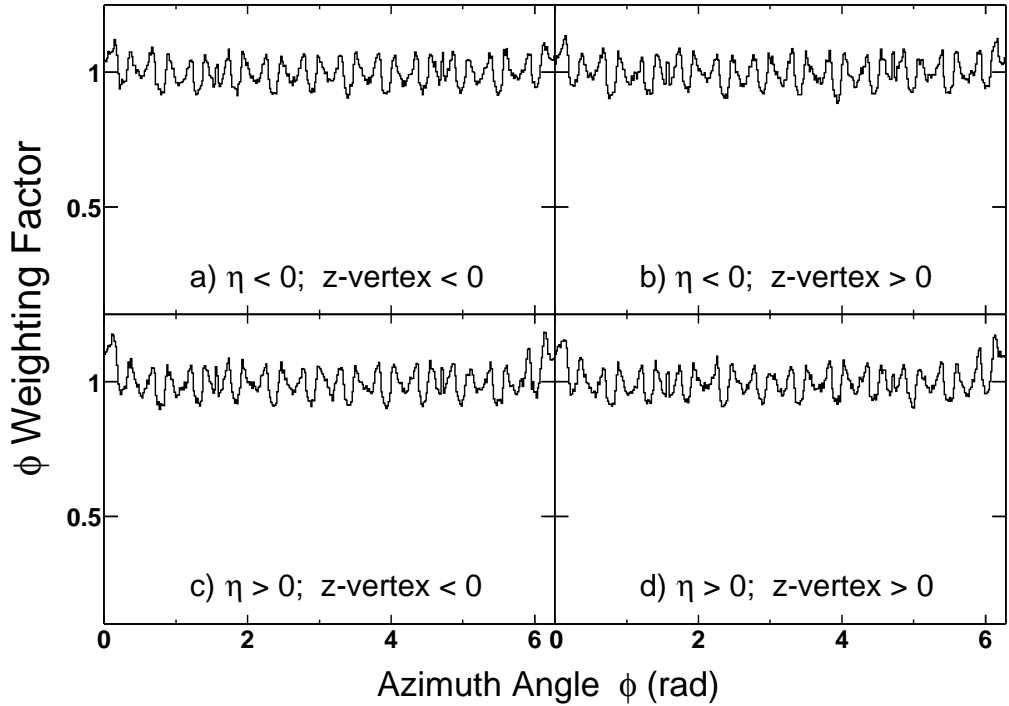}}
\caption[Track weights]{ An example of the weights used to correct
the reaction plane calculation for acceptance is shown. These
weights are for the centrality bin corresponding to 30--40\% of
the Au + Au collision cross-section and forward field polarity
with the RDO board 3 in sector 9 and RDO board 4 in sector 21
where masked out. } \label{fig:phiweights}
\end{figure}

Figure~\ref{fig:phiweights} shows examples of the $\phi$-weights.
The 12-fold periodic structure is caused by the change in the
detectors acceptance near the sector boundaries. The 24-fold
periodic structure arises because a track that starts near the
edge of a sector and curves toward the middle of the sector has a
high probability of being reconstructed. As a consequence, the
efficiency for detecting a positive particle is enhanced near the
one side of a sector while the efficiency for detecting a negative
particle is enhanced on the other side: two maxima are seen for
each sector.

\begin{figure}[htbp]
\centering\mbox{
\includegraphics[width=1.0\textwidth]{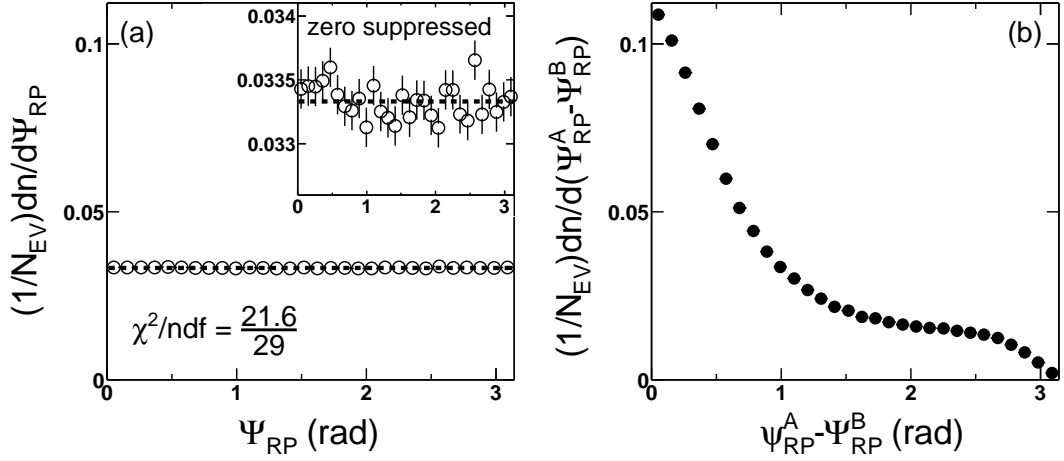}}
\caption[Event plane distributions]{ Left: Azimuthal distribution
of the event plane fit to a constant. The event plane distribution
should be `flat' (azimuthally isotropic). Right: Distribution of
the difference between the event-plane angles for sub-events of
randomly assigned tracks. } \label{fig:eventplane}
\end{figure}

When the tracks in the event plane calculation are re-weighted
properly the distribution of the azimuthal angle of the event
plane is isotropic. Figure~\ref{fig:eventplane} (a) shows the
event plane's azimuthal distribution fit to a constant. When the
event plane distribution is flat, the acceptance for the particles
being compared to the event plane will not bias the measurement of
$v_2$: as long as either the distribution of the event plane or
the distribution of tracks is isotropic, acceptance effects
introduce no bias. Poor acceptance will, however, negatively
impact the reaction plane resolution. The decrease in the reaction
plane resolution caused by the imperfect acceptance can be
accounted for by applying a resolution correction factor to the
measured $v_2$.


A correction to the observed $v_2$ is introduced to account for
the uncertainty in the determination of the reaction plane
(\textit{i.e.} the \textbf{reaction plane resolution}). The event
plane (calculated from Equation~\ref{eq:subevent2}) is used to
estimate the reaction plane: the accuracy of the estimation
depends on the number of tracks used and the magnitude of the true
$v_2$ signal. With an infinite number of tracks, a finite $v_2$,
and a perfect detector the event plane could be a perfectly
accurate estimator of the true reaction plane. With a limited
number of tracks detected in a real detector, however, we cannot
assume that the true reaction plane has been accurately estimated.
Poor resolution leads to a decrease in the calculated value of the
flow parameters because, the correlation of the particles with the
reaction plane is partially lost.  The correction factor necessary
to compensate for the resolution is found as follows:
\begin{gather}
v_2^{\Re} = \langle e^{2i(\phi-\Psi_{RP}^{\Re})} \rangle, v_2^{obs} = \langle e^{2i(\phi-\Psi_{EV})} \rangle \\
\frac{v_2^{\Re}}{v_2^{obs}} = \langle
\exp^{2i(\Psi_{EP}-\Psi_{RP}^{\Re})} \rangle,
\label{eq:correction}
\end{gather}
where $v_2^{\Re}$, $v_2^{obs}$, $\Psi_{RP}^{\Re}$, and $\Psi_{EP}$
are the real $v_2$, the observed $v_2$, the real reaction-plane
angle, and the reconstructed event-plane angle. From
Equation~\ref{eq:correction} the proper correction factor is found
to be $\langle \cos2(\Psi_{EP}-\Psi_{RP}^{\Re}) \rangle$. This
quantity can be calculated by reconstructing event planes from two
random sub-sets of tracks within the same event (sub-events). The
difference between the two sub-event-plane angles
($\Psi_{EP}^A-\Psi_{EP}^B$) is shown in
Figure~\ref{fig:eventplane} (b). The resolution measured from two
sub-events with equal numbers of tracks is given by
Equation~\ref{eq:subevent1}
\begin{equation}
\langle \cos \left[ 2 \left( \Psi_{EP}^{A} - \Psi^{\Re}_{RP}
\right) \right] \rangle = \sqrt{\langle \cos \left[ 2 \left(
\Psi_{EP}^{A} - \Psi_{EP}^{B} \right) \right] \rangle}.
\label{eq:subevent1}
\end{equation}
The second harmonic sub-event-plane angles are calculated from
Equation~\ref{eq:subevent2}:
\begin{equation}
\tan \left( 2 \Psi_{EP} \right) = \frac{\sum_{i} w_i \sin \left( 2
\phi_i \right)}{\sum_{i} w_i \cos \left( 2 \phi_i \right)}.
\label{eq:subevent2}
\end{equation}
The $w_{i}$'s in Equation~\ref{eq:subevent2} are weights used to
maximize the resolution.  In our case we use the particle's
transverse momentum and the $\phi$-weights as the weighting
factor. An interpolation formula is then used with an iterative
routine to calculate its roots and find the correction factor for
the full event plane $\langle \cos2(\Psi_{EP}-\Psi_{RP}^{\Re})
\rangle$. We estimate the errors in the correction factor by
varying the input by a small amount and calculating the change in
the result.

\begin{figure}[htb]
\centering\mbox{
\includegraphics[width=1.0\textwidth]{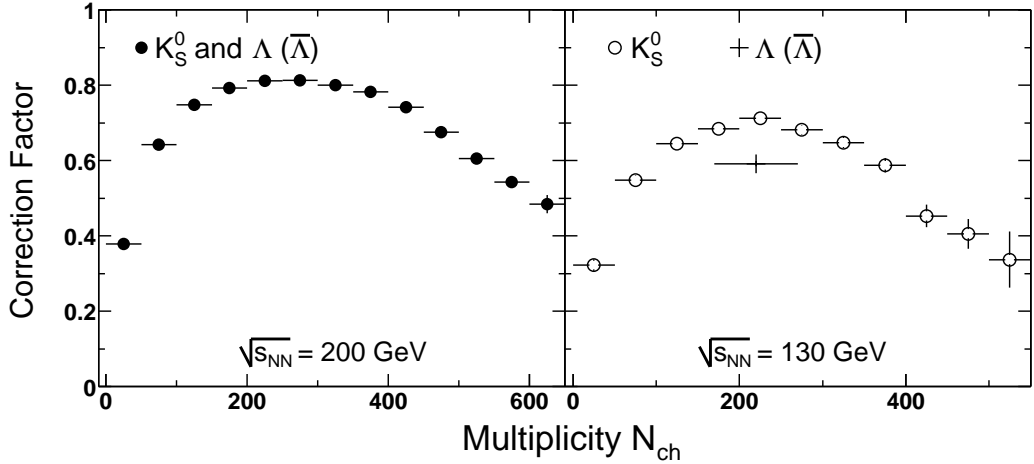}}
\caption[Event plane resolution]{ Left: Resolution correction
factor for the $\sqrt{s_{_{NN}}} = 200$~GeV analysis. Right:
Resolution correction factor for the $\sqrt{s_{_{NN}}}=130$~GeV
analysis. The increase in the resolution from Run-1 to Run-2
reflects the implementation of a new method (discussed in the
text) for calculating the event plane. The resolution for the
Run-1 $\Lambda$ analysis was lower than the $K^{0}_{S}$ analysis,
only reaching a maximum of $0.58 \pm 0.007$. With the new event
plane calculation the resolution for $\Lambda$ and $K^{0}_{S}$
$v_2$ are the same and reach a maximum of $0.813 \pm 0.003$. }
\label{fig:resolution}
\end{figure}
The resolution correction factor is shown in
Figure~\ref{fig:resolution}. The resolution depends on the number
of tracks used and the magnitude of the event asymmetry. For the
most peripheral events the small number of tracks available
reduces the resolution while for the most central events the
symmetry of the collision overlap region degrades it. As a result,
the resolution is greatest at a centrality corresponding to
roughly 20--30\% of the collision cross-section.

%



Since $v_2$ is calculated from the distribution of ($\phi_i -
\Psi_{RP}$), if particle $i$ is included in the evaluation of
$\Psi_{RP}$, \textbf{auto-correlations} are introduced. For the
measurement of $K_S^0$ $v_2$ in Run-1, these unwanted correlations
are eliminated by calculating the event plane from charged
particle tracks with a DCA to the primary vertex less than 1~cm
while we use only tracks with a DCA greater than 1~cm to
reconstruct the $K_{S}^{0}$.  For $\Lambda$ we exclude from the
event plane calculation all tracks identified as proton candidates
by their energy loss in the TPC. This method has one important
disadvantage: the reduction in the number of tracks used degrades
the resolution of the reaction plane.  This in turn leads to
larger errors on the final $v_2$ measurement.

To maximize the number of tracks available for our event plane
calculation, we implemented a new method to remove
auto-correlations for neutral vertex particles.  Rather than
dividing tracks into two sets---one to be used for reconstructing
neutral vertices and the other to be used for calculating the
event plane---we calculate a separate event-plane angle for every
$K^0_S$, $\Lambda$, or $\overline{\Lambda}$ candidate.
Auto-correlations are removed by excluding only the two tracks
associated with a specific $K_S^0$, $\Lambda$, or
$\overline{\Lambda}$ from the event plane calculation. With a
larger track sample the reaction plane resolution increases and
the statistical and systematic uncertainty on $v_2$ decreases.
Figure~\ref{fig:resolution} shows the reaction plane resolution
correction factor from the method used for the
$\sqrt{s_{_{NN}}}=130$~GeV data and the method used for the
$\sqrt{s_{_{NN}}}=200$~GeV data.


\section{Calculating the $v_{2}$ of $K_{S}^{0}$ and $\Lambda +
\overline{\Lambda}$} \indent

We use the measured yield within multiple intervals of
$(\phi_{i}-\Psi_{RP}^{i})$ to calculate $v_2 = \langle \cos\left
[2(\phi_{i}-\Psi_{RP}^{i})\right ]\rangle$, where $\phi_{i}$ is
the azimuth angle of the momentum vector of particle $i$ and
$\Psi_{RP}^{i}$ is the reaction-plane angle for the event that
particle $i$ was observed in. To remove autocorrelations, the
contributions from the decay daughter tracks associated with
particle $i$ are subtracted from the right hand side of
Equation~\ref{eq:subevent2}. The candidates are categorized by
invariant mass, $p_T$ and $(\phi_{i}-\Psi_{RP}^{i})$.  For each
$p_T$ interval, twenty $(\phi-\Psi^{RP})$ intervals from 0 to
2$\pi$ are created (all events are combined). The yield in
interval $j$ ($dn_j$) is calculated by fitting a smooth function
to the mass region outside the candidate mass window and
integrating the number of counts above this background. To
minimize the systematic errors associated with fitting the
background, the same background shape is used for every $j$
$(\phi-\Psi^{RP})$ interval: only the relative amplitude of the
background is allowed to change. Once the yields are known, we
calculate $v_2(p_T)$ using
\begin{equation}
v_2(p_T) = \frac{\sum_{j} dn_{j}\cos 2\left[\left (\phi-\Psi^{RP}
\right )_j \right]}{\sum_{j} dn_{j}}. \label{eq:mean}
\end{equation}

\begin{figure}[htbp]
\centering\mbox{
\includegraphics[width=0.9\textwidth]{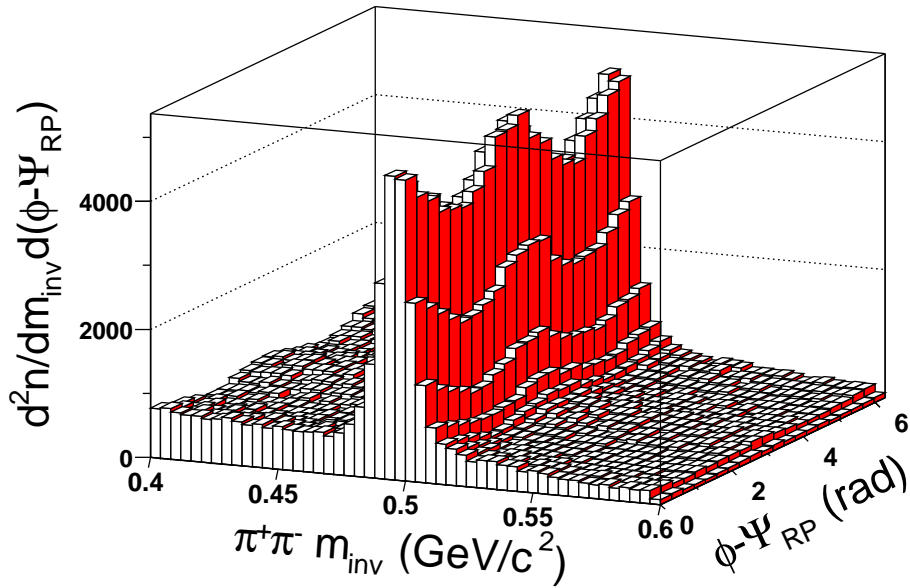}}
\caption[Invariant mass--($\phi-\Psi^{RP}$) distribution
($K_S^0$)]{ Distribution of $K_S^0$ candidates in the invariant
mass versus ($\phi-\Psi^{RP}$) plane.} \label{fig:v2minv}
\end{figure}
\begin{figure}[htbp]
\centering\mbox{
\includegraphics[width=0.9\textwidth]{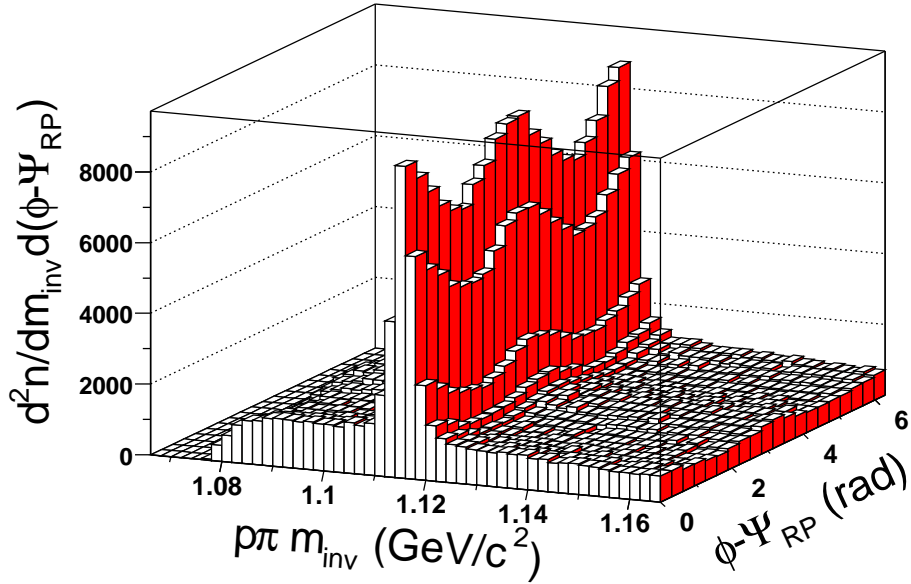}}
\caption[Invariant mass--($\phi-\Psi^{RP}$) distributions
($\Lambda+\overline{\Lambda}$)]{ Distribution of $\Lambda$ and
$\overline{\Lambda}$ candidates in the invariant mass versus
($\phi-\Psi^{RP}$) plane.} \label{fig:v2minv2}
\end{figure}
Figure~\ref{fig:v2minv} shows the $p_T$ inclusive distribution of
the $K_S^0$ candidates in the invariant mass versus
$(\phi-\Psi_{RP})$ plane. The $v_2$ signal---an enhancement near
zero, $\pi$ and $2\pi$ radians---is clearly visible. We find our
method of calculating $v_2$ is insensitive to the background
contamination, so, to maximize our statistical sample, we use
relatively loose selection criteria for the identification of
$K_S^0$, $\Lambda$, and $\overline{\Lambda}$ candidates.
Table~\ref{tab:v0cuts} lists the criteria used for the final
analysis. Figure~\ref{fig:v2minv2} shows the same distribution for
$\Lambda$ and $\overline{\Lambda}$ candidates.

\begin{table}[hbt]
\centering\begin{tabular}{l|ccc|cc} \hline \hline
 & \multicolumn{3}{c|}{$K_S^0$} & \multicolumn{2}{c}{$(\Lambda)$} \\
\hline
{$p_T$ (GeV/c)}        & {$< 0.6$} & {$0.6-2.0$} & {$> 2.0$} & {$< 2.0$} & {$> 2.0$} \\
\hline
{$\pi^+(p)$--$\pi^-$ DCA}    & {$< 0.70$} & {$< 0.70$} & {$< 0.80$} & {$< 0.70$} & {$< 0.70$} \\
{$\pi^+(p)$--$V_{prim}$ DCA} & {$> 1.50$} & {$> 1.50$} & {$> 0.35$} & {$> 0.50$} & {$> 0.25$} \\
{$\pi^-$--$V_{prim}$ DCA}    & {$> 1.50$} & {$> 1.50$} & {$> 0.35$} & {$> 1.00$} & {$> 1.00$} \\
{Decay Length}               & {$> 3.50$} & {$> 5.50$} & {$> 7.00$} & {$> 4.50$} & {$> 4.50$} \\
{$V^0-V_{prim}$ DCA}         & {$< 0.70$} & {$< 0.70$} & {$< 0.80$} & {$< 0.70$} & {$< 0.70$} \\
\hline \hline
\end{tabular}
\caption[Candidate selection criteria ($v_2$)] { The vertex and
daughter track selection criteria for $K_{S}^{0}$, and $\Lambda$.
For $\overline{\Lambda}$ criteria are the same as $\Lambda$ with
the $p(\pi^-)$ exchanged with $\overline{p}(\pi^+)$. All units are
centimeters unless indicated otherwise.} \label{tab:v0cuts}
\end{table}

\subsection{Systematic Uncertainties and Correlations Unrelated to the Reaction Plane} \indent

\begin{table}[hbt]
\centering\begin{tabular}{l|r@{.}lr@{.}lr@{.}l|r@{.}lr@{.}lr@{.}l}
\hline\hline
~ & \multicolumn{6}{c}{$K_S^0$} \vline & \multicolumn{6}{c}{$\Lambda + \overline{\Lambda}$} \\
\hline
$p_T$ (GeV/c) & 1&0 & 2&5 & 4&0 & 1&0 & 2&5 & 4&0 \\
\hline \multirow{2}{*}{$v_2$ (bg)}
    & $+0$&$000$ & $+0$&$001$ & $+0$&$003$ & $+0$&$001$ & $+0$&$005$ & $+0$&$005$ \\
    & $+0$&$001$ & $-0$&$007$ & $-0$&$018$ & $-0$&$007$ & $-0$&$001$ & $-0$&$001$ \\
\multirow{2}{*}{$v_2$ (n-f)}
    & $+0$&$00$ & $+0$&$00$ & $+0$&$00$ & $+0$&$00$ & $+0$&$00$ & $+0$&$00$ \\
    & $-0$&$01$ & $-0$&$04$ & $-0$&$03$ & $-0$&$01$ & $-0$&$04$ & $-0$&$04$ \\
\hline $\delta p_T/p_T$
    & $\pm0$&$012$ & $\pm0$&$027$ & $\pm0$&$030$ & $\pm0$&$016$ & $\pm0$&$027$ & $\pm0$&$037$ \\
\hline\hline
\end{tabular}
\caption[Systematic errors ($v_2$)]{The systematic errors from
background (bg) and non-flow effects (n-f) are listed for $v_2$
(0--80\%) at three $p_T$ values along with the $p_T$ resolution
($\delta p_T/p_T$). The values listed are absolute errors. }
\label{v2syserr}
\end{table}
Sources of systematic error in the calculation of $v_{2}$ are
correlations unrelated to the reaction plane (non-flow effects),
estimation of the yield from the invariant mass distributions, the
$p_T$ resolution ($\delta p_T/p_T$), and biases introduced by the
cuts used in the analysis. Table~\ref{v2syserr} lists the dominant
systematic errors for $K_S^0$ and $\Lambda+\overline{\Lambda}$
$v_2$.
The systematic error in $v_2$ associated with the yield extraction
is found to be small and the non-flow systematic error is
dominant.
%
%

\begin{figure}[htb]
\centering\mbox{
\includegraphics[width=1.00\textwidth]{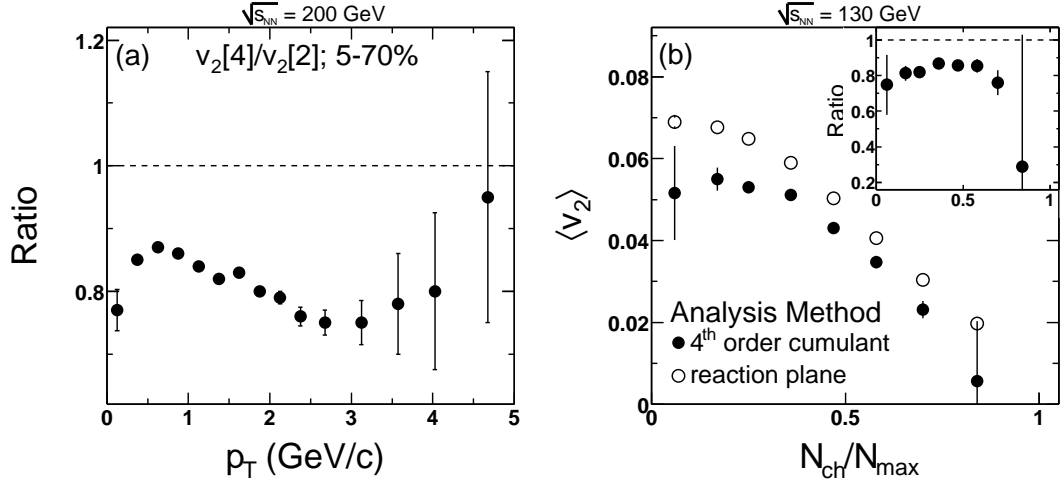}}
\caption[Non-flow]{ Left: Ratio of $v_2$ calculated by the
fourth-order cumulant method and the reaction plane
method~\cite{aihong}. Right: Integrated $v_2$ as a function of
centrality for each method~\cite{aihong}.} \label{fig:nonflow}
\end{figure}

The magnitude of charged particle $v_2$ absent of non-flow effects
has been estimated using a four-particle cumulant
analysis~\cite{aihong}: a method thought to be less sensitive to
non-flow correlations. Figure~\ref{fig:nonflow} (a) shows the
ratio of $v_2$ from a four particle cumulant analysis and a two
particle cumulant analysis. The two particle cumulant $v_2$ result
has been shown to be similar to---but slightly larger than---the
$v_2$ from a reaction plane analysis~\cite{aihong}.
Figure~\ref{fig:nonflow} suggests that for minimum-bias
collisions, non-flow correlations may account for 10--20\% of the
charged particle $v_2$.

The four-particle cumulant method can be adapted to study $v_2$
for $\Lambda+\overline{\Lambda}$ and $K_S^0$ absent of non-flow
contributions but, to be decisive, will require a large data
sample. Nuclear modification of jet production and fragmentation
could lead to a particle-type dependence in the relative fraction
of the non-flow contribution to $v_2$. At $p_T > 3$~GeV, jet
production is thought to be a likely source of non-flow
correlations. The effect of standard jet fragmentation on $v_2$
was examined using superimposed p+p collisions generated with
PYTHIA~\cite{pythia}. Within the measured $p_T$ region, no
significant difference is seen between
$\Lambda+\overline{\Lambda}$ and $K_S^0$ non-flow from this
source.
As such, in this analysis, we assume a similar magnitude for the
non-flow contribution to the $v_2$ of $\Lambda+\overline{\Lambda}$
and $K_S^0$.

Figure~\ref{fig:nonflow} (b) shows the integrated $v_2$ versus
centrality ($\mathrm{N_{ch}/N_{max}}$) from
Reference~\cite{aihong}, where $v_2$ calculated using a four
particle cumulant analysis is compared to $v_2$ from a reaction
plane analysis. Table~\ref{tab:nch} lists the values of
$\mathrm{N_{ch}}$ and the collision cross-sections corresponding
to the $x$-axis in Figure~\ref{fig:nonflow} (b)~\cite{aihong}. The
difference between the $v_2$ calculated from these methods is used
to estimate the centrality dependence of non-flow effects. The
cumulant analysis indicates that non-flow effects are largest in
the most central and most peripheral events. Some or all of the
difference between these methods could, however, arise from
event-by-event fluctuations in the initial source shape: these
fluctuations would reduce the value of $v_2$ calculated from the
cumulant analysis. As such, the cumulant analysis and the reaction
plane analysis are often taken as estimates of the upper and lower
limits on the true $v_2$.
\begin{table}[hbt]
\centering\begin{tabular}{l|cccccccc} \hline\hline
$\mathrm{N_{ch}/N_{max}}$ & 0.849 & 0.708 & 0.590 & 0.472 & 0.363 & 0.258 & 0.160 & 0.060 \\
X-section (\%) & 0--5 & 5--10 & 10--16 & 16--24 & 24--31 & 31--41 & 41--53 & 53--77 \\
\hline \hline
\end{tabular}
\caption[$\mathrm{N_{ch}/N_{max}}$ centrality intervals] { Percent
of the collision cross-section corresponding to
$\mathrm{N_{ch}/N_{max}}$. The value of $\mathrm{N_{max}}$ is
approximately 878. } \label{tab:nch}
\end{table}

\chapter{Results}
\indent\label{chp:results}

As shown in Figure~\ref{fig:pid}, $K_S^0$, $\Lambda$ and
$\overline{\Lambda}$ particles were identified during Run-2 across
a broader $p_T$ range than any other particle. In this chapter we
present the measurement of $v_2$ for $K_S^0$ and
$\Lambda+\overline{\Lambda}$ at mid-rapidity from Au+Au collisions
at $\sqrt{s_{_{NN}}}=130$\footnote{The $v_2$ measurements at
$\sqrt{s_{_{NN}}}=130$~GeV were made in collaboration with J. Fu
and were published in
References~\cite{Sorensen:2001fm,kslam130:Adler:2002pb,Sorensen:2002ci,jhfu}.}
and 200~GeV. The $p_T$ spectra are shown for 0--5\%, 40--60\%, and
60--80\% centrality intervals. The centrality dependence is
studied via the nuclear modification factor $R_{CP}$ which is
derived from the spectra.

\section{Elliptic Flow}
\indent

\begin{figure}[htbp]
\centering\mbox{
\includegraphics[width=1.00\textwidth]{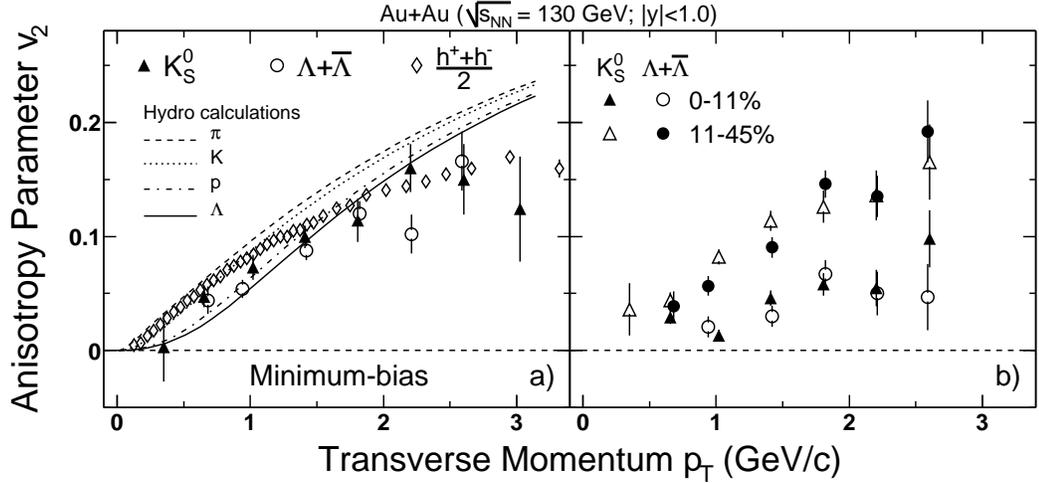}}
\caption[Elliptic flow $v_2$ ($\sqrt{s_{_{NN}}}=130$~GeV) ]{
Elliptic flow for $K_S^0$ and $\Lambda+\overline{\Lambda}$
particles at mid-rapidity in Au+Au collisions at
$\sqrt{s_{_{NN}}}=130$~GeV.} \label{fig:v2_130}
\end{figure}

Elliptic flow at mid-rapidity as a function of transverse momentum
for minimum-bias (a), 0--11\% central and 11--45\% central (b)
Au+Au collisions at $\sqrt{s_{_{NN}}}=130$~GeV is shown in
Figure~\ref{fig:v2_130}.
The $v_2$ of $K_S^0$ and the $v_2$ of $\Lambda+\overline{\Lambda}$
both increase monotonically with $p_T$ in the 11--45\% centrality
interval.
Throughout the measured $p_T$ range, the $v_2$ for both particles
is larger in more peripheral collisions than in the central
collisions.
A similar dependence was observed for charged particles in Au + Au
collisions at the same RHIC energy~\cite{v2:Ackermann:2000tr}.
Also shown in the figure is $v_2(p_T)$ for charged
hadrons~\cite{Adler:2002ct} and hydrodynamic model calculations.
Within statistical uncertainty, the minimum-bias $K_S^0$ results
are in agreement with the $v_2$ of charged kaons~(not
shown)~\cite{Adler:2001nb}. We observe that $v_2$ for both strange
particles increases as a function of $p_T$ up to about 1.5~GeV/c,
similar to the hydrodynamic model prediction.
For $p_T~\ge~2$~GeV/c however, the values of $v_2$ seem to be
saturated.
It has been suggested that the shape and height of $v_2$ above
2--3~GeV/c in a pQCD model is related to energy loss in an early,
high-parton-density, stage of the
evolution~\cite{Gyulassy:highptv2:2000gk}.

These are the first measurements of $v_2$ for $K_S^0$ and
$\Lambda+\overline{\Lambda}$ at RHIC energy and the first
measurement of $v_2$ for any identified particle above $p_T \sim
1.0$ GeV/c. The statistical sample for Run-1, however, is limited
and several important details remain to be studied---how well do
the hydrodynamic models reproduce the mass dependence for $K_S^0$
and $\Lambda+\overline{\Lambda}$, does $v_2$ for all particles
saturate with the same magnitude at the same $p_T$, and does the
relative strangeness content of the particle affect its elliptic
flow? For Run-2, improvements were made to the analysis technique
and a larger data set became available.

\begin{figure}[htbp]
\centering\mbox{
\includegraphics[width=1.00\textwidth]{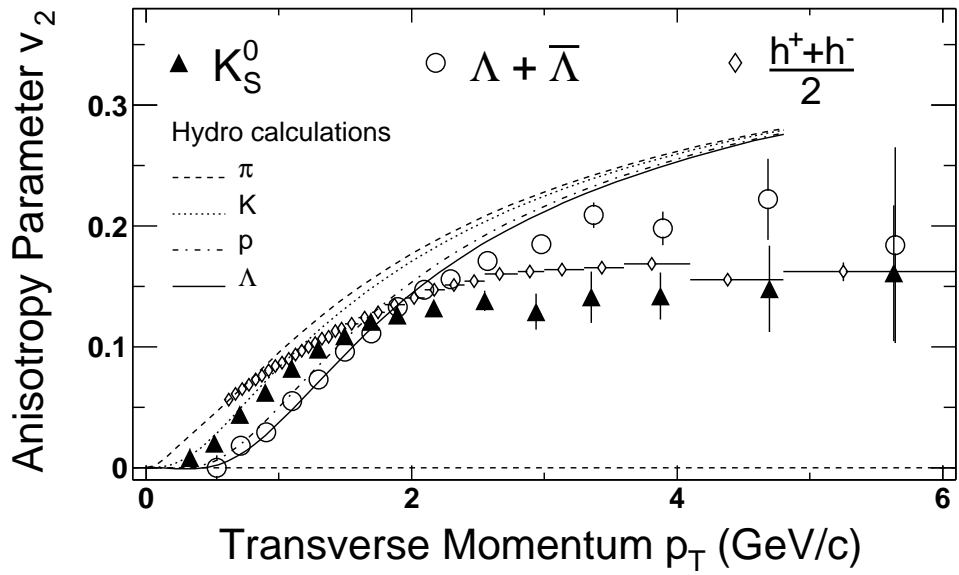}}
\caption[Minimum-bias $v_2$ ($\sqrt{s_{_{NN}}}=200$~GeV)]{ The
minimum-bias (0--80\% of the collision cross-section) $v_{2}(p_T)$
for $K_{S}^{0}$, $\Lambda + \overline{\Lambda}$ and $h^{\pm}$ at
mid-rapidity in Au+Au collisions at $\sqrt{s_{_{NN}}}=200$~GeV.
The error bars shown are statistical only. Hydrodynamical
calculations of $v_2$ for pions, kaons, protons and lambdas are
also plotted~\cite{hydro:Huovinen:2001cy}.} \label{fig:v2_200}
\end{figure}

Figure~\ref{fig:v2_200} shows minimum-bias $v_{2}$ at
$\sqrt{s_{_{NN}}}=200$~GeV for $K_{S}^{0}$, $\Lambda +
\overline{\Lambda}$ and charged hadrons $h^++h^-$. The analysis of
the charged hadron $v_2$ is described in
Reference~\cite{Adler:2002ct}.
The curves in the figure represent hydrodynamic model calculations
of $v_2$ for pions, kaons, protons, and
lambdas~\cite{hydro:Huovinen:2001cy}.
At low $p_T$, the model calculations are in good agreement with
the mass and $p_T$ dependence of $v_2$.
At intermediate $p_T$ however, we find
$v_{2,\Lambda+\overline{\Lambda}}
> v_{2,K}$ in contradiction to hydrodynamical
calculations: where at a given $p_T$, heavier particles have
smaller $v_{2}$ values.
The $p_T$-scale where $v_{2}$ deviates from the hydrodynamical
prediction is $\sim 2.5$~GeV/c for $\Lambda+\overline{\Lambda}$
and $\sim 1$~GeV/c for $K_{S}^{0}$.
Our measurement at $\sqrt{s_{_{NN}}}=200$~GeV establishes the
particle-type dependence of the $v_2$ saturation at intermediate
$p_T$ ($1.5<p_T<4.0$~GeV/c).

\begin{figure}[htbp]
\centering\mbox{
\includegraphics[width=1.00\textwidth]{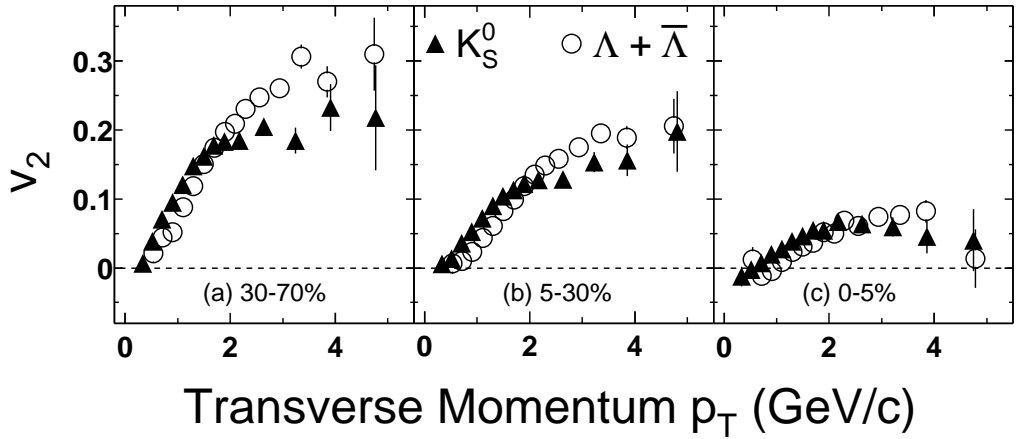}}
\caption[Differential $v_2$ versus centrality]{The $v_2$ of
$K_{S}^{0}$ and $\Lambda + \overline{\Lambda}$ at
$\sqrt{s_{_{NN}}}=200$~GeV as a function of $p_T$ for 30--70\%,
5--30\% and 0--5\% of the collision cross-section. The error bars
represent statistical errors only. The non-flow systematic errors
for the 30--70\%, 5--30\% and 0--5\% centralities are -25\%, -20\%
and -80\% respectively. } \label{fig:v2_cent_200}
\end{figure}

Figure~\ref{fig:v2_cent_200} shows $v_{2}$ of $K_{S}^{0}$ and
$\Lambda + \overline{\Lambda}$ at mid-rapidity for Au+Au
collisions at $\sqrt{s_{_{NN}}}=200$~GeV as a function of $p_T$
for three centrality intervals: 30--70\%, 5--30\%, and 0--5\%
of the geometrical cross-section. 
The $p_T$ dependence of $v_{2}$ from all three centrality bins has
a similar trend: a monotonic rise with $p_T$ at low $p_T$ and a
saturation at intermediate $p_T$.
The saturation of $v_2$ in the intermediate $p_T$ region from
minimum-bias trigger data in Figure~\ref{fig:v2_200} is not due to
the superposition of drastically different $p_T$ dependencies for
various centrality bins.
The values of $v_2$ at saturation show a particle-type and
centrality dependence.

\section{Spectra}
\indent

\begin{figure}[htbp]
\centering\mbox{
\includegraphics[width=1.00\textwidth]{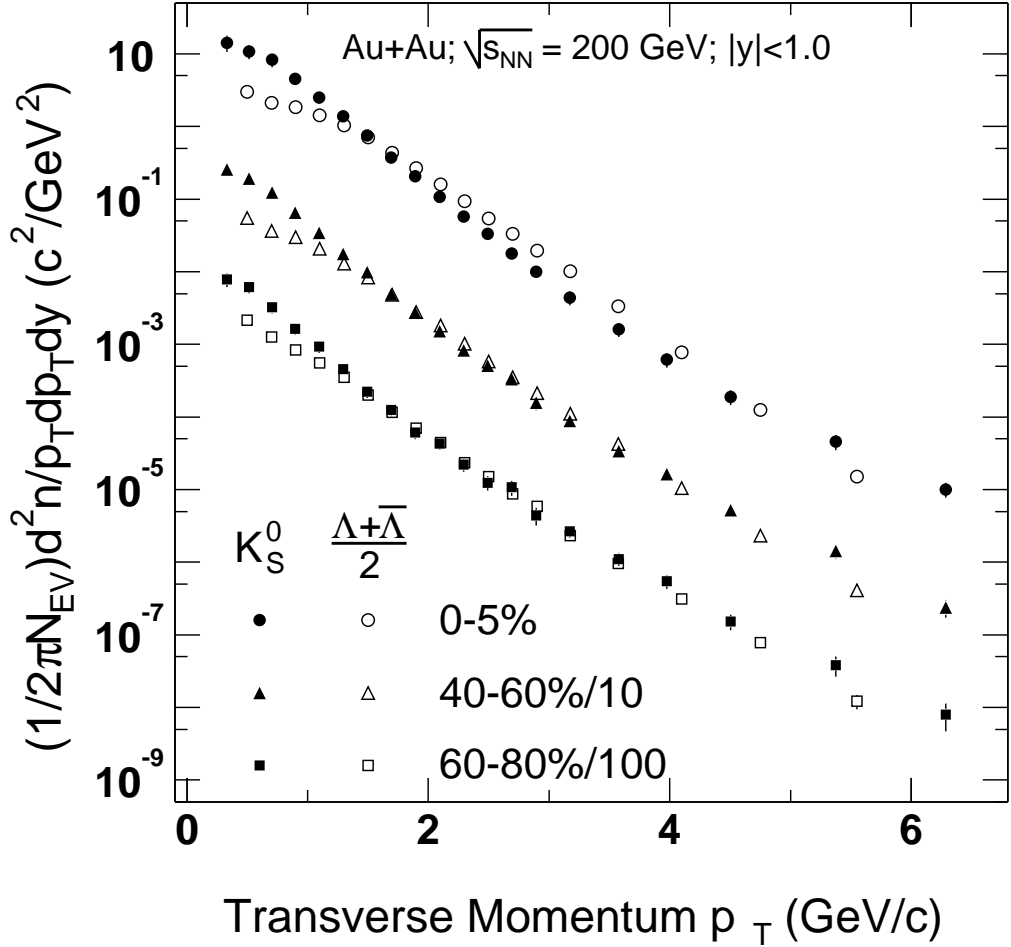}}
\caption[$K_S^0$ and $\Lambda+\overline{\Lambda}$ spectra]{
Spectra for $K_S^0$ and
$\Lambda+\overline{\Lambda}$~\cite{hlongsqm} at mid-rapidity from
central and peripheral Au+Au collisions at $\sqrt{s_{_{NN}}} =
200$~GeV.} \label{fig:spectra}
\end{figure}

In Figure~\ref{fig:spectra} we show the spectra for $K_S^0$ and
$\Lambda+\overline{\Lambda}$~\cite{hlongsqm} at mid-rapidity in
Au+Au collisions at $\sqrt{s_{_{NN}}}=200$~GeV. At $p_T \sim 1.5$
and 4.5~GeV/c, the $K_S^0$ and $\Lambda+\overline{\Lambda}$ yields
coincide. In the intermediate $p_T$ region, a greater number of
$\Lambda+\overline{\Lambda}$ particles are produced than $K_S^0$
particles. The $K_S^0$ spectra show a clear hardening at higher
$p_T$ while the $\Lambda+\overline{\Lambda}$ spectra appears to
remain soft within most of the measured $p_T$ range. As a result,
the number of $K_S^0$ particles produced becomes larger than the
number of $\Lambda+\overline{\Lambda}$ particles again for $p_T >
4.5$~GeV/c. For peripheral collisions the separation at
intermediate $p_T$ between the $K_S^0$ and
$\Lambda+\overline{\Lambda}$ yields appears much smaller than in
central collisions. The nuclear modification factor $R_{CP}$ is a
useful measure for studying the relative centrality dependencies.

\section{Nuclear Modification $R_{CP}$}
\indent

\begin{figure}[htbp]
\centering\mbox{
\includegraphics[width=1.00\textwidth]{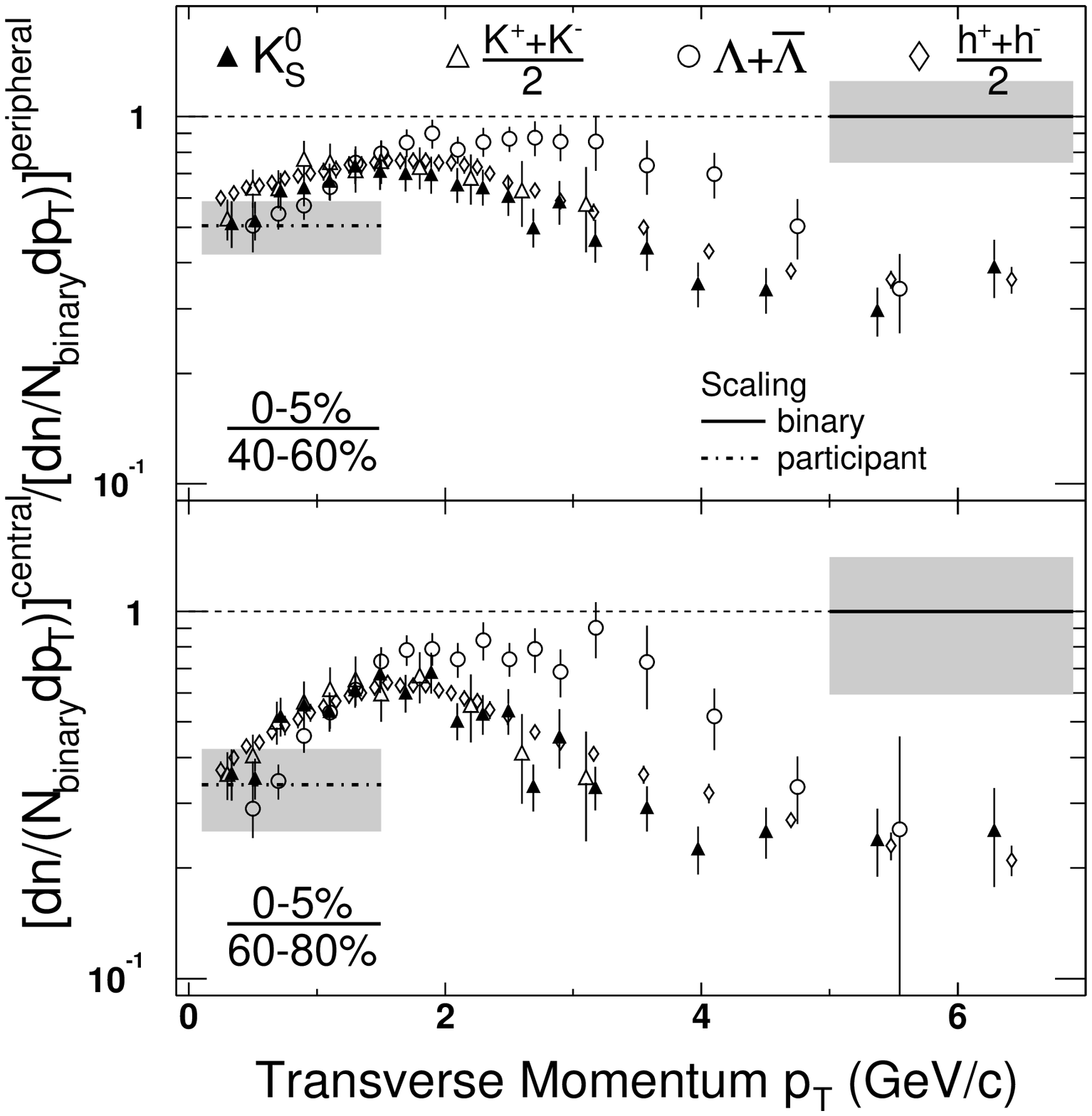}}
\caption[Nuclear modification factor $R_{CP}$]{ Nuclear
modification of $K_S^0$ and $\Lambda+\overline{\Lambda}$
production at mid-rapidity for Au+Au collisions at
$\sqrt{s_{_{NN}}}=200$~GeV. For comparison, the $R_{CP}$ for
charged kaons is also shown where the kaons have been identified
from single-prong decays (kinks) as discussed in
Reference~\cite{kaon:Adler:2002wn}.} \label{fig:Rcp}
\end{figure}

Figure~\ref{fig:Rcp} shows $R_{CP}$ for $K_{S}^{0}$, and $\Lambda
+ \overline{\Lambda}$ using the 5\% most central collisions,
normalized by peripheral collisions (40--60\% and 60--80\%). For
reference, the charged kaon and charged hadron $R_{CP}$ are also
shown.
For charged hadrons, these $40$-$60\%$ and $60$-$80\%$ bins have
been shown to approximately follow binary collision scaling
(relative to p+p collisions) without medium
modification~\cite{highpt:Adams:2003}.
The bands in Figure~\ref{fig:Rcp} represent the expected values of
$R_{CP}$ for binary and participant ($\mathrm{N_{part}}$) scaling
including systematic variations from the
calculation~\cite{highpt:Adams:2003}. Table~\ref{tab:nbin} gives
the values for $\mathrm{N_{bin}}$ and $\mathrm{N_{part}}$

\begin{table}[hbt]
\centering\begin{tabular}{l|ccc} \hline\hline
Cross-section (\%) & 0--5 & 40--60 & 60--80 \\
\hline
$\langle\mathrm{N_{bin}}\rangle$  & $990^{+67}_{-67}$ & $91.8^{+22}_{-23}$ & $20.0^{+7}_{-9}$ \\
$\langle\mathrm{N_{part}}\rangle$ & $352^{+6}_{-7}$  & $61.0^{+10}_{-10}$  & $19.8^{+5}_{-6}$ \\
\hline \hline
\end{tabular}
\caption[$\mathrm{N_{bin}}$ and $\mathrm{N_{part}}$] { Monte-Carlo
Glauber model calculations of the number of participating nucleons
$\mathrm{N_{part}}$ and the number of binary nucleon-nucleon
collisions $\mathrm{N_{bin}}$ for three centrality
intervals~\cite{highpt:Adams:2003}. } \label{tab:nbin}
\end{table}

The kaon and (anti-)lambda yields are suppressed by different
magnitudes and the $p_T$-scales associated with the onset of the
high $p_T$ suppression are different.
$R_{CP}$ for kaons increases with $p_T$ from the participant
scaling limit, reaches a maximum of approximately 0.6 at $p_T \sim
1.6$~GeV/c and then decreases with $p_T$.
The $R_{CP}$ for $\Lambda + \overline{\Lambda}$, however, rises to
a maximum of approximately 0.9 at $p_T \sim 2.0$~GeV/c, remains
near that value up to $p_T \sim 3.5$~GeV/c and then decreases with
$p_T$.
For most of the intermediate $p_T$ region, $\Lambda +
\overline{\Lambda}$ $R_{CP}$ within errors coincides with binary
collision scaling, while the kaon $R_{CP}$ is significantly below
unity.
For both species, the $p_T$ where $R_{CP}$ begins to decrease
approximately coincides with the $p_T$ where $v_2$ in
Figure~\ref{fig:v2_200} saturates.
At high $p_T$ ($p_T > 5.0$~GeV/c), $R_{CP}$ values for $K_{S}^{0}$
and $\Lambda + \overline{\Lambda}$ are approaching the value of
the charged hadron $R_{CP}$.
The apparent disappearance of the particle-type dependence of
$R_{CP}$ may signify that single parton fragmentation dominates
the features of $R_{CP}$ above $p_T \sim 5$~GeV/c.

\chapter{Discussion}
\indent

Event-by-event azimuthal anisotropy in particle production is
thought to probe the early stages of relativistic heavy-ion
collisions~\cite{hydro:Ollitrault:1992bk}. For $p_T < 1.0$~GeV/c,
hydrodynamic models~\cite{hydro:Huovinen:2001cy} describe
$v_2$~\cite{v2:Ackermann:2000tr,Adler:2001nb}, and the particle
spectra~\cite{hydro:Kolb:2003dz} well. These models predict that
$v_2$ will rise monotonically with $p_T$. It is expected however,
that for particles with large $p_T$ the model assumptions will
break down. Measurements using charged particles and our
measurements using $K_S^0$ and $\Lambda + \overline{\Lambda}$ do
indeed indicate a saturation of $v_2$---well below the
hydrodynamic calculations---at intermediate $p_T$. The nuclear
modification factor $R_{CP}$ for charged particles also shows a
large $p_T$-independent suppression of particle production at $p_T
> 4.5$~GeV/c. Models based on parton energy
loss~\cite{Gyulassy:highptv2:2000gk,Surf:Shuryak:2001me} and
transport opacity~\cite{trans:Molnar:2001ux} have been discussed
in relation to the saturation and centrality dependence of $v_2$
at intermediate and high $p_T$. The authors of Reference
\cite{MesonBaryon:Gyulassy:2001kr} propose that the saturation of
charged particle $v_2$ could be a consequence of the transition
from soft to hard production processes occurring at different
$p_T$-scales for pions and protons.

We show here that the extent of the $p_T$ region where
hydrodynamic like processes (or other soft processes) dominate the
spectrum is particle species dependent. In
Chapter~\ref{chp:results}, we reported the measurement of $v_2$
and $R_{CP}$ for $K_S^0$ and $\Lambda+\overline{\Lambda}$. These
measurements show that for $p_T$ up to 3~GeV/c
$\Lambda+\overline{\Lambda}$ $v_2$ continues to rise---similar to
hydrodynamic model calculations---while the $K_S^0$ $v_2$
saturates at $p_T \sim 1.5$~GeV/c. We also find that there is a
particle-type dependence to the onset of the suppression---as
measured by $R_{CP}$---of $K_S^0$ and $\Lambda+\overline{\Lambda}$
production. In addition, for each particle the lower $p_T$ bound
of the suppressed region of $R_{CP}$ coincides with the lower
bound of the saturated region of $v_2$.

It has been suggested that if a partonic state exists prior to
hadronization, the process of particle formation at intermediate
$p_T$, by string fragmentation, parton
fragmentation~\cite{coal:Lin:2001zk} or quark
coalescence~\cite{coal:Lin:2002rw,coal:Lin:2003jy,coal:Hwa:2003bn,coal:Greco:2003xt,coal:Fries:2003vb,reco:Best:Fries:2003kq},
may lead to a dependence of $v_2$ and $R_{AA}$ on particle type.
In this case, it is possible that these measurements will provide
information on the existence and nature of an early partonic
state. In this chapter we investigate the interplay between the
apparently soft and hard components of the $K_S^0$ and
$\Lambda+\overline{\Lambda}$ spectrum and explore possible sources
for the particle-type dependence of our measurements. We begin
with a brief description of hydrodynamic and energy loss models
and we compare their predictions to our data. In
Section~\ref{sec:regimes} we use hydrodynamical inspired and pQCD
inspired fits to estimate the $p_T$ where soft,
hydrodynamical-type processes are no longer appreciable.

\section{Describing Heavy-Ion Collisions with Hydrodynamics}

A system can be described within a hydrodynamical formalism when
the time scales of its microscopic processes are sufficiently
smaller than the time scale for its macroscopic evolution. For
heavy-ion collisions this means the time between interactions
amongst the constituents---partonic and/or hadronic---must be much
smaller than the lifetime of the system. When this condition is
met, the constituents can interact enough times to equilibrate.
The space-time evolution of the system can then be described in
the framework of relativistic fluid dynamics. The equations of
motion are derived from the conservation of energy and momentum
$\partial_{\mu}T^{\mu\nu}=0$. The energy-momentum tensor
$T^{\mu\nu}$ in the ideal fluid approximation is given by:
\begin{equation}
T^{\mu\nu}=(\epsilon+p)u^{\mu}u^{\nu}-pg^{\mu\nu},
\end{equation}
where $\epsilon$, $p$, and $u^{\mu}$ are respectively the energy
density, pressure, and four velocity.

Prior to the onset of local equilibrium, the hydrodynamical
equations are invalid. As such, they can only describe heavy-ion
collisions from an initial time $\tau_0$ until the time when
interaction rates in the system become too small and local thermal
equilibrium can no longer be maintained (\textit{i.e} the
freeze-out time). The initialization of the hydrodynamic evolution
requires that the pre-thermalization stage be modelled so that the
initial conditions can be estimated. Given a set of initial
conditions and the hydrodynamic equations, all that remains to be
specified is the nuclear equation-of-state EOS which relates the
thermodynamic quantities of the system. The EOS can be modelled or
calculated using lattice QCD. Finding the EOS that governs nuclear
matter at high temperature and density is the primary objective of
heavy-ion physics and hydrodynamical model calculations may
provide insight into its form.

\begin{figure}[htpb]
\resizebox{.536\textwidth}{!}{\includegraphics{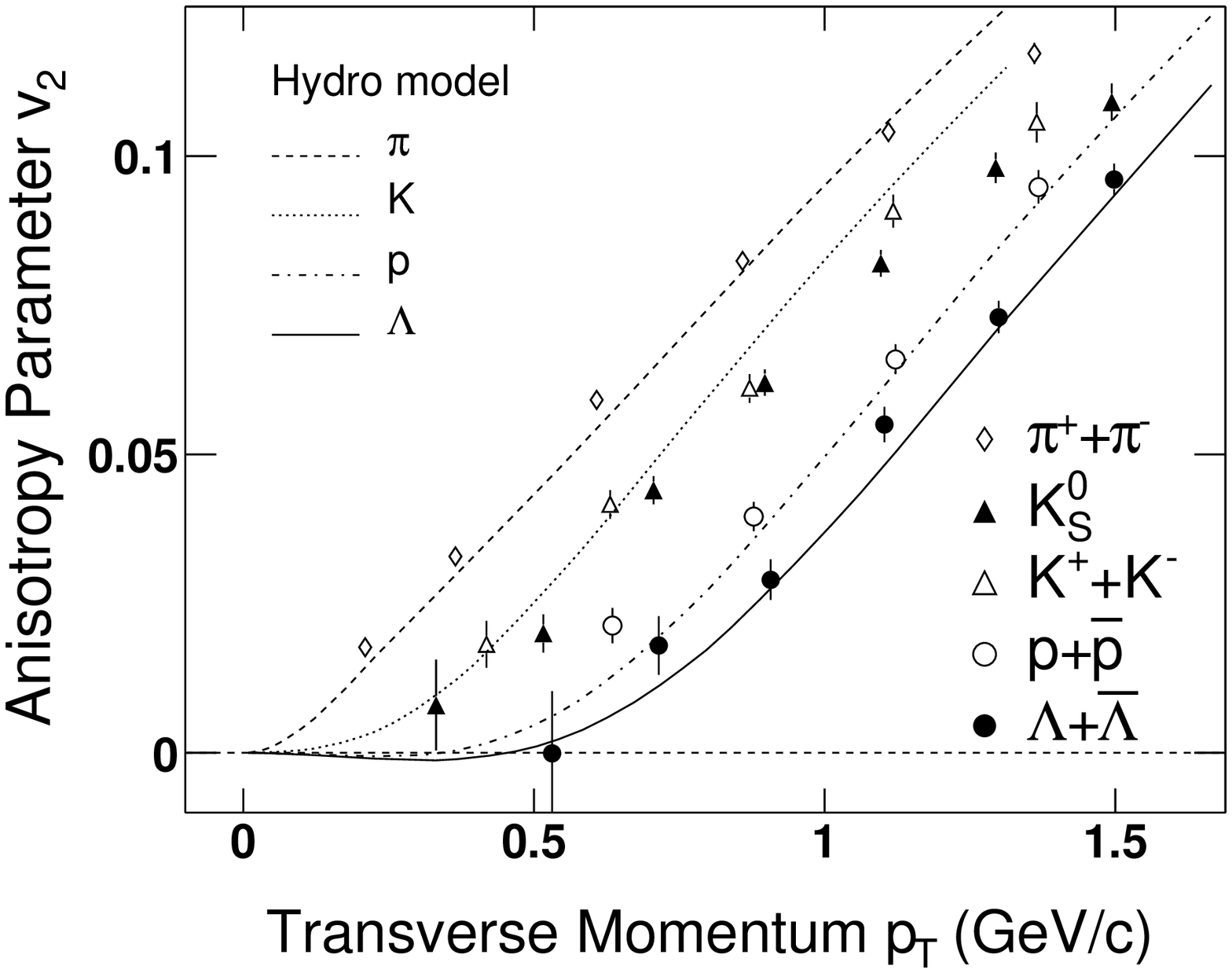}}
\resizebox{.464\textwidth}{!}{\includegraphics{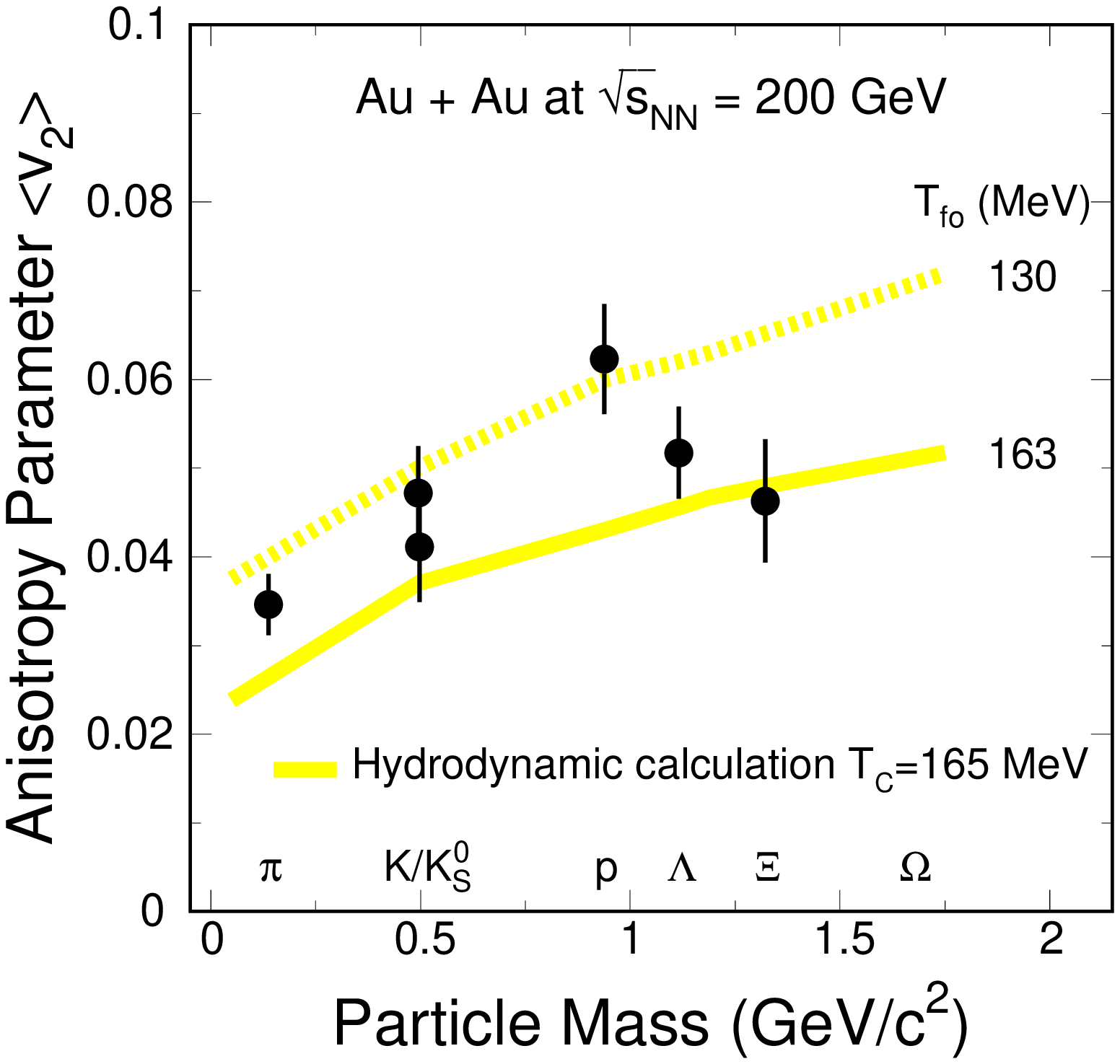}}
\caption[Hydrodynamical $v_2$] {Left: Differential $v_2$ at low
$p_T$ for pions, kaons, protons, and lambdas at mid-rapidity from
Au+Au collisions at $\sqrt{s_{_{NN}}}=200$~GeV. The charged pion,
charged kaon, and the proton $v_2$ were measured by the PHENIX
collaboration~\cite{PhenixPidv2:Esumi:2002vy}. Right: Integrated
$v_2$ versus mass compared to hydrodynamic calculations with
different freeze-out temperatures. The integrated $v_2$ is
calculated by weighting the measured $v_2$ with the particle yield
extracted from fits to the spectra.} \label{fig:allv2}
\end{figure}

The hydrodynamic equations for an ideal fluid describe the
velocity and pressure fields for thermalized fluid elements. The
Cooper-Frye formula~\cite{Cooper:1974mv} is used to calculate the
momentum distribution for the hadrons created from the fluid
elements on the freeze-out hyper-surface $\Sigma$:
\begin{equation}
E\frac{dn_i}{d^3p} = \frac{d_i}{(2\pi)^3} \int_{\Sigma}
\frac{p^{\mu}d\sigma_{\mu}}{\exp[(p^{\mu}u_{\mu}-\mu_i)/T^{th}]\mp1},
\label{eq:cooperfrye}
\end{equation}
where $d_i$ is a degeneracy factor, $\mu_i$ are the chemical
potentials for the hadrons, $p^{\mu}$ are their four momentum, and
$d\sigma_{\mu}$ is the outward normal vector on $\Sigma$.
Hydrodynamic models have been particularly successful in
reproducing the mass dependence of $v_2$ at low $p_T$.
Figure~\ref{fig:allv2} shows $v_2$ versus $p_T$ (left) and the
integrated $v_2$ versus particle mass for identified particles
(right) compared to hydrodynamical model calculations. In these
models the increase in the integrated $v_2$ with mass is a
consequence of the collective motion of the fluid elements built
up as the system interacts and expands. When particles with more
mass freeze-out from fluid elements flowing with a given velocity,
they will carry greater momenta. In this way, an anisotropic
collective flow velocity at freeze-out leads to an increase of the
integrated $v_2$ with particle mass. This increase and the mass
dependence of the differential $v_2$ both indicate that a
significant collective motion is established---perhaps early in
the collision---and that the source eccentricity is efficiently
transferred to momentum-space anisotropy.

The results of a full hydrodynamical model calculation can be
approximated with a simple analytical model: the blast wave
model~\cite{hydro:Huovinen:2001cy}. The equations of the blast
wave model describe particle emission from a thin cylindrical
shell of thermalized matter with temperature $T$. The particle
production is approximated by a boosted Boltzmann distribution so
that the particle spectra can be calculated from the equations:
\begin{equation}
\frac{d^3n}{dp^3} \sim
\int_{0}^{2\pi}d\phi_sK_1(\beta(\phi_s))e^{\alpha(\phi_s)\cos(\phi_s-\phi_p)},
\label{eq:blastwave}
\end{equation}\begin{gather*}
\alpha(x,p)=\frac{p_T}{T\sinh(\rho(x))}, \\
\beta(x,p)=\frac{m_T}{T\cosh(\rho(x))},
\end{gather*}
where $\phi_p=\tan^{-1}(p_y/p_x)$, and $\phi_s=\tan^{-1}(y/x)$ are
respectively the momentum and coordinate space azimuthal
freeze-out angles, $K_1$ is the modified Bessel's function and
$\rho(x)$ is the transverse flow rapidity. In
Section~\ref{sec:regimes} we'll use the blast wave equations to
parameterize the soft part of the $K_S^0$ and
$\Lambda+\overline{\Lambda}$ spectrum\footnote{The author thanks
F. Retiere for his help with these fits.}.

\section{Energy Loss}
For p+p collisions, when scatterings involve sufficiently large
momentum transfer, pQCD calculations describe hadron production
well (see Figure~\ref{fig:jets}). To relate the partonic and
hadronic invariant cross-sections ($E_a\frac{d^3n_a}{dp_a^3}$ and
$E\frac{d^3n_h}{dp^3}$ respectively) it is assumed the calculation
of the partonic cross-section and the hadronization process for
production of hadron $h$ with momentum $p$ can be
factorized~\cite{Factorize:Owens:1987mp}:
\begin{equation}
E\frac{d^3n_h}{dp^3} = \sum_a\int_0^1\frac{dz}{z^2}D_{a\rightarrow
h}(z)E_a\frac{d^3n_a}{dp_a^3}. \label{eq:frag}
\end{equation}
The probability that parton $a$ with momentum $p_a$ fragments into
hadron $h$ with momentum $p=z \times p_a$ is expressed in terms of
fragmentation functions $D_{a\rightarrow h}(z)$. The fragmentation
functions are typically taken to be universal: Once measured they
can be used to describe hadron production for other hard
processes.

For heavy-ion collisions however, neither the validity of
factorization nor the universality of the fragmentation functions
can be assumed a priori. Fast partons---presumably produced from
hard interactions between two colliding nuclei---may need to
traverse hot, dense matter before escaping from the system. Over
two decades ago Bjorken estimated that these secondarily produced
quarks and gluons could lose tens of GeV of their initial
transverse momentum via elastic scattering with quanta in the
medium~\cite{Bjorken:1982tu}. In this publication Bjorken also
proposed what would later be called \textit{surface emission} as a
signature of \textit{jet-quenching}.
\begin{quote}
An interesting signature may be events in which the hard collision
occurs near the edge of the overlap region, with one jet escaping
without absorption and the other fully absorbed.
\end{quote}
Recent observations at RHIC confirm that in central Au+Au
collisions, while near-angle jet-like correlations exist,
away-side jet-like correlations are suppressed
\cite{highptpi0:Adler:2003qi,highpt:Adams:2003}. Jet-quenching has
been studied extensively throughout the last decade (see for
example References
\cite{dedx2:Thoma:1991fm,dedx2:Wang:1992xy,dedx2:Gyulassy:1994hr,dedx2:Zakharov:1997uu,dedx2:Baier:1997kr,dedx2:Gyulassy:2000fs,dedx2:Wiedemann:2000za,dedx2:Baier:2001yt,dedx2:Muller:2002fa})
and although energy loss from elastic scattering has been shown to
be small, radiative energy loss may be large
\cite{dedx:Gyulassy:1990ye,dedxmore:Baier:1995bd,dedx2:Wang:1992xy}.
The magnitude of the energy loss is thought to depend on the gluon
density of the medium being traversed, and should therefore be
sensitive to the creation of hot, dense, and perhaps deconfined
matter.

Measurements of the suppression of neutral pions and charged
hadrons in central Au+Au collisions confirm many of the
expectations of energy loss
\cite{hightptPh130:Adcox:2000sp,highptStar130:Adler:2002xw,highpt:Adams:2003,highptpi0:Adler:2003qi},
with the suppression accompanied by the disappearance of
back-to-back jet-like
correlations~\cite{Adler:2002ct,btob200:Adler:2002tq}. It's
possible (even likely) however, that the interactions of the fast
partons with the matter, will induce not only energy loss but also
changes to the hadronization process
\cite{recoa:Das:1977cp,recob:Roberts:1979ku,recoc:Aitala:1996hf,recod:Anjos:2001jr,recoe:Braaten:2002yt,recof:Gupt:1983rq,recog:Ochiai:1986eh,recoh:Biro:1995mp,recoi:Biro:2001zj,recoj:Hwa:2002zu,recok:Greco:2003mm,recol:Rapp:2003wn,reco:Best:Fries:2003kq}.
In this case, even if factorization still proves to be valid, we
can't assume the fragmentation functions measured in $e^++e^-$
collisions will be relevant to Au+Au collisions. Measurements of
the production of identified particles are needed to study the
possible evolution of the fragmentation functions with system size
and to understand the processes that may govern those changes. In
so doing, it may be possible to not only reach a better
understanding of heavy-ion collisions but to also develop a much
deeper understanding of hadronization in general. This would
constitute a major advance relevant to all particle physics, where
previously hadronization has only been dealt with
phenomenologically.

Surface emission has been discussed in relation to the large,
$p_T$-independent $v_2$ measured for charged
hadrons~\cite{Surf:Shuryak:2001me} and the large $p_T$-independent
suppression of charged particle production at high $p_T$.
Within this scenario, more energy loss will lead to a larger $v_2$
\textbf{and} a greater suppression of particle production. This is
inconsistent, however, with our measurements of $v_2$ and $R_{CP}$
for kaons and $\Lambda + \overline{\Lambda}$ at intermediate
$p_T$: we find that kaons have a smaller $v_2$ but a larger
suppression.
These calculations do not, however, account for how the process of
hadronization may change the observed $v_2$.

\begin{figure}[htbp]
\centering\mbox{
\includegraphics[width=1.00\textwidth]{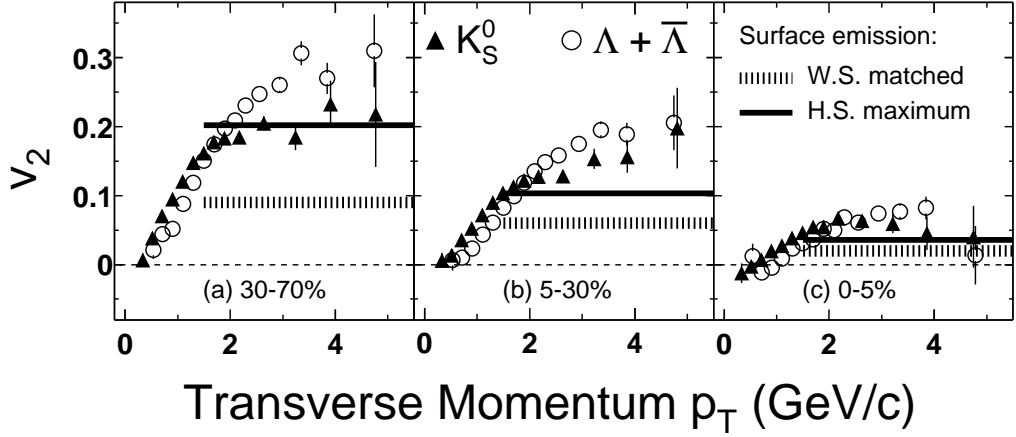}}
\caption[Comparison of $v_2$ and surface emission]{ Comparison of
$K_S^0$ and $\Lambda+\overline{\Lambda}$ $v_2$ with expectations
from surface emission models (see text). } \label{fig:surf}
\end{figure}
Figure~\ref{fig:surf} shows $v_2$ for $K_S^0$ and
$\Lambda+\overline{\Lambda}$ from the 30--70\%, 5--30\%, and
0--5\% centrality intervals along with calculations of $v_2$ from
two surface emission scenarios: ``W.S. matched'' uses a
Woods-Saxon distributions in the calculation of the nuclear
overlap function and matches the energy loss to the observed
suppression of neutral pion production~\cite{jia} while ``H.S.
maximum'' uses an analytic function representing pure surface
emission---infinite energy loss---from a hard sphere overlap
geometry~\cite{Voloshin:2002wa}. The models are not in good
agreement with our measurements. We do not, however, rule out the
surface emission scenario for $p_T$ above 4.5~GeV/c. We also note
that for $p_T$ from 1.5--4.5~GeV/c, it may be that kaons are
produced predominantly from hard processes in a surface volume
while (anti-)lambdas in the same range are produced by soft
processes. The magnitude of the kaon $v_2$ in this region,
however, is much larger than would be expected from the surface
emission model with a realistic nuclear overlap density. Stronger
conclusions on this point can be drawn from a larger data sample
and more extensive studies of possible non-flow contributions to
identified particle $v_2$.

In the case that particles are produced from hard
processes---either near the surface or throughout the entire
volume---their spectra are expected to be well represented by a
power-law function~\cite{powerlaw:Albajar:1990an}. In
Section~\ref{sec:regimes} we use the power-law function
\begin{equation}
\frac{d^3n}{dp^3} \sim C(1+\frac{p_T}{p_0})^{-\alpha}
\label{eq:powerlaw}
\end{equation}
to parameterize the hard part of the $K_S^0$ and
$\Lambda+\overline{\Lambda}$ spectrum.

\section{Transverse Momentum Regimes}
\indent\label{sec:regimes}

In what follows, we study the two component nature of the $K_S^0$
and $\Lambda+\overline{\Lambda}$ $p_T$ spectra.  We seek to
delineate the boundary between the soft and hard region of the
spectra.

\begin{figure}[htbp]
\centering\mbox{
\includegraphics[width=1.00\textwidth]{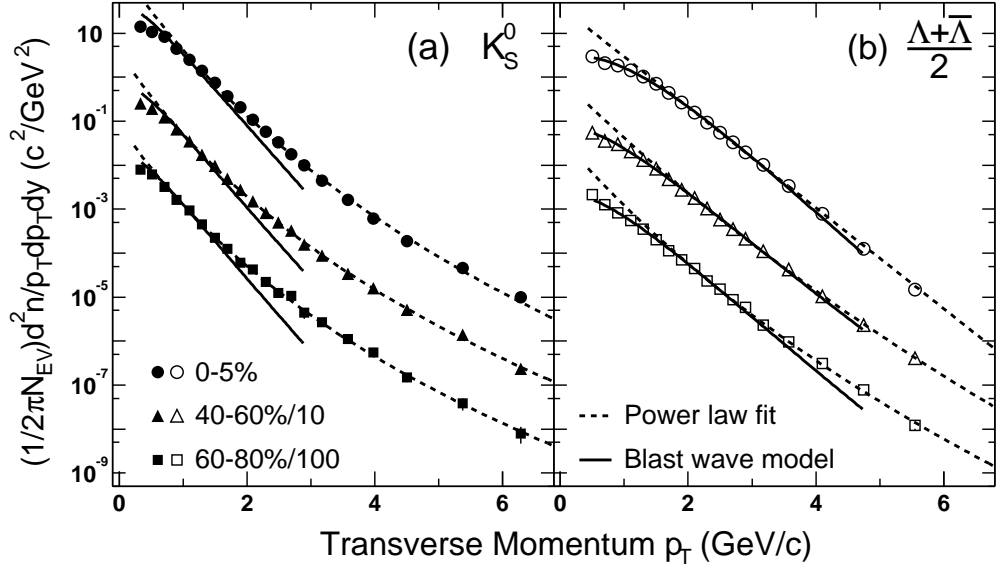}}
\caption[Power-law and blast-wave fits to spectra]{ Blast wave and
power-law fits to the $K_S^0$ and $\Lambda+\overline{\Lambda}$
spectra.  Fitting parameters are listed in Table~\ref{tab:fits}.}
\label{fig:specfit}
\end{figure}

In Reference~\cite{Sorensen:2003wi} we stressed the effectiveness
of the combination of $R_{CP}$ and $v_2$ for mapping out the
transition between the region dominated by soft processes and the
region dominated by hard processes. More detailed calculations
making use of these ideas in
Reference~\cite{ptcross:Hirano:2003pw} are in relatively good
agreement with the available data. We use hydrodynamical model
inspired blast wave functions and pQCD motivated power-law
functions to fit the $K_S^0$ and $\Lambda+\overline{\Lambda}$
spectra with the intention of extracting the value of $p_T$ where
the soft-to-hard crossover occurs. Figure~\ref{fig:specfit} shows
blast wave and power-law fits to the $K_S^0$ and
$\Lambda+\overline{\Lambda}$ spectra for three centrality
intervals. The fit parameters are listed in Table~\ref{tab:fits}.
\begin{table}[hbt]
\centering\begin{tabular}{l|c|c|c|c|c|c} \hline \hline
\multicolumn{1}{c|}{Fit} & \multicolumn{3}{c|}{$K_S^0$} & \multicolumn{3}{c}{$\Lambda(\overline{\Lambda})$} \\
\multicolumn{1}{c|}{Parameters} & \multicolumn{1}{c}{\;~0--5\%~\;} & \multicolumn{1}{c}{40--60\%} & \multicolumn{1}{c|}{60--80\%} & \multicolumn{1}{c}{~0--5\%~} & \multicolumn{1}{c}{40--60\%} & \multicolumn{1}{c}{60--80\%} \\
\hline
$C$           & 330.11 & 3.6785 & 0.0829 & 40.914  & 0.9832 & 0.0424 \\
$p_0$~(GeV/c) & 3.280  & 2.786  & 2.717  & 2.12e+5 & 10.38  & 5.691  \\
$\alpha$           & 16.43  & 13.96  & 13.41  & 5.58e+5 & 34.25  & 21.91  \\
\hline
$T$~(MeV)     & 140.7  & 179.1  & 175.5  & 136.3   & 258.7  & 223.2  \\
$\rho_0$      & 0.686  & 0.499  & 0.489  & 0.866   & 0.499  & 0.412  \\
\hline \hline
\end{tabular}
\caption[Spectra fits] {Fitting parameters for power-law (first
three rows) and blast wave (last two rows) parameterizations of
the $K_S^0$ and $\Lambda+\overline{\Lambda}$
spectra.}\label{tab:fits}
\end{table}

Radial flow, as established in hydrodynamical models for example,
can lead to a mass dependence in the soft-to-hard $p_T$ crossover
$p_{T,cross}$. Reference~\cite{ptcross:Hirano:2003pw} uses the
$p_T$ where the yields from hard and soft production processes
become equivalent to define $p_{T,cross}$. Using hydrodynamic
model calculations for the soft contribution, and a pQCD parton
model---incorporating energy loss, gluon shadowing, and initial
state rescattering---for the hard component, they calculate
$p_{T,cross} = 1.8$, 2.7, and 3.7~GeV/c for pions, kaons, and
protons respectively. From our blast-wave and power-law fits,
however, judging by the applicability of the fit functions we find
that $p_{T,cross}$ for kaons in central events is closer to
1.8~GeV/c, and consistent with the pion $p_{T,cross}$ apparent in
the pion spectrum measured by
PHENIX~\cite{PhenixPid:Adler:2003cb}. Given the $p_T$ reach of our
measurements it's difficult to determine $p_{T,cross}$ for lambdas
from spectra fits alone. We can, however, conclude that the lambda
$p_{T,cross}$ is greater than 3 GeV/c: within a range that would
be consistent with the proton $p_{T,cross}$ quoted in
Reference~\cite{ptcross:Hirano:2003pw}.

\begin{figure}[htb]
\centering\mbox{
\includegraphics[width=0.60\textwidth]{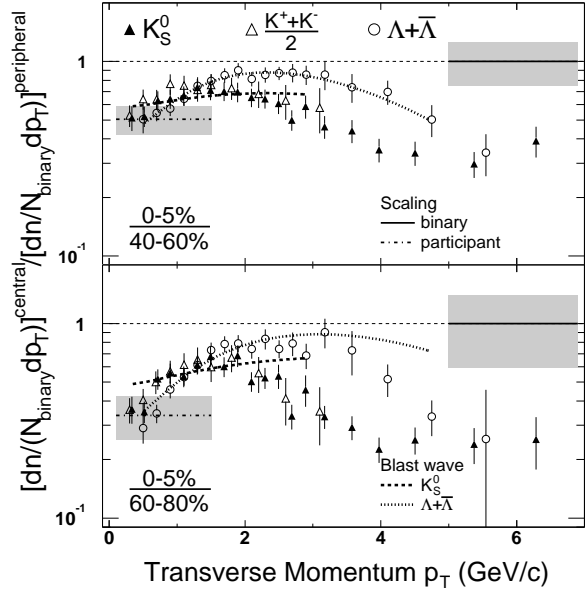}}
\caption[Blast-wave fits to $R_{CP}$]{ The $R_{CP}$ of $K_S^0$ and
$\Lambda+\overline{\Lambda}$ with blast wave curves extracted from
fits to the particles spectra. } \label{fig:Rcpblast}
\end{figure}
The $p_T$ where the $\Lambda+\overline{\Lambda}$ $R_{CP}$
(0--5\%/60--80\%) begins to decrease in Figure~\ref{fig:Rcpblast}
(bottom panel) provides clearer evidence for the cross-over and
suggest that for (anti-)lambdas $p_{T,cross} \sim 3.5$~GeV/c. We
note, however, that the blast wave fits to the
$\Lambda+\overline{\Lambda}$ spectra for the 0--5\% and 40--60\%
centrality intervals are in good agreement throughout the fit
range so that we cannot exclude the possibility that the blast
wave parameterization describes lambda production well throughout
the measured $p_T$. This is also apparent in the blast wave curves
for $R_{CP}$ using those same centralities (the top panel of
Figure~\ref{fig:Rcpblast}). To determine whether or not this
indicates that the assumptions of the hydrodynamical models are
valid from 0--60\% centrality but break down in the 80--100\%
interval will require further study.

The similarity of the $p_{T,cross}$ for pions and kaons may
indicate that the number of quarks rather than the mass of the
hadron provides the relevant scale for the transition between
regions of predominantly soft and predominantly hard production
processes. This empirical observation is consistent with a picture
where at intermediate $p_T$, hadronization occurs through the
coalescence of co-moving partons. The recent observation that
$R_{CP}$ for the $\phi$ meson (m=1.019~GeV/c$^2$) is closer to the
pion and kaon $R_{CP}$ than the proton or lambda
$R_{CP}$~\cite{phipaper} supports this picture.

\begin{figure}[htb]
\centering\mbox{
\includegraphics[width=0.60\textwidth]{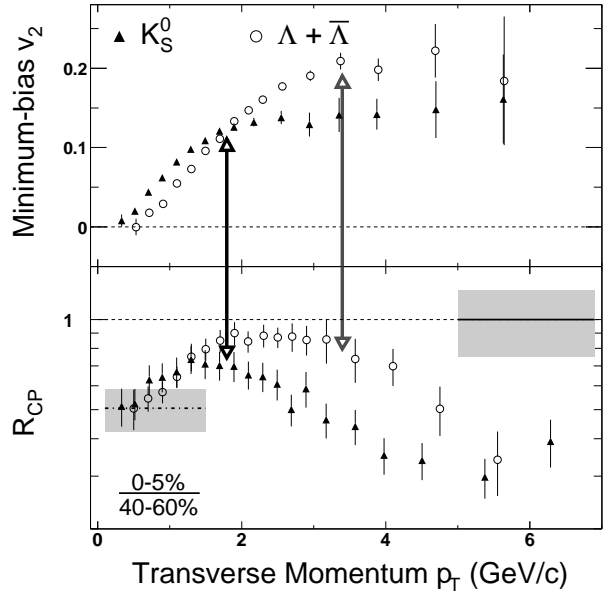}}
\caption[Similarity of $R_{CP}$ and $v_2$]{ The $v_2$ and $R_{CP}$
of $K_S^0$ and $\Lambda+\overline{\Lambda}$. The vertical lines
emphasize the correlation between the saturation of $v_2$ and the
decline of $R_{CP}$.} \label{fig:scales}
\end{figure}
To emphasize the correlation between the behavior of identified
particle $v_2$ and identified particle $R_{CP}$ we plot $v_2$ and
$R_{CP}$ together in Figure~\ref{fig:scales}. Although $R_{CP}$
depends only on the yield in the central and peripheral bins, and
the $v_2$ in Figure~\ref{fig:scales} is from a minimum-bias
centrality interval, the two parameters may still be intimately
related. Differential elliptic flow $v_2(p_T)$ measures the
difference between the $p_T$ spectrum of particles emitted in the
direction of the reaction plane (in-plane) to that of particles
emitted perpendicular the reaction plane (out-of-plane). In
hydrodynamic models, we expect the pressure gradient to be larger
in the in-plane direction than the out-of-plane direction. Since
the average pressure gradient in central collisions is expected to
be larger than in peripheral collisions in a the hydrodynamical
picture the ratio $R_{CP}$ and $v_2$ \textbf{should} have a
similar $p_T$ and particle-type dependencies. Coalescence or
recombination models should also give a similar correlation
between $v_2$ and $R_{CP}$. Both measures reflect relative
probabilities for forming hadrons: The probabilities depend on the
phase-space density of partons, and the phase-space density varies
with centrality and azimuthal angle. In Appendix~\ref{app:var} we
introduce a formalism more suited to studying the centrality
dependence of particle yields in heavy-ion collisions across a
broad $p_T$ range and for a variety of system sizes.



\section{Initial State Effects}

Nuclear enhancement of hadron yields (relative to
number-of-binary-collisions scaling) for lower beam energies has
been observed at intermediate $p_T$ in p+A collisions---with a
larger enhancement for baryons than mesons~\cite{Straub:1992xd}.
The empirical observation of an enhancement in the scaled yield is
known as the Cronin effect~\cite{Cronin:1973fd,Antreasyan:1979cw}.
The Cronin effect is generally attributed to multiple scatterings
between the projectile partons and the cold nuclear matter in the
target~\cite{Accardi:2002ik}. The ratio of the scaled yields in
d+Au and p+p collisions $R_{dAu}$ has been measured for neutral
pions~\cite{dA:Adler:2003ii} and inclusive charged hadrons at
RHIC~\cite{dA:Adams:2003im,dA:Back:2003ns,dA:Adler:2003ii,dA:Arsene:2003yk}.
At intermediate $p_T$, the neutral pion $R_{dAu}$ is consistent
with one (no Cronin effect) or a small enhancement. The inclusive
charged hadron $R_{dAu}$ shows an enhancement of roughly 35\%
however, indicating the presence of a particle-type dependent
Cronin effect at RHIC energy.
The existence of an enhancement has not been established for
kaons, lambdas, or antilambdas, but the Cronin effect cannot be
ruled out as the origin of the dependence of $R_{CP}$ on particle
type.

Theoretical calculations, such as those involving initial parton
scatterings off cold nuclear matter (e.g.~\cite{Lev:1983hh}),
don't reproduce the particle-type dependence of the enhancement
factor observed in p+A collisions. The inadequate particle
dependence in these calculations may arise from the fact that
these models only deal with initial parton scatterings while the
observed hadrons are formed at the late stage of the collision. It
is argued in Reference~\cite{Accardi:2003jh}, that the
fragmentation process can distort the features of the parton level
Cronin effect. As such, the strong particle-type dependence in
$R_{p(d)A}$ may indicate a nuclear modification of the parton
fragmentation into baryons and mesons or alternatively, the
presence in p(d)+A collisions of a multi-parton particle formation
mechanism such as coalescence~\cite{coal:Greco:2003xt} or
recombination~\cite{reco:Best:Fries:2003kq}. These mechanisms are
beyond the framework of many existing theoretical models for the
Cronin effect.

\section{Hadronization of Dense Matter}
\indent

The varied initial conditions in heavy-ion collisions may provide
important phenomenological clues to the nature of hadron
formation---particularly baryon formation. Observing how the
presence of dense nuclear matter influences hadronization will
almost certainly help clarify by what processes three quarks can
come together to form a baryon.

Modification of the fragmentation functions $D_{a \rightarrow
h}(z)$ has been offered as a strategy to account for the influence
of the surrounding matter on the fragmentation
process~\cite{FragMod:Guo:2000nz,FragMod:Wang:2001if}. When this
strategy leads to a rescaling of $z$ it should affect all hadrons
in a similar way~\cite{reco:Best:Fries:2003kq}---a behavior that
is inconsistent with the measurements presented in this thesis. To
account for species dependence observed in $v_2$ and $R_{CP}$ a
new understanding of hadronization may be necessary.  In addition,
we note that in the complex and highly interacting systems created
in heavy-ion collisions the assumption of factorization implicit
in Equation~\ref{eq:frag} may no longer be valid.

Although the observables of heavy-ion collisions---$R_{CP}$,
particle ratios, and $v_2$ for example---and their variation with
particle-type challenge current theoretical models, they also hint
at possible resolutions.

We note for example, that the absence of a significant suppression
with respect to binary scaling of the $\Lambda+\overline{\Lambda}$
yield at intermediate $p_T$ in central Au+Au collisions may
indicate the presence of dynamics beyond parton energy loss and
standard fragmentation.
The larger $\Lambda+\overline{\Lambda}$ $R_{CP}$ at intermediate
$p_T$ means that the (anti-)lambda yield increases with the parton
density of the collision fireball much faster than the meson
yield. The rate of increase of the proton
yield~\cite{proton:Adler:2003kg} and the multi-strange baryon
yield ($\Xi+\overline{\Xi}$)~\cite{hlongsqm} are found to be
similar to that of (anti-)lambda. In addition, at intermediate
$p_T$, the centrality dependence of the $\phi$-meson production is
more similar to that of kaons than (anti-)lambdas~\cite{phipaper}.
Stronger dependence on parton density for baryon production is
naturally expected from multi-parton production mechanisms such as
gluon junctions~\cite{Vance:1999pr}, quark
coalescence~\cite{coal:Molnar:2003ff}, or
recombination~\cite{coal:Fries:2003vb}.

\begin{figure}[htbp]
\centering\mbox{
\includegraphics[width=0.80\textwidth]{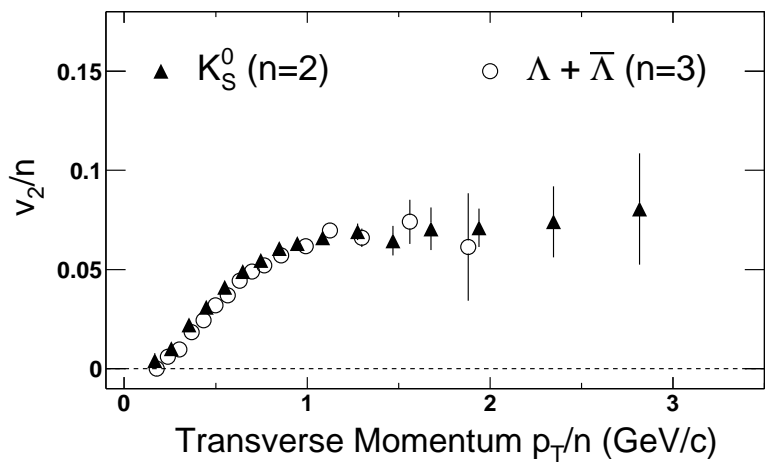}}
\caption[Constituent-quark-number scaling (i)]{ The $v_2$
parameter for $K_{S}^{0}$ and $\Lambda + \overline{\Lambda}$
scaled by the number of constituent quarks (n) and plotted versus
$p_{T}/$n. } \label{fig:v2scaling}
\end{figure}

As discussed in Section~\ref{sec:regimes}, it is apparently the
number of constituent quarks in a hadron that predominantly
determines the characteristics of it's spectra at intermediate
$p_T$. Figure~\ref{fig:v2scaling} shows $v_2$ of $K_{S}^{0}$ and
$\Lambda+\overline{\Lambda}$ as a function of $p_T$, where the
$v_2$ and $p_T$ values have been scaled by the number of
constituent quarks (n).
While $v_2$ is significantly different for $K_S^0$ and
$\Lambda+\overline{\Lambda}$, within errors, $v_2$/n vs $p_T$/n is
the same for both species above $p_T$/n $\sim 0.8$~GeV/c.
This behavior is consistent with a scenario where hadrons at
intermediate $p_T$ are formed from bulk partonic matter by
coalescence of co-moving quarks: in this case $v_2/$n vs $p_T/$n
reveals the momentum space azimuthal anisotropy that partons
develop from the collision ellipsoid, e.g.
Reference~\cite{coal:Molnar:2003ff}.
Hadronization by coalescence and the large partonic anisotropy
subsequently inferred from our empirical observations would both
argue strongly for the existence of a strongly interacting early
partonic stage.

\begin{figure}[htbp]
\centering\mbox{
\includegraphics[width=0.80\textwidth]{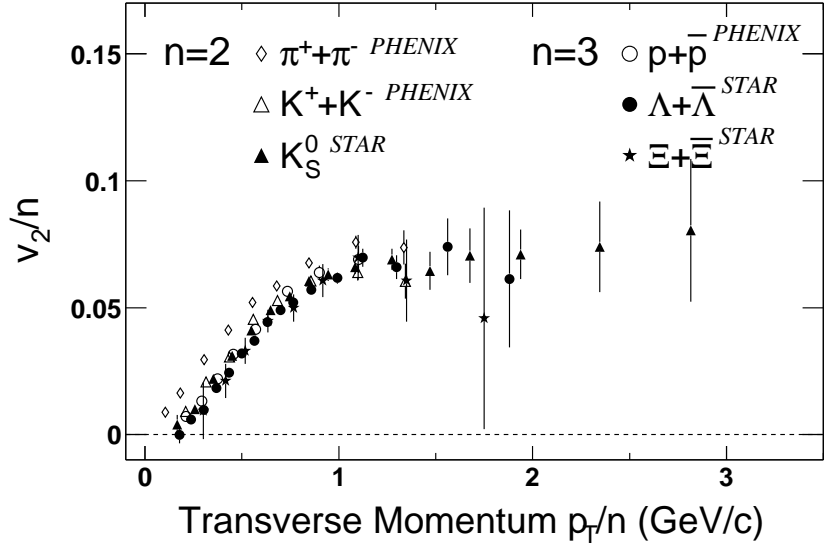}}
\caption[Constituent-quark-number scaling (ii)]{ The $v_2$
parameter for $K_{S}^{0}$ and $\Lambda + \overline{\Lambda}$
scaled by the number of constituent quarks (n) and plotted versus
$p_{T}/$n. } \label{fig:v2scaling2}
\end{figure}

The suggestive scaling behavior if Figure~\ref{fig:v2scaling} can
be tested by including measurements of $v_2$ for other identified
particles that extend into the intermediate $p_T$ region.
Figure~\ref{fig:v2scaling2} shows $v_2/$n versus $p_T/$n for all
identified particles currently available from RHIC
experiments~\cite{PhenixPidv2:Esumi:2002vy,msbv2}. At low $p_T$
($p_T < 1$~GeV/c), where hydrodynamic calculations were already
seen to reproduce the measured $v_2$ well, and at $p_T > 6$~GeV/c,
the signatures of coalescence are not expected to be prominent.
The deviation of the scaled pion $v_2$ may prove to be problematic
or it may just reflect the break-down of coalescence at low $p_T$.
Otherwise, the results in Figure~\ref{fig:v2scaling2} are
consistent with the expectations of constituent-quark-number
scaling and provide strong evidence for coalescence and the
existence of a quark-gluon plasma in the early stage of the
collision system.

For n$\times p_{T,hard}$, where $p_{T,hard}$ is a momentum for
which the spectra of the underlying partons follows a power-law,
the yield of hadrons from coalescence of n quarks will no longer
dominate the yield of hadrons from fragmentation of partons with
momentum $p_{T,hard}/z$. Based on this picture, we would expect
the $v_2$ of all hadrons at high $p_T$ ($p_T > 6$) to take on the
unscaled value of the parton $v_2$. Assuming the $v_2/$n values
from Figure~\ref{fig:v2scaling2} reflect the parton $v_2$ we would
conclude that at high $p_T$ (where we expect $v_2^{meson} =
v_2^{baryon} = v_2^{quark}$), $v_2$ of all light-flavored
particles (particles having only u, d, or s constituent quarks)
will take on the same value $v_{2,hard} \sim 0.07$. This assumes a
saturated parton $v_2$ at high $p_T$: an assumption that is
consistent with expectations from
transport~\cite{trans:Molnar:2001ux}, and surface
emission~\cite{Surf:Shuryak:2001me} models.

\section{Conclusions}
\indent

In summary, we have reported the measurement of $v_2$ and $R_{CP}$
up to $p_T$ of $6.0$~GeV/c for kaons and $\Lambda +
\overline{\Lambda}$ from Au+Au collisions at
$\sqrt{s_{_{NN}}}=200$~GeV.
At low $p_T$, hydrodynamic model calculations agree well with
$v_2$ for $K_S^0$ and $\Lambda + \overline{\Lambda}$.
At intermediate $p_T$, however, hydrodynamics no longer describes
the particle production well.
For $K_{S}^{0}$, $v_2$ saturates earlier and at a lower value than
for $\Lambda + \overline{\Lambda}$.
In addition, $R_{CP}$ shows that the kaon yield in central
collisions is suppressed more than the (anti-)lambda yield.
At intermediate $p_T$, the $\Lambda+\overline{\Lambda}$ yield in
central Au+Au collisions is close to expectations from binary
scaling of peripheral Au+Au collisions.
At high $p_T$, the $R_{CP}$ of $K_{S}^{0}$,
$\Lambda+\overline{\Lambda}$ and charged hadrons are approaching
the same value.
The measured features in the kaon and (anti-)lambda $v_2$ and
$R_{CP}$ may indicate the presence of dynamics beyond the
framework of parton energy loss followed by fragmentation.
The particle- and $p_T$-dependence of $v_2$ and $R_{CP}$,
particularly at intermediate $p_T$, provides a unique means to
investigate the anisotropy and hadronization of the bulk dense
matter formed in nucleus-nucleus collisions at RHIC.

In Au+Au collisions that copiously populate phase space, it would
be naive to expect a single parton description of hadronization to
remain valid. Our measurements verify that multi-parton dynamics
significantly alter production mechanisms: giving rise to a
number-of-constituent-quark dependence for particle spectra,
$R_{CP}$, and $v_2$ at intermediate $p_T$. From the mass
dependence of $v_2$ at low $p_T$ and the partonic $v_2$ inferred
from the scaled $v_2$ at intermediate $p_T$ it appears that the
existence of a thermalized partonic state at RHIC is not just
likely but perhaps unavoidable.

\section{Future Directions}
\indent


The future direction of our studies at RHIC are clear:
verification of the creation of a QGP, then characterization of
the QGP. The measurement of identified particle $v_2$ and $R_{CP}$
will play a prominent role in both of these endeavors. The full
implication of the measurements presented in this thesis will
become clearer when $v_2$ and $R_{CP}$ have been measured well
into the hard region ($p_T \approx 8$~GeV/c) for $\pi^0$, $K_S^0$,
and $\phi$ mesons and for $p$, $\Lambda$, $\Xi$, and $\Omega$
baryons. Several of these measurements are already available and
most of the others will become available in the near future. The
measurements already made argue strongly for the existence of a
thermalized partonic state that at least partly hadronizes by
coalescence. With the larger data sets collected during future
RHIC runs $v_2$ should be measured for identified particles using
a cumulant analysis. The cumulant analysis can easily be adapted
for statistically identified particles when only one particle
candidate in a given $p_T$ interval is considered from each event.

Once a suite of identified meson and baryon measurements has been
made to sufficiently high $p_T$, the first priority should be to
conduct a system-size scan. RHIC is already operating at its top
energy for Au+Au collisions and running with lower energy will
drastically reduce high $p_T$ particle yields---essentially
eliminating several of our most promising QGP signatures. Instead,
it will be more fruitful to study collisions with fewer
participants. This will provide three exciting opportunities: A
chance to search for a region where the QGP state is turned on or
off, a chance to run at even higher energy with the existing RHIC
facility (Si+Si top energy would be $\sqrt{s_{_{NN}}}=250$~GeV)
and the chance to search for rare/exotic states (\textit{i.e.}
glue-balls, penta-quarks, etc.) in collisions that generate less
combinatorial background. The reduction in the combinatorial
background will benefit many identified particle measurements.
Cu+Cu collisions may be preferable since they should still allow
reliable measurements of $v_2$ while probing smaller system sizes.

\appendix
\chapter{Collision Geometry and the Source Eccentricity}
\label{app:geo}

Here we provide information about the geometry and coordinate
systems of relativistic heavy-ion collisions.

\begin{figure}[htpb]
\centering\mbox{
\includegraphics[width=1.00\textwidth]{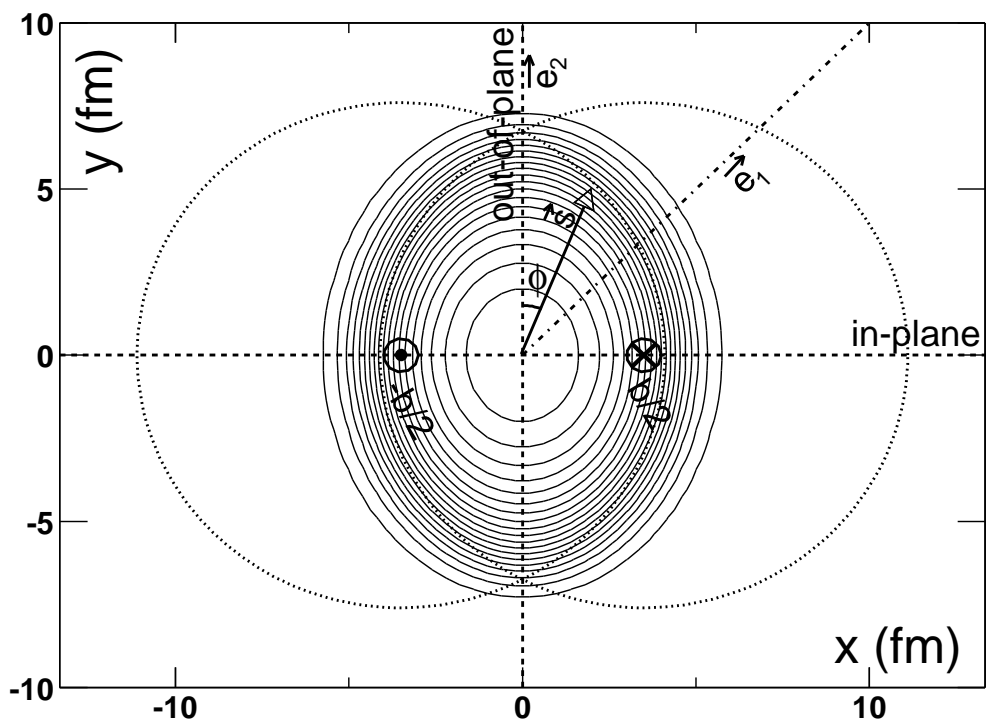}}
\caption[Coordinate system of the transverse plane]{ Coordinate
system for the transverse plane in heavy-ion collisions. Contours
of the overlap density are also shown. Here $\phi$ is the
azimuthal angle with respect to the $y$-axis. Also shown are
generalized coordinates $\vec{e_1}$ and $\vec{e_2}$ used to
calculate the fourth-harmonic eccentricity where the
($y$-)$\vec{e_2}$-axis is the only uniquely defined direction in
the transverse plane.} \label{fig:geoazimuth}
\end{figure}
Figure~\ref{fig:geoazimuth} gives the coordinates in the azimuthal
plane of the collisions. The beam direction ($z$-axis) is in or
out of the page as indicated by the $\otimes$ or $\odot$
respectively. The ($y$-)$\vec{e_2}$-axis is uniquely defined
(using the right-hand-rule) by the collision axis and the
directions of the colliding beams. When calculating $v_2 = \langle
\cos(2\phi) \rangle$, $\phi$ is defined with respect to the
reaction plane (the $x$-axis). For our eccentricity calculations
we will define $\phi$ with respect to the $y$-axis. In this way,
$v_2$ and the $2^{nd}$-harmonic eccentricity will have the same
sign.

The density of Au nucleus is represented by a Woods-Saxon
distribution,
\begin{equation}
\rho_A = \rho_0\frac{1}{1+\exp(\frac{r-R_A}{\xi})}
\end{equation}
where $r=\sqrt{s^2+z^2}$, $R_A = 1.12 \times A^{1/3}$, with
$\rho_0 = 0.159$~GeV/fm$^{3}$ and $\xi=0.535$ fm. The thickness
function of nucleus $A$ ($T_A$) is the nuclear density integrated
over the $z$ direction: $T_A(\vec{s})=\int dz\rho_A(a,\vec{s})$.
The density of nucleons participating in the collision
(\textit{wounded nucleons}) is given by:
\begin{equation}
n_{part}(x,y) = T_A \times (1-e^{-T_B\sigma_{NN}}) + T_B \times
(1-e^{-T_A\sigma_{NN}}),
\end{equation}
where for collisions at $\sqrt{s_{_{NN}}}=200$~GeV we use
$\sigma_{NN} \approx 42$. The density of binary nucleon-nucleon
collisions is taken as: $n_{bin}(x,y) = \sigma_{NN} \times T_A
\times T_B$. In the plots that follow, we use $n_{part}$ as the
density in the overlap region.

\begin{figure}[htpb]
\centering\mbox{
\includegraphics[width=1.00\textwidth]{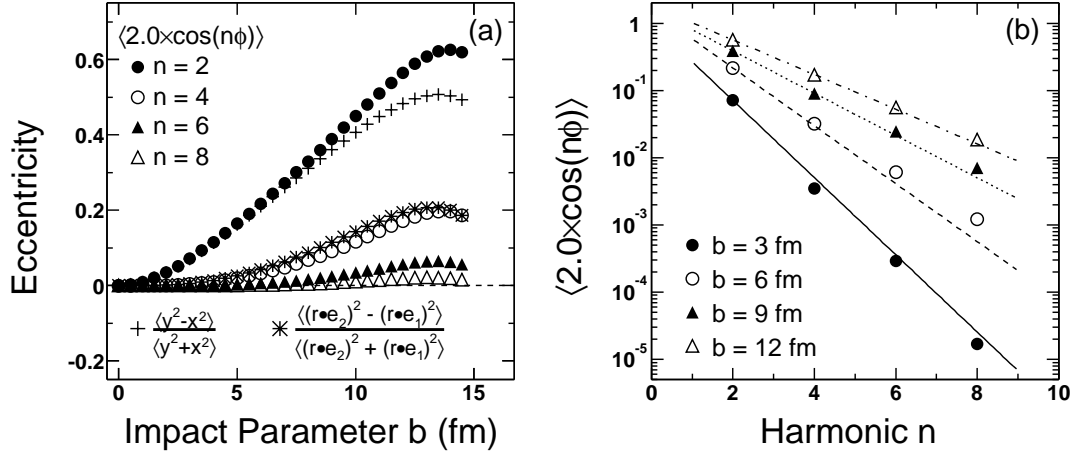}}
\caption[Eccentricity estimates (i)]{ Left: Analytic calculations
of the even n$th$-harmonic eccentricities versus impact parameter.
Right: Relative magnitudes of harmonics for various impact
parameters.  For more central collisions the higher harmonics fall
off more quickly.} \label{fig:epsilons}
\end{figure}
\begin{figure}[htpb]
\centering\mbox{
\includegraphics[width=1.00\textwidth]{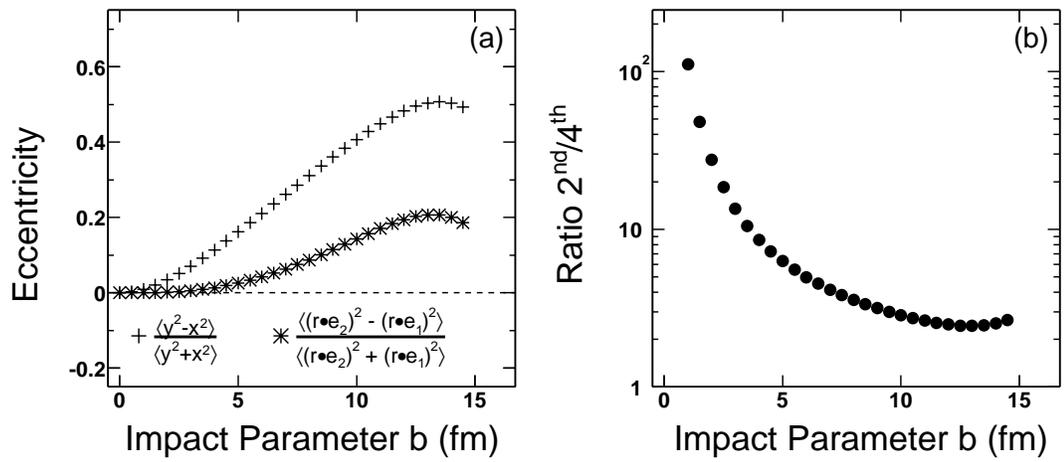}}
\caption[Eccentricity estimates (ii)]{ Left: Analytic calculations
of the $2^{nd}$-harmonic and $4^{th}$-harmonic eccentricities.
Right: The ratio of the second and fourth harmonic
eccentricities.} \label{fig:epsilons2}
\end{figure}

To better understand the initial spatial asymmetry of the
collision regions in Figure~\ref{fig:epsilons} we show
calculations of eccentricities defined for even harmonics (the odd
harmonic eccentricities are all zero). The most commonly used
definition of the second harmonic eccentricity is $\epsilon_2 =
\langle y^2-x^2\rangle/\langle y^2+x^2 \rangle$.  In the figure we
also include calculations of eccentricities from $2\langle
\cos(n\phi)\rangle$ and a calculation of the $4^{th}$-harmonic
eccentricity using the appropriate coordinate system as shown in
Figure~\ref{fig:geoazimuth}. We note that the magnitude of higher
harmonic eccentricities falls off more quickly in central
collisions than peripheral collisions (panel b). For peripheral
collisions the higher harmonics are appreciably larger. The
different methods for calculating the eccentricity do not yield
the same result.  In Figure~\ref{fig:epsilons2} we show the
eccentricity from the more standard calculation. We also show the
ratio of the $2^{nd}$- and $4^{th}$-harmonic eccentricities versus
impact parameter (panel b).
\clearpage

\begin{figure}[htpb]
\centering\mbox{
\includegraphics[width=1.00\textwidth]{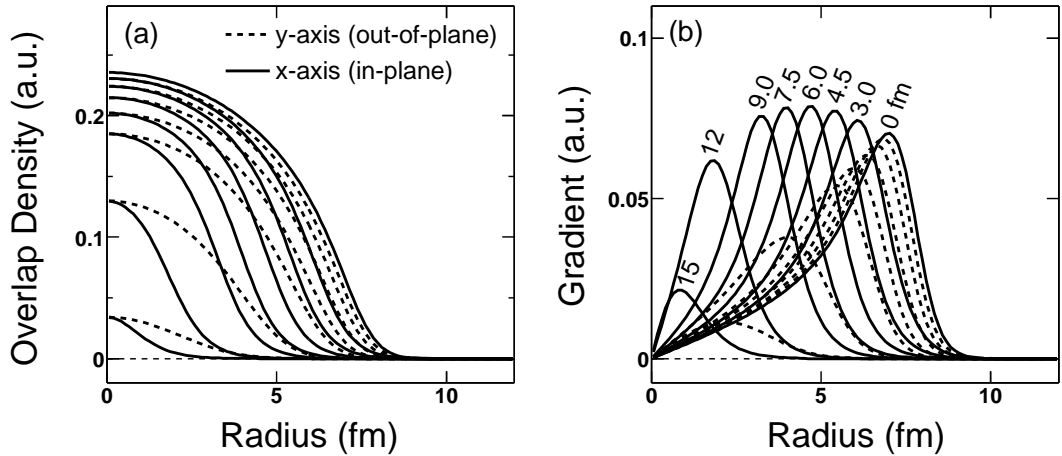}}
\caption[Density profiles]{ Left: Collision overlap densities
calculated for various impact parameters (listed in panel b) and
using a Woods-Saxon nuclear density profile and a wounded nucleon
model. Right: The gradients of the density profiles in panel a. We
take the magnitude of the gradient to estimate of the initial
pressures in the system.} \label{fig:gradient}
\end{figure}

In Figure~\ref{fig:gradient} (a) we plot profiles (along the x and
y axis) of the overlap density for the impact parameters listed in
panel b. In Figure~\ref{fig:gradient} (b) we plot the gradient of
the profiles in panel a. In a hydrodynamical picture---once a
system has thermalized---the density changes will generate
pressure gradients that drive the flow of matter and induce
collective motion. We note that Figure~\ref{fig:gradient} (b)
suggests that the greatest initial pressure gradients will be
found in the in-plane direction of collisions having impact
parameter $b = 6$~fm. This impact parameter corresponds roughly to
collisions within the 10--20\% centrality interval.

\clearpage

\begin{figure}[htpb]
\centering\mbox{
\includegraphics[width=1.00\textwidth]{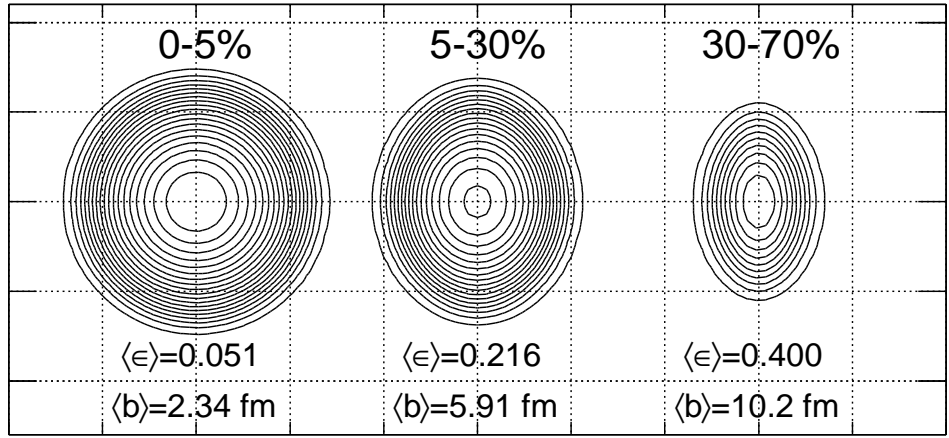}}
\caption[Elliptic flow geometry]{ Overlap densities, mean
eccentricities, and mean impact parameters for the centrality
intervals used for our $v_2$ analysis.  } \label{fig:geov2}
\end{figure}

In Figure~\ref{fig:geov2} we show the geometries corresponding to
the centrality intervals used in our $v_2$ measurements. The mean
eccentricities and impact parameters in this figure were
calculated as in Reference~\cite{aihong}\footnote{The author
thanks A. Tang for his help with these calculations.}.

\clearpage

\begin{figure}[htpb]
\resizebox{.5\textwidth}{!}{\includegraphics{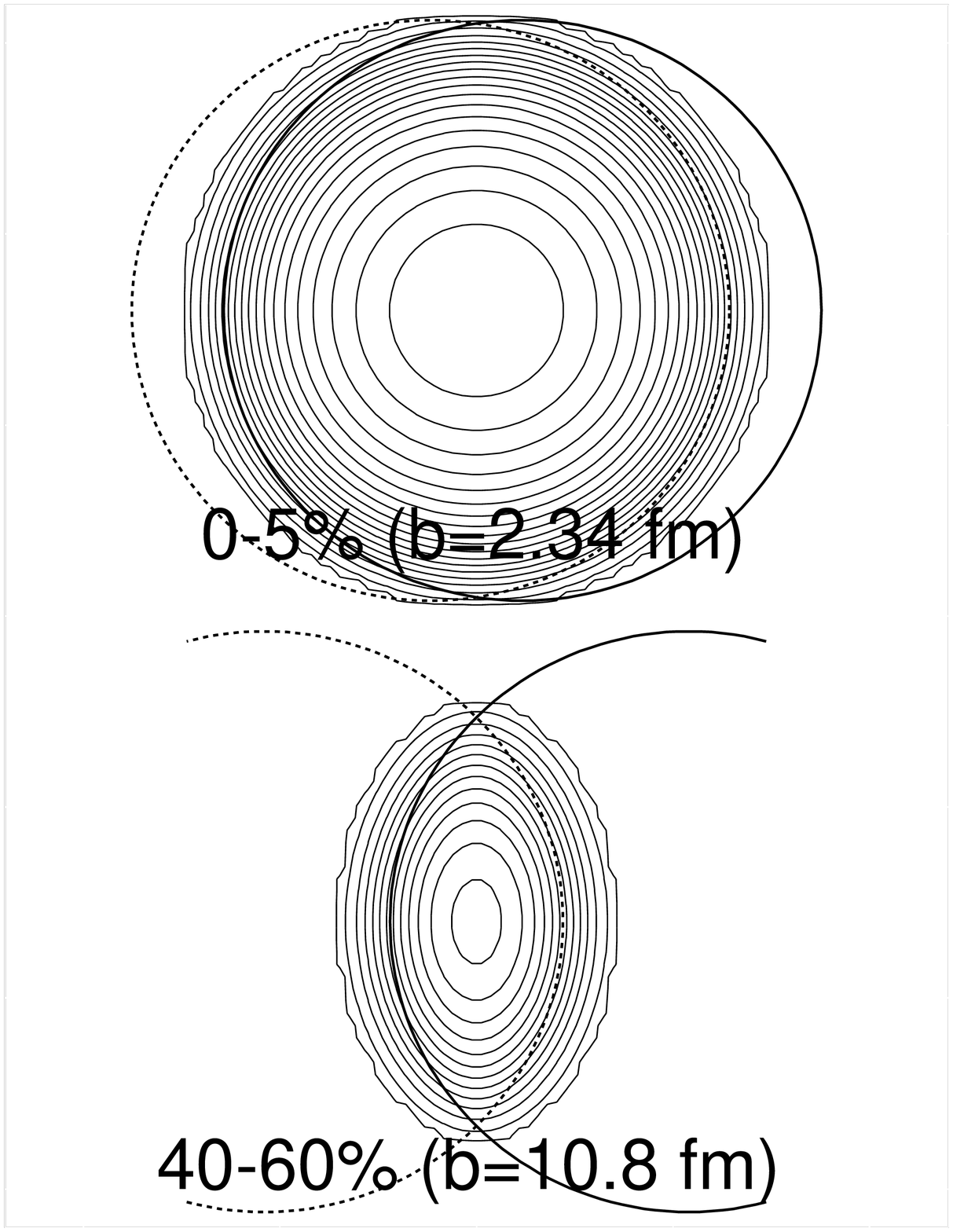}}
\resizebox{.5\textwidth}{!}{\includegraphics{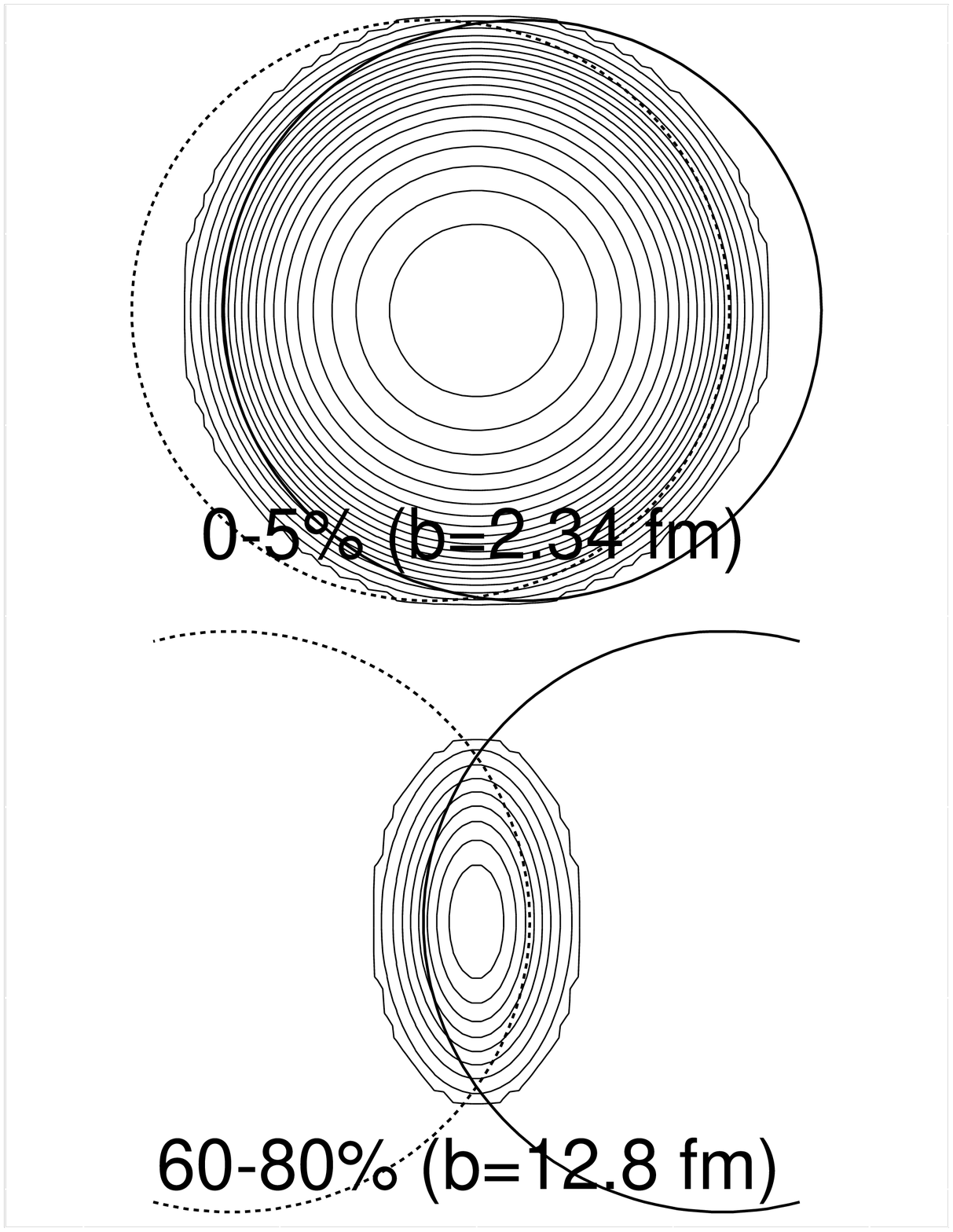}}
\caption[Nuclear modification geometry] { Configurations for the
collision centrality intervals used to calculate $R_{CP}$. }
\label{fig:georcp}
\end{figure}
In Figure~\ref{fig:georcp} we show the geometry of the centrality
intervals used in the calculation of $R_{CP}$.

\chapter{Kinematic Variables}
\label{app:kinematic}

The azimuthal components of a particles momentum $p_x$ and $p_y$
are used to define its transverse momentum,
\begin{equation}
p_T \equiv \sqrt{p_x^2 + p_y^2}.
\end{equation}
The transverse mass/energy of a particle having mass $m_0$ is
\begin{equation}
m_T \equiv \sqrt{p_T^2 + m_0^2},
\end{equation}
so that the transverse kinetic energy of the particle is $m_T -
m_0$. The transverse kinetic energy is commonly used in place of
$p_T$.

In the lab frame, the azimuthal angle of a particles momentum is
simply $\phi_{lab} = \tan^{-1}(p_y/p_x)$.  When the reaction plane
can be measured the azimuthal angle can be measured with respect
to the transverse angle of the reaction plane $\Psi_{RP}$:
\begin{equation}
    \phi = \phi_{lab} - \Psi_{RP}.
\end{equation}
This is the $\phi$ shown in Figure~\ref{fig:geoazimuth} where the
reaction plane is aligned with the $x$-axis.

With the transverse coordinates defined all that remains is to
define a longitudinal variable. The rapidity $y$ is defined as
\begin{equation}
y \equiv \frac{1}{2}\ln \left(\frac{E + p_z}{E - p_z} \right)
\label{eq:rapidity}
\end{equation}
and is boost invariant. In the case that the momentum of a
particle is known but not its energy---typically because it's mass
is unknown---then the pseudo-rapidity $\eta=-\ln \tan\left(
\cos^{-1}(p_z/p)/2 \right)$ can be used. For $p \gg m$, $\eta
\approx y$.

The differential cross-section for particle production, is found
by counting the number of particles $d^3n$ produced in a phase
space element. It's advantageous to define the phase-space element
to be Lorentz invariant. The usual choice is the element
$dp_xdp_ydp_z/E$, so that the invariant cross-section is
\begin{equation}
E\frac{d^3n}{dp^3}=\frac{d^3n}{p_Tdp_Td\phi dy},
\end{equation}
where we've expressed the cross-section in terms of the variables
defined previously.

\chapter{An Improved Formalism for Studying the System Size Dependence of Nucleus-Nucleus Collisions}
\label{app:var}

Up to now the centrality dependence of particle yields in
heavy-ion collision has primarily been studied either by plotting
the spectra from different centralities in the same panel or by
forming the ratio
\begin{equation}
R_{CP}(p_T) = \frac{\left [dn/\left (\mathrm{N_{binary}}dp_T
\right ) \right ]^{\mathrm{central}}}{\left [dn/\left
(\mathrm{N_{binary}}dp_T \right ) \right ]^{\mathrm{peripheral}}}.
\end{equation}
This ratio is formed in order to test how different high $p_T$
particle production in central collisions is from peripheral
collisions where yields are expected to scale by the number of
binary collisions. Here we suggest another formalism that we
believe is better suited to the study of the centrality dependence
of particle production in a variety of heavy-ion collision systems
and across the entire $p_T$ range.

Particle yields scale predominantly with the number of
participating nucleons $\mathrm{N_{part}}$, not
$\mathrm{N_{binary}}$ and $\mathrm{N_{part}}$ is less model
dependent than $\mathrm{N_{binary}}$. For this reason we choose to
consider the yield of a particle $Y$ from collisions with impact
parameter $b$ scaled by $\mathrm{N_{part}}$:
\begin{equation}
y(b) \equiv Y(b)/\mathrm{N_{part}}.
\end{equation}
We can write $y(b)$ in terms of a Taylor expansion around a given
impact parameter $b_0$:
\begin{equation}
y(b) = y(b_0)\left(1 + \frac{1}{y(b_0)}\frac{\partial y}{\partial
b}(b-b_0) + \frac{1}{2y(b_0)}\frac{\partial^2 y}{\partial
b^2}(b-b_0)^2 + \dots \right).
\end{equation}
We label the terms $s_1 = \frac{1}{y(b_0)}\frac{\partial
y}{\partial b}$, $s_2 = \frac{1}{2y(b_0)}\frac{\partial^2
y}{\partial b^2}$, etc. so that $s_n =
\frac{1}{(n-1)!y(b_0)}\frac{\partial^n y}{\partial b^n}$. In this
case if the yield scales with $\mathrm{N_{part}}$ then $s_n = 0$
for all $n$. The term $s_1$ represents the linear dependence of
$y(b)$ and $s_2$ represents the quadratic dependence.

These measures have several benefits over the scaled ratio
$R_{CP}$. First we note that if one of the centrality parameters
$s_n$ is dominant then coalescence scaling would give
$s_n^{meson}(p_T) = 2\times s_n^{quark}(p_T/2)$ and
$s_n^{baryon}(p_T) = 3\times s_n^{quark}(p_T/3)$. Even if several
parameters must be taken into account trivial scaling rules
relating meson and baryon $s$ parameters can still be attained.
Second, we can easily define a minimum-bias $s_n$ measurement or
an $s_n$ measurement within a given centrality so that the
centrality dependence at different intervals and for different
systems can be more easily compared. For example, we can expand
$y(b)$ about $b_0=b_{max}/2.0$ and calculate the minimum-bias $s_1
= \frac{12 \langle b - b_0 \rangle}{b_{max}^2}$ (where the scaled
yield $y$ has been used to calculate the mean). These variables
may be particularly valuable for studying changes to the
systematic variation of the yield with system size. One would hope
to see a notable change in the $s$ parameters as the system size
increases to a size necessary to form a QGP.

A correction to the $s_n$ variables can be introduced to account
for the impact parameter resolution.

\chapter{The STAR Collaboration}
\begin{singlespace}
\begin{center}
\small{ J.~Adams$^3$, C.~Adler$^{11}$, Z.~Ahammed$^{24}$,
C.~Allgower$^{12}$, J.~Amonett$^{14}$, B.D.~Anderson$^{14}$,
M.~Anderson$^5$, D.~Arkhipkin$^{10}$, G.S.~Averichev$^{9}$,
J.~Balewski$^{12}$
% ...
\\
}

\end{center}
\begin{center}
\small{
% ...
$^3$University of Birmingham, Birmingham, United Kingdom\\
% ...
$^5$University of California, Davis, California 95616\\
% ...
$^{9}$Laboratoryfor High Energy (JINR), Dubna, Russia\\
$^{10}$Particle Physics Laboratory (JINR), Dubna, Russia\\
$^{11}$University of Frankfurt, Frankfurt, Germany\\
$^{12}$Indiana University, Bloomington, Indiana 47408\\
% ...
$^{14}$Kent State University, Kent, Ohio 44242\\
% ...
$^{24}$Purdue University, West Lafayette, Indiana 47907\\
% ...
}
\end{center}
\end{singlespace}

\bibliography {thesis}
\bibliographystyle {uclathes}
\end {document}